\documentclass[12pt]{article}

\usepackage{amsmath,amsthm,amssymb,amsfonts,bbm,framed}
\usepackage{graphicx}
\usepackage{subcaption}
\usepackage{hyperref}
\usepackage{tikz}
\usepackage{tabularx}
\usepackage[group-separator={,}]{siunitx}
\usepackage{comment}
\usepackage{url} %

\usepackage[inline, shortlabels]{enumitem}
\setlist[enumerate]{
  itemjoin={{, }},
  itemjoin*={{, and }},
}

\usepackage{algorithm,algpseudocode}

\usepackage{natbib}

\usepackage[capitalise]{cleveref}

\newcommand{\M}[1]{\ensuremath{#1}} %

\begin{document}

  \title{Computational Inference for Directions in Canonical Correlation Analysis}
  \author{Daniel Kessler\\
    Departments of Statistics and Psychiatry, University of Michigan\\
    and \\
    Elizaveta Levina \\
    Department of Statistics, University of Michigan}
  \maketitle
\begin{abstract}

  Canonical Correlation Analysis (CCA) is a method for analyzing pairs of random vectors; it learns a sequence of paired linear transformations such that the resultant canonical variates are maximally correlated within pairs while uncorrelated across pairs.
  CCA outputs both canonical correlations as well as the canonical directions which define the transformations.
  While inference for canonical correlations is well developed, conducting inference for canonical directions is more challenging and not well-studied, but is key to interpretability.
  We propose a computational bootstrap method (combootcca) for inference on CCA directions.
  We conduct thorough simulation studies that range from simple and well-controlled to complex but realistic and validate the statistical properties of combootcca while comparing it to several competitors.
  We also apply the combootcca method to a brain imaging dataset and discover linked patterns in brain connectivity and behavioral scores.

\end{abstract}

\section{Introduction}
\label{sec:combootcca-intro}

Canonical Correlation Analysis (CCA) is a classical technique \citep{hotelling1935MostPredictableCriterion} for identifying linear relationships among two sets of variables.
Informally, it learns a linear transformation for each set of variables such that the transformed variables are maximally correlated with one another.
CCA has seen a recent resurgence of interest with the growing popularity of multi-modal datasets, which have two (or more) sets of variables collected on the same individual.  
For example, brain imaging studies often capture various brain-related metrics as well as phenotypic and behavioral measures on the same individuals, and it is natural to ask how these are related; see, e.g., \citet{wang2020FindingNeedleHighdimensional} for a review of CCA in neuroscience applications.     In many of these applications, at least one of the two datasets is high-dimensional, i.e., with more variables than observations.   In those settings, data reduction can be applied upstream
\citep[e.g., with principal components analysis (PCA) as in][]{smith2015PositivenegativeModePopulation,goyal2022PositiveNegativeMode}
to render the problem low-dimensional, or regularized forms of CCA which seek sparsity can be used \citep{witten2009PenalizedMatrixDecomposition,xia2018LinkedDimensionsPsychopathology}.

In order to build intuition for CCA, it is useful to view it as a generalization of regression.
Consider the classical regression model
\begin{equation*}
  Y = X \beta + \epsilon,
\end{equation*}
where $Y \in \mathbb{R}^{N}, X \in \mathbb{R}^{N \times p}, \epsilon \in \mathbb{R}^{N}$.
We can obtain an estimator $\hat{\beta}$ by solving an optimization problem, with the classic least squares estimator given by 
\begin{equation}
  \label{eq:OLS}  
  \hat{\beta}_{\text{OLS}} = \operatorname*{argmin}_{\beta} \lVert Y - X \beta \rVert_2^2.
\end{equation}
However, we can obtain a closely related estimator by solving a different optimization problem:
\begin{gather}
  \label{eq:CCAregression}
  \hat{\beta}_{\text{CCA}} = \operatorname*{argmax}_{\beta} \operatorname{Corr} \left( X \beta, Y \right), \\
  \text{s.t. } \operatorname{Var} \left( X \beta \right) = 1, \nonumber
\end{gather}
where the constraint serves only to make the solution unique by requiring that the predictions have unit variance.
The estimators obtained from \eqref{eq:OLS} and \eqref{eq:CCAregression} will satisfy $\hat{\beta}_{\text{OLS}} \propto \hat{\beta}_{\text{CCA}}$,
i.e., we will have found the \emph{direction} in $\mathbb{R}^p$ of our regression coefficients.
Now suppose we measure $q$ different responses for each observation, so that $Y \in \mathbb{R}^{N \times q}$ is now a matrix rather than a vector, with $q > 1$.
A natural analogue to the problem in \eqref{eq:CCAregression} is the following optimization problem: 
\begin{gather}
  \label{eq:CCAcorrelation}
  \left( \hat{\beta}_{\text{CCA}}, \hat{\gamma}_{\text{CCA}} \right) = \operatorname*{argmax}_{\beta, \gamma} \operatorname{Corr} \left( X \beta , Y \gamma \right), \\
  \text{s.t. } \operatorname{Var} \left( X \beta \right) = \operatorname{Var} \left( Y \gamma \right) = 1, \nonumber
\end{gather}
where again the constraint serves to make the solution unique up to sign flipping.
The solution to this problem gives exactly the first pair of canonical directions defined by CCA.

Despite its long history in the statistical literature, CCA is generally deployed as an exploratory tool, without a readily available set of tools for inference.
Indeed, in a recent review of CCA aimed at neuroscientists \citep{wang2020FindingNeedleHighdimensional},
CCA is categorized as a method focused on estimation (in contrast to prediction or inference);
the authors go on to emphasize that if inference does occur, it is often constrained to testing a global null corresponding to no correlation between the datasets.
While exploratory analysis is useful, there is growing appreciation in applications that the discoveries of CCA analyses may be illusory \citep{dinga2019EvaluatingEvidenceBiotypes}.
The development of valid inferential tools for this setting is vital in order to appropriately characterize uncertainty so that truly interesting phenomena may be distinguished from optimistic over-fitting.

A natural starting point is to assess whether the estimated correlation among the canonical variates is significantly different from zero.
While parametric tests for this hypothesis have been developed
\citep[for a review, see \S 11.3.6 of][]{muirhead1982AspectsMultivariateStatistical},
in practice they are sensitive to violations of assumptions \citep{winkler2020PermutationInferenceCanonical},
and so non-parametric approaches have become more popular,
e.g., as used in \citet{witten2009PenalizedMatrixDecomposition}.
While the use of permutation-based procedures to assess the canonical correlations is quite common in the neuroimaging literature,
the recent work \citet{winkler2020PermutationInferenceCanonical} shows that a simple permutation procedure yields inflated Type I for all canonical correlations beyond the first;
they also introduce a more nuanced permutation procedure that addresses this issue and also allows for the incorporation of nuisance covariates.
Inference for the canonical correlations in the high dimensional setting remains an active research area, too \citep{mckeague2022SignificanceTestingCanonical}.

If the hypothesis of zero canonical correlations can be rejected, the next natural question to ask is which variables have significant coefficients in the canonical directions; as in regression, one may only be interested in inference on individual coefficients if the $F$-test rejects the global null of all coefficients being zero.  
However, inference on canonical directions remains an elusive goal in CCA.
For example,  \citet{rosa2015EstimatingMultivariateSimilarity} applied 
sparse, non-negative CCA  to pairs of brain images obtained via arterial spin labeling under different pharmacological challenges, and performed a permutation-based procedure to assess the significance of the canonical correlations,
but the authors acknowledge that they are unable to perform inference at the level of individual features.
Indeed, a recent review of CCA intended for neuroscientists \citep{zhuang2020TechnicalReviewCanonical} concludes by acknowledging that,
at the time of writing,
inferential tools are only available for the canonical correlations rather than the canonical directions, and that the development of inference for directions at the level of individual features would benefit future neuroscience research.

The importance of developing these tools is highlighted in a recent paper that studies the stability of CCA  \citep{helmer2020StabilityCanonicalCorrelation}.
In this work, the authors consider the sampling error of the estimated CCA directions,
but rather than considering individual coordinates,
they focus on the angle between an estimated direction and the true direction.
While stability in this sense is important if a canonical direction is to be interpreted holistically,
it does not afford inference at the level of individual coefficients.
Indeed, there may be cases where an overall canonical direction is ``unstable,''
but a small number of coordinates of interest can be reliably differentiated from 0
\citep[for related discussion, see]{mcintosh2021ComparisonCanonicalCorrelation}.
The authors provide guidance for what sort of stability can be expected as the ratio of samples to features varies.
While \citet{helmer2020StabilityCanonicalCorrelation}'s approach is generally numerical and relies on the generation of synthetic data,
recent theoretical developments in \citet{bykhovskaya2023HighdimensionalCanonicalCorrelation} echo these empirical findings, 
characterizing limiting angles between true and estimated canonical variates in terms of these ratios.

In the absence of rigorous statistical tests,
practitioners have developed ad hoc methods to characterize uncertainty about canonical directions.
These assessments generally involve some form of resampling, such as bootstrap or permutation tests,
and the statistical properties of these approaches have not been well studied.
These procedures do not always take the form of hypothesis testing.
For example, \citet{alnaes2020PatternsSociocognitiveStratification,linke2021SharedAnxietySpecificPediatric} performed CCA to obtain canonical directions,
but subjected the resultant directions to ICA to aid interpretability \citep{miller2016MultimodalPopulationBrain}.
They then used a resampling procedure in order to assess the stability of their final results,
but they do not perform inference for individual coordinates of the canonical directions.
In other cases, variants of the bootstrap are used in order to construct confidence intervals for individual coordinates of the directions \citep{xia2018LinkedDimensionsPsychopathology}.
One shortcoming of these approaches is that their statistical properties (e.g., control of Type I error) are not well-studied.

In this work, we help to fill this gap and provide concrete guidance regarding statistical inference for canonical directions.
We propose a method, combootcca, and provide evidence for its validity.
We also review several other approaches based on our review of the literature and compare their performance to combootcca empirically in terms of coverage, statistical validity (control of Type I error), and power (control of Type II error), in a variety of simulation studies that range from simple but carefully controlled to complex but realistic.
We then illustrate our recommended methodology in an application to neuroimaging data.

\subsection{CCA: Population Model and Estimation}
\label{sec:combootcca-cca-pop-model}

Let $x \in \mathbb{R}^p$ and $y \in \mathbb{R}^q$ be random vectors with covariances $\Sigma_{x}$ and $\Sigma_{y}$, respectively,
and let $\Sigma_{x y} = \operatorname{Cov} \left( x, y \right)$ denote their cross-covariance.
Informally, the initial goal of CCA is to identify a pair of vectors $\beta \in \mathbb{R}^p, \gamma \in \mathbb{R}^q$
such that $x^{\intercal} \beta$ is maximally correlated with $y^{\intercal} \gamma$.
In order to fix the scale of $\beta$ and $\gamma$,
we require that $\beta^{\intercal} \Sigma_x \beta = \gamma^{\intercal} \Sigma_y \gamma = 1$,
i.e., the transformed variables have unit variance.
We can then proceed to find additional pairs of vectors subject to an orthogonality constraint.
Formally, CCA involves solving the following sequence of optimization problems for $k = 1, 2, \ldots, K$,
where $K$ is the rank of $\Sigma_{xy}$:
\begin{equation}
  \label{eq:cca-obj-pop}
  \begin{aligned}
    \left( {\beta}_k, {\gamma}_k \right) &= \operatorname*{argmax}_{\left( b_k, g_k \right)} b_k^{\intercal} {\Sigma}_{xy}  g_k, \\
    \text{s.t. } b_k^{\intercal} {\Sigma}_{x} b_l &= g_k^{\intercal} {\Sigma}_{y} g_l = \mathbb{I} \left( k = l \right).
  \end{aligned}
\end{equation}
We shall refer to
$\beta_1, \beta_2, \ldots, \beta_K$ and
$\gamma_1, \gamma_2, \ldots, \gamma_K$ as the \emph{canonical directions} associated with $x$ and $y$, respectively.
It is often convenient to gather the canonical directions into a matrix, writing 
\begin{align*}
  B &=
      \begin{bmatrix}
        \beta_1 & \beta_2 & \ldots & \beta_K
      \end{bmatrix}
      \in \mathbb{R}^{p \times K}, \\
  \Gamma &=
           \begin{bmatrix}
             \gamma_1 & \gamma_2 & \ldots & \gamma_K
           \end{bmatrix}
           \in \mathbb{R}^{q \times K},
\end{align*}
which permits us to rewrite the constraint in \eqref{eq:cca-obj-pop} as
$B^{\intercal} \Sigma_x B = \Gamma^{\intercal} \Sigma_{y} \Gamma = I_K$.
We refer to the transformed variables as the \emph{canonical variates} and denote them by
\begin{equation*}
  \begin{aligned}
    c =
    \begin{bmatrix} c_1 & c_2 & \ldots c_K \end{bmatrix}^{\intercal} =
    \begin{bmatrix} x^{\intercal} \beta_1 & x^{\intercal} \beta_2 & \ldots & x^{\intercal} \beta_K \end{bmatrix}^{\intercal} \in \mathbb{R}^{K}, \\
    d =
    \begin{bmatrix} d_1 & d_2 & \ldots d_K \end{bmatrix}^{\intercal} =
    \begin{bmatrix} y^{\intercal} \gamma_1 & y^{\intercal} \gamma_2 & \ldots & y^{\intercal} \gamma_K \end{bmatrix}^{\intercal} \in \mathbb{R}^{K}.
  \end{aligned}
\end{equation*}
The correlations between the canonical variates are the \emph{canonical correlations} and are denoted by
\begin{equation*}
  \rho = \begin{bmatrix} \rho_1 & \rho_2 & \ldots & \rho_K \end{bmatrix}^{\intercal} \in \left[ 0, 1 \right]^K,  \ \ R = \mathrm{diag}(\rho) , 
\end{equation*}
where 
$R \in \mathbb{R}^{K \times K}$ is a diagonal matrix with $R_{k, k} = \rho_k$.

If the population cross-covariance $\Sigma_{xy}$ is known and the covariances $\Sigma_x$ and  $\Sigma_y$ are known and non-singular,
then all of the components of the CCA solution can be obtained as follows as presented in \citet{muirhead1982AspectsMultivariateStatistical}.
First, perform the singular value decomposition (SVD)
\begin{equation*}
  \Sigma_x^{-1/2} \Sigma_{xy} \Sigma_y^{-1/2}  = U S V^{\intercal}.
\end{equation*}
The diagonal entries of $S$ are the canonical correlations, i.e., $R = S$.
The canonical directions can be obtained as
$B = \Sigma_{x}^{-1/2} U$ and
$\Gamma = \Sigma_{y}^{-1/2} V$.
This formulation will be especially convenient for our numerical studies discussed in Section \ref{sec:combootcca-empirical-results},
as for a fixed generative covariance structure we can recover the true CCA solution.

In practice, we observe the data matrices
$X \in \mathbb{R}^{N \times p}$ and
$Y \in \mathbb{R}^{N \times q}$,
where $N$ is the number of observations;
without loss of generality, suppose that both $X$ and $Y$ have been column-centered.  
We will use the data matrices to construct CCA estimators $\hat{R}, \hat{B}$ and, $\hat{\Gamma}$.
Classical CCA is based on the empirical covariances, which are the maximum likelihood estimators,
although there are many options available for estimating the covariance matrices \citep{fan2016OverviewEstimationLarge}.
Replacing the covariances
$\Sigma_x, \Sigma_y,$ and $\Sigma_{xy}$
with their estimated counterparts
$\hat{\Sigma}_x, \hat{\Sigma}_y,$ and $\hat{\Sigma}_{xy}$
an estimated CCA solution can be obtained using the SVD approach described above,
but a more popular approach (also used by R's \texttt{cancor} function)
is due to \citet{bjorck1973NumericalMethodsComputing}
and is briefly summarized in \citet[p.~331]{golub2013MatrixComputations}.
In our notation, it first performs (thin) QR decompositions
\begin{align*}
  X  = Q_X R_X , \   Y = Q_Y R_Y, 
\end{align*}
followed by the SVD decomposition
\begin{equation*}
  Q_X^{\intercal} Q_Y = U S V^{\intercal}.
\end{equation*}
The estimated canonical correlations are given by the diagonal of $S$,
and the estimated canonical directions can be obtained as
$\hat{B} = R_X^{-1} U$ and
$\hat{\Gamma} = R_Y^{-1} V$.
Because of the special form of $R_X$ and $R_{Y}$,
it is not necessary to explicitly invert them,
and the canonical directions can instead be found by back-solving.
However, we must note that this approach satisfies a related, but distinct, constraint from that presented in \eqref{eq:cca-obj-pop}: the resulting 
canonical variates are empirically uncorrelated with one another,
but rather than having unit variance, they have unit norm,
which we can see by observing
\begin{align*}
  \lVert \M{X} \hat{\beta}_k  \rVert_2^2
  &= \lVert Q_X R_X \hat{\beta}_k  \rVert_2^2 = \lVert Q_X R_X R_X^{-1} U e_k  \rVert_2^2 
  = e_k^{\intercal} U^{\intercal} Q_X^{\intercal} Q_X U e_k
  = 1,
\end{align*}
with an analogous result for $\hat{\Gamma}$.
Recalling that we assume $\M{X}$ and $\M{Y}$ are column-centered,
the empirical variance of these canonical variates will be $(N-1)^{-1}$.
This is problematic for inference on the canonical directions,
but can be readily remedied by multiplying
$\hat{B}$ and $\hat{\Gamma}$ by
$\sqrt{N-1}$,
which will set the canonical variates' empirical variance to 1
and put the estimated canonical directions on a scale free of $N$.

\begin{comment}
#+ORGTBL: SEND notation orgtbl-to-latex :splice nil :skip 0
| Object   | Dimension    | Description                              |
|----------+--------------+------------------------------------------|
| $x$      | $p$          | Random Vector                            |
| $y$      | $q$          | Random Vector                            |
| $X$      | $N \times p$ | Data matrix                              |
| $Y$      | $N \times q$ | Data matrix                              |
| $\rho$   | $K$          | Canonical Correlations (vector)          |
| $R$      | $K \times K$ | Canonical Correlations (diagonal matrix) |
| $B$      | $p \times K$ | Canonical Directions                     |
| $\Gamma$ | $q \times K$ | Canonical Directions                     |
\end{comment}

\begin{table}
  \centering
\begin{tabular}{lll}
Object & Dimension & Description\\[0pt]
\hline
\(x\) & \(p\) & Random Vector\\[0pt]
\(y\) & \(q\) & Random Vector\\[0pt]
\(X\) & \(N \times p\) & Data matrix\\[0pt]
\(Y\) & \(N \times q\) & Data matrix\\[0pt]
\(\rho\) & \(K\) & Canonical Correlations (vector)\\[0pt]
\(R\) & \(K \times K\) & Canonical Correlations (diagonal matrix)\\[0pt]
\(B\) & \(p \times K\) & Canonical Directions\\[0pt]
\(\Gamma\) & \(q \times K\) & Canonical Directions\\[0pt]
\end{tabular}
  \caption{CCA notation.}
  \label{tab:notation}
\end{table}

\subsection{Inverting the CCA Model}
\label{sec:combootcca-cca-inverse-model}

Given the population covariance matrix and assuming that both $\Sigma_x$ and $\Sigma_y$ are non-singular,  there is a straightforward mapping from the covariance $\Sigma$ to the CCA solution $R, B, \Gamma$ as given in Section \ref{sec:combootcca-cca-pop-model}.
However, in some settings (for example, Simulation III discussed in Section \ref{sec:combootcca-sim-data-based}),
it is more convenient to directly specify the CCA parameters $\left( R, B, \Gamma \right)$.   Unfortunately, these parameters do not typically uniquely identify $\Sigma$, but given a set of CCA parameters $R, B, \Gamma$, one can define a covariance matrix $\Sigma$ to match them.  
Assume that both $B$ and $\Gamma$ have full column rank and that the diagonal of $R$ is a descending sequence of unique and strictly positive canonical correlations.  Without loss of generality, assume that $p \ge q = K$.  
Recall that the covariances $\Sigma_{x}$ and $\Sigma_{y}$  satisfy $B^{\intercal} \Sigma_{x} B =\Gamma^{\intercal} \Sigma_{y} \Gamma = I_K$.
Since $\Gamma$ is a square matrix of full column rank, we can just solve for $\Sigma_{y} = \left( \Gamma^{-1} \right)^{\intercal} \Gamma^{-1}$.
If $p = q$, we can do the same for $B$, but if $p > q$, then
$B$ is a rectangular matrix with full column rank, so let $B^+$ denote the Moore-Penrose inverse of $B$, satisfying $B^+ B = I_{K}$.  
Then we can find \emph{a} solution $\Sigma_{x} = \left( B^+ \right)^{\intercal} B^+$.
However, $\Sigma_{x}$ is a $p \times p$ square matrix of rank at most $K < p$.
This rank deficiency may be undesirable, %
and so in simulation studies we remedy this by inflating the trailing eigenvalues of $\Sigma_{x}$ to make it full rank.  
Specifically, we replace the $(K + 1)$th through $p$th eigenvalues of $\Sigma_x$ by linearly interpolating between the $K$th eigenvalue and $0$ (without including the endpoints).
This $\Sigma_{x}$ has full rank and will still satisfy $B^{\intercal} \Sigma_{x} B = I_K$.
Following \citet{chen2013SparseCCAPrecision}, we can take $\Sigma_{xy} = \Sigma_x B R \Gamma^{\intercal} \Sigma_y$ and $\Sigma_{yx} = \Sigma_{xy}^{\intercal}$.   Finally, we can assemble $\Sigma$ as
\begin{equation*}
  \Sigma =
  \begin{bmatrix}
    \Sigma_{x} & \Sigma_{xy} \\
    \Sigma_{yx} & \Sigma_{y}
  \end{bmatrix}.
\end{equation*}

\subsection{Inference for CCA}
\label{sec:combootcca-cca-inference}

In practice, inference for both the correlations $\rho$ and the directions $B$ and $\Gamma$ may be of interest.  Inferential tools for the correlations $\rho$ are relatively better developed, especially when the number of observations $N$ exceeds both $p$ and $q$.    Hotelling's classic \citeyear{hotelling1936RelationsTwoSets} paper discusses two different tests for ``complete independence'' which are equivalent to testing $\rho_1 = 0$ (and therefore all subsequent canonical correlations as well).
Anderson's \citeyear{anderson2003IntroductionMultivariateStatistical} textbook,
in Section 12.4.1,
reviews approaches to inference both on $\rho$ globally and on its individual elements.
Inference for $\rho$ in the high-dimensional case is an area of current research, see for example \citet{mckeague2022SignificanceTestingCanonical}.

Permutation tests are also a popular approach for testing $\rho$, applied by \citet{witten2009PenalizedMatrixDecomposition} and appearing in subsequent applied work, e.g., \citet{alnaes2020PatternsSociocognitiveStratification}.   Recently, \citet{winkler2020PermutationInferenceCanonical} carefully studied the use of permutation tests for the canonical correlations, noting the practical difficulties of parametric inference, demonstrating shortcomings of permutation tests as typically applied and proposing a remedy.

Inference for the canonical directions, however, has received less recent attention in the statistical literature.
\citet{anderson1999AsymptoticTheoryCanonical} reviews several decades of work on the limiting distributions of the estimated canonical directions, finds faults with all past results, and derives the limiting distribution of $\hat{B}$ and $\hat{\Gamma}$ when $x$ and $y$ are jointly normal {\em and} $p = q$.  The latter is an especially significant limitation for the neuroimaging applications we have in mind.  

\citet{laha2021StatisticalInferenceHigh} developed asymptotically exact inference for the first canonical direction and its associated correlation in the high-dimensional setting under the assumption of sparsity.  
Their approach is based on a debiasing argument that yields an asymptotically normal distribution.
However, their method is unable to provide results for any canonical directions except for the first,
and moreover it is not clear that their method is applicable in low-dimensional settings with moderate $N$.
We will compare our proposal to this approach in simulations.

Recently, the permutation method of \citet{winkler2020PermutationInferenceCanonical},
originally proposed for canonical correlations,
was extended to perform inference on the canonical loadings \citep{zhang2022SharedBrainGenetic}.
The canonical loadings are related to the canonical directions but are not the same:
they are the correlations between the original variables and the canonical variates.
This approach optionally obtains a permutation distribution corresponding to the maximal absolute value over all coordinates for each direction, resulting in ``built-in'' family-wise error control, if desired,
 achieved by comparing the empirical values to the distribution of the maxima.
However, this may be more stringent control than necessary and result in the loss of power.
In addition, it only tests whether a given coordinate is nonzero and does not provide confidence intervals, which limits interpretability of this inference approach. 

\section{Bootstrap Inference for CCA}
\label{sec:combootcca-bootstrap}

In the absence of analytical tools for inference, practitioners of CCA often apply the bootstrap \citep{efron1993IntroductionBootstrap} to characterize the uncertainty in CCA estimates, especially the canonical directions.  
While at a high level applying bootstrap sounds straightforward, it has been implemented in a variety of ad hoc ways in practice, without any clarity about the relative merits of different approaches.   Next, we discuss two important aspects of designing a reliable bootstrap algorithm for CCA and compare multiple alternatives for each.   In Section \ref{sec:combootcca-bootstrap-consequences},
we will demonstrate the empirical consequences of these choices, and show that they can have substantial impacts on the statistical properties of the procedure.     Our newly proposed algorithm for bootstrapping CCA, informed by these results, is stated formally at the end of this section.    We call it combootcca (COMputational BOOTstrap for CCA).

\subsection{Alignment of Bootstrap Replicates}
One significant hurdle when using resampling-based methods with CCA is the issue of alignment.    There is a fundamental sign ambiguity in CCA, just like with any estimated direction, since  $\operatorname{Corr} \left( X^{\intercal} \beta, Y^{\intercal} \gamma \right) = \operatorname{Corr} \left( X^{\intercal} \left( -\beta \right), Y^{\intercal} \left( -\gamma \right) \right)$.  
When considering a single CCA solution, this ambiguity is of little consequence,
but when multiple bootstrap realizations are drawn,  there is a need to ``align'' the estimates obtained from the resampled data so that they can be meaningfully combined and compared to the estimate obtained from original data.  In addition to sign ambiguity, the canonical directions may change order, especially if the associated canonical correlations are not well separated, or be rotated in some way.
Without alignment, all of these can substantially inflate our estimates of variance, leading to conservative inference with little power.   On the other hand, a ``strict" alignment strategy may lead to underestimating variance and result in invalid inference, so a balance is needed.

The general goal of alignment is to learn and apply,
for each resampled estimate $\left( \tilde{R}^{\star}, \tilde{B}^{\star}, \tilde{\Gamma}^{\star} \right)$,
a transformation $f$ to obtain an aligned version $\left( \hat{R}^{\star}, \hat{B}^{\star}, \hat{\Gamma}^{\star} \right)$.
We will learn $f$ by comparing $\left( \tilde{R}^{\star}, \tilde{B}^{\star}, \tilde{\Gamma}^{\star} \right)$ to $\left( \hat{R}, \hat{B}, \hat{\Gamma} \right)$, where the latter is a ``reference'' solution which typically corresponds to the values estimated on the full data.
While the canonical correlations are invariant to the scale of the predictors,
the canonical directions are not.
Because our original variables may be on different scales which may unduly influence alignment,
prior to learning $f$ we multiply\footnote{
  At first glance, division may seem more appropriate, but if we increase the variance associated with say the first coordinate of $x$, then its associated canonical direction coordinate must \emph{decrease} to offset this, so the remedy is indeed to \emph{multiply} the canonical directions (which is then tantamount to having divided the variables in the first place).}
the rows of $\left( \tilde{B}^{\star}, \tilde{\Gamma}^{\star} \right)$ and $\left(\hat{B}, \hat{\Gamma} \right)$
by the standard deviations of their corresponding samples, which
transforms the canonical direction matrices to what they would have been if all variables had been standardized prior to CCA and prevents coordinates corresponding to variables with low empirical variance from dominating the alignment.
We consider several possible alignment strategies, described below.   The empirical results on the significant impact of alignment on statistical validity will be presented
in Section \ref{sec:combootcca-bootstrap-consequences}.

\paragraph{Identity (no alignment). }  This alignment ``strategy'' is included for baseline assessment of the need for alignment.   Here $f$ is the identity operator, and given that it does not even correct for the sign ambiguity, we expect it to perform poorly.

\paragraph{Sign Flip.}
This alignment strategy deals with sign ambiguity by flipping the signs of canonical directions and has been used in practice \citep[e.g., in][]{mcintosh2021ComparisonCanonicalCorrelation,nakua2023ComparingStabilityReproducibility}.
That is, the transformation $f$ right-multiplies the matrices $\tilde{B}^{\star}$ and $\tilde{\Gamma}^{\star}$ by a signature matrix $H$ (i.e., a diagonal matrix with entries in $\pm 1$).
To decide which directions need to be flipped,
we construct the similarity matrix $G_B$ by calculating pairwise cosine similarity between the columns of the matrices $\hat{B}$ and $\tilde{B}^{\star}$ (after they have been standardized as described above),
where the cosine similarity between a vector $u$ and $v$ is
$u^{\intercal} v \left\lVert u \right\rVert_2^{-1} \left\lVert v \right\rVert_2^{-1}$.
We similarly obtain $G_{\Gamma}$ as the pairwise cosine similarities between the columns of the standardized matrices $\hat{\Gamma}$ and $\tilde{\Gamma}^{\star}$.
Finally, we average them together and obtain $G = \frac{1}{2} \left( G_B + G_{\Gamma} \right)$.
We then construct $H$ by setting $H_{k, k} = \operatorname{sign} \left( G_{k, k} \right)$,
i.e., if the averaged cosine similarity is negative, we flip the sign,
and if it is positive, we do not.
The aligned solution is given by
$\left( \hat{R}^{\star}, \hat{B}^{\star}, \hat{\Gamma}^{\star}  \right)
= \left( \tilde{R}^{\star}, \tilde{B}^{\star} H, \tilde{\Gamma}^{\star} H \right)$.

\paragraph{Assignment via Weighted Hungarian Algorithm.}
This alignment strategy is allowed to both change the ordering as well as the associated signs of the directions.
Thus, we find a transformation matrix $T$ that can be written as the product of a permutation matrix $P$ and a signature matrix $H$.
To the best of our knowledge, this approach is novel and has not been considered in the literature.
We treat alignment as an assignment problem,
where the task is to optimally assign the columns of
$\left( \tilde{R}^{\star}, \tilde{B}^{\star}, \tilde{\Gamma}^{\star}  \right)$ to the columns of
$\left( \hat{R}, \hat{B}, \hat{\Gamma} \right)$ while allowing sign flipping.
After adjusting for scaling as described above,
we construct the similarity matrix $G$ based on the cosine similarity in the same manner that we did for the ``Sign Flip'' alignment strategy.
In order to incorporate information about the (empirical) canonical correlations into the alignment strategy,
we weight the matrix of cosine similarities by the square roots of the canonical correlations.
That is, we construct $G_{\text{w}} = (\hat{R})^{\frac{1}{2}} G (\tilde{R}^{\star})^{\frac{1}{2}}$.
We then take the entry-wise absolute value to obtain $G_{\text{wPos}} =  \operatorname{abs} \left( G_{\text{w}} \right)$.
Then, we find a permutation matrix $P$ that maximizes
$\operatorname{trace} \left( G_{\text{wPos}} P \right)$
by using the Hungarian Algorithm \citep{kuhn1955HungarianMethodAssignment} as implemented in the \texttt{R} package \texttt{RcppHungarian} \citep{silverman2022RcppHungarianSolvesMinimum}.
We then apply this permutation to the original matrix and extract the signs of the diagonal entries as
$\operatorname{diag} \left( H \right) = \operatorname{sign} \left( \operatorname{diag} \left( G_{\text{w}} P \right) \right)$.
Our transformation can then be written as $T = P H$,
and our aligned solution for this bootstrap realization is
$\left( \hat{R}^{\star}, \hat{B}^{\star}, \hat{\Gamma}^{\star}  \right)
= \left( \tilde{R}^{\star} P, \tilde{B}^{\star} T, \tilde{\Gamma}^{\star} T \right)$.
This is the approach that is used in combootcca,
our recommended approach that is described in Section \ref{sec:combootcca-combootcca},
and it is an integral part of Algorithm \ref{alg:cap}.

\paragraph{Rotation via Procrustes.}
This alignment strategy involves finding orthogonal matrices that can be applied to the directions of $\tilde{B}^{\star}$ and $\tilde{\Gamma}^{\star}$, respectively.
Since CCA is symmetric, we learn separate transformations for $\tilde{B}^{\star}$ and $\tilde{\Gamma}^{\star}$,
although in principle one could learn the transformation for $\tilde{B}^{\star}$ and apply it to both $\tilde{B}^{\star}$ and $\tilde{\Gamma}^{\star}$, or vice versa.
Formally, we find
\begin{align*}
  T_B = \operatorname*{argmin}_{Q: Q^{\intercal} Q = I}\left\lVert \hat{B} - \tilde{B}^{\star} Q \right\rVert_{F}.
\end{align*}
The solution to this problem is well-known \citep{schonemann1966GeneralizedSolutionOrthogonal},
and a concise treatment is given in \citet[p.\ 328]{golub2013MatrixComputations}:
take the SVD $\left( \tilde{B}^{\star} \right)^{\intercal} \hat{B} = U S V^{\intercal}$,
and the optimal solution is given by $T_B = U V^{\intercal}$;
an analogous approach can be used to find $T_{\Gamma}$.
Once we have obtained the orthogonal matrices $T_B$ and $T_{\Gamma}$,
we align through right multiplication, i.e.,
taking
$\left( \hat{B}^{\star}, \hat{\Gamma}^{\star} \right) = \left( \tilde{B}^{\star} T_B, \tilde{\Gamma}^{\star} T_{\Gamma} \right)$.
At first glance, this may suggest that it will serve to rotate the canonical directions, but this will generally not be the case.
While it is generally true that \emph{left} multiplication by an orthogonal matrix will apply a rotation to the columns of a given matrix,
\emph{right} multiplication by an orthogonal matrix will not generally perform a rotation of the columns.
Instead, it will apply a rotation to the rows.
In the special case where the matrix being transformed is itself orthogonal, then this will also effect a rotation of the columns, but in CCA, the canonical directions are generally not orthogonal matrices.
Recall that the population quantities and related estimates satisfy $B^{\intercal} \Sigma_x B =  \Gamma^{\intercal} \Sigma_y \Gamma = I_{K}$,
i.e., they have orthonormal columns with respect to the inner product induced by $\Sigma_x$ and $\Sigma_y$, respectively.
$B$ (or $\Gamma$) will be orthogonal in the usual sense only in the special case that $\Sigma_x$ (or $\Sigma_y$) is equal to $I$.
One additional consequence is that the matrix of canonical correlations will generally no longer be diagonal, i.e., $\tilde{R}^{\star} T_{B}$ or $\tilde{R}^{\star} T_{\Gamma}$ may have non-zero entries that are not on the diagonal.
There are examples in the literature where it seems a Procrustes alignment is used.
For example,
\citet{xia2018LinkedDimensionsPsychopathology} describes a matching procedure and cites \citet{misic2016NetworklevelStructurefunctionRelationships}, which in turn refers to \citet{mcintosh2004PartialLeastSquares}, which proposes a Procrustes alignment.
Notably, \citet{xia2018LinkedDimensionsPsychopathology} uses a version of sparse CCA \citep{witten2009PenalizedMatrixDecomposition} which assumes that the covariances $\Sigma_x$ and $\Sigma_y$ are identity,
in which case the associated directions are orthogonal.

The preceding strategies have been described in an order that proceeds from the least to most strict alignment.
In general, if we are too ``gentle'' with our alignment, we will sacrifice power, as much of our apparent variability will simply be due to ambiguities that arise due to poor alignment.
On the other hand, if we are too strict, we will sacrifice control of Type I error as we will underestimate variance.
A balance must be struck, and as we shall see later, the weighted Hungarian approach appears to do just this.

\subsection{Constructing Confidence Intervals from the Bootstrap Distribution}
There are multiple options available for how to construct confidence intervals from (aligned) bootstrap replicates.  
One option is the so-called ``normal bootstrap.''
In this approach, one assumes that the distribution of the estimator is normally distributed and centered at its true value.
Then, a confidence interval centered at the estimate can be constructed using the variance of the bootstrap replicates and the quantiles of the normal distribution.
This approach was used to construct confidence intervals for the coefficients in \citet{nakua2023ComparingStabilityReproducibility,misic2016NetworklevelStructurefunctionRelationships,kebets2019SomatosensoryMotorDysconnectivitySpans} (the latter two analyses used Partial Least Squares [PLS], a method closely related to CCA).
It is also the approach offered by the \texttt{RGCCA} package \citep{rgcca} (but which does not offer options regarding alignment strategy).

An alternative is the so-called ``percentile'' bootstrap \citep{efron1993IntroductionBootstrap}, which directly uses quantiles of the empirical distribution of the bootstrapped replicates in order to construct a confidence interval, without relying on normality.  
As a consequence, this interval is not guaranteed to be centered at the original estimate.
As we shall see, this approach will generally perform better as it does not explicitly assume that the initial estimator is an unbiased proxy for the truth, nor does it make strong assumptions about the normality of the sampling distribution.
Although a fairly common approach in general, it does not seem popular in CCA applications.

\subsection{The Combootcca Algorithm}
\label{sec:combootcca-combootcca}

Based on the careful investigation  of the different options discussed above and empirical comparisons between them  in  Section \ref{sec:combootcca-bootstrap-consequences}, we present our final algorithm for a computational bootstrap approach to inference for CCA directions (combootcca).     To the best of our knowledge,
this is a novel algorithm which has not been previously considered in the literature for inference on CCA directions.
We present this approach in Algorithm \ref{alg:cap}.
In brief, given data matrices
$X \in \mathbb{R}^{N \times p}$ and
$Y \in \mathbb{R}^{N \times q}$,
we first fit the CCA model using the approach described in Section \ref{sec:combootcca-cca-pop-model}
and obtain estimates $\left( \hat{R}, \hat{B}, \hat{\Gamma} \right)$.
These quantities will subsequently be used as a ``reference solution'' for alignment.
Then we draw bootstrap samples from the rows of the data matrices (with replacement),
to obtain $X^{\star}$ and $Y ^{\star}$, and the corresponding bootstrapped CCA estimates
$\left( \tilde{R}^{\star}, \tilde{B}^{\star}, \tilde{\Gamma}^{\star} \right)$.
Then, we align the solutions using the weighted Hungarian strategy described above to obtain
$\left( \hat{R}^{\star}, \hat{B}^{\star}, \hat{\Gamma}^{\star}  \right)
= \left( \tilde{R}^{\star} P, \tilde{B}^{\star} T, \tilde{\Gamma}^{\star} T \right)$.
We record the values, 
and we repeat the procedure \texttt{nBoots} times (we use \num{1e4} repetitions for all results presented below),
each time drawing a new sample with replacement $\left( X^{\star}, Y^{\star} \right)$.
To obtain $1 - \alpha$ level confidence intervals for $\left( \beta_i \right)_j$ or $\left( \gamma_i \right)_j$,
we find the empirical $\frac{\alpha}{2}$ and $1 - \frac{\alpha}{2}$ quantiles of
the bootstrapped estimates at that coordinate
and set these as the end points of our confidence intervals.
We provide the \texttt{R} \citep{rcoreteamLanguageEnvironmentStatistical} package \texttt{combootcca} at
\url{https://github.com/dankessler/combootcca};
it implements the combootcca approach and other competing methods considered in our simulation studies below.
It depends upon the \texttt{boot} package \citep{davison1997BootstrapMethodsTheir,canty2022BootBootstrapSplus}
for its bootstrap implementation.

\begin{algorithm}
  \caption{The combootcca algorithm for confidence intervals for CCA directions.}
  \label{alg:cap}
  \begin{algorithmic}
    \Require $\M{X} \in \mathbb{R}^{N \times p}, \M{Y} \in \mathbb{R}^{N \times q}, \alpha \in (0, 1), \mathtt{nBoots} \in \mathbb{N}$
    \State $K \gets \Call{Minimum}{p, q}$
    \State $\left( \hat{R}, \hat{B}, \hat{\Gamma} \right) = \Call{CCA}{\M{X}, \M{Y}}$
    \State $\hat{B}^{\star} \gets 0^{p \times K \times \mathtt{nBoots}}$
    \State $\hat{\Gamma}^{\star} \gets 0^{q \times K \times \mathtt{nBoots}}$
    \For{$k \gets 1, \mathtt{nBoots}$}
      \State $(\M{X}^{\star}, \M{Y}^{\star}) \gets \Call{Resample with Replacement}{\M{X}, \M{Y}}$
      \State $\left( \tilde{R}^{\star}, \tilde{B}^{\star}, \tilde{\Gamma}^{\star} \right) \gets \Call{CCA}{\M{X}^{\star}, \M{Y}^{\star}}$
      \State $G_{\M{B}} \gets \Call{CosineSim}{\hat{B}, \tilde{B}^{\star}}$ \Comment{Compute column-wise cosine similarity}
      \State $G_{\M{\Gamma}} \gets \Call{CosineSim}{\hat{\Gamma}, \tilde{\Gamma}^{\star}}$
      \State $G \gets \frac{1}{2} \left( G_B + G_{\Gamma} \right)$ \Comment{Average cosine similarity for $B$ and $\Gamma$}
      \State $G_{\text{w}} \gets (\hat{R})^{\frac{1}{2}} G (\tilde{R}^{\star})^{\frac{1}{2}}$ \Comment{Weight by canonical correlations}
      \State $G_{\text{wPos}} \gets \operatorname{abs} \left( G_{\text{w}} \right)$ \Comment{Take entry-wise absolute value}
      \State $P \gets \Call{Hungarian}{G_{\text{wPos}}}$ \Comment{Permutation $P$ maximizes $\operatorname{trace} \left( G_{\text{wPos}} P \right)$}
      \State $H \gets \Call{SignDiag}{G_{\text{w}} P}$ \Comment{$H$ reflects any negative cosine similarities}
      \State $\hat{B}^{\star} {\left[ :, :, k \right]} \gets \tilde{B}^{\star} P H$
      \State $\hat{\Gamma}^{\star} {\left[ :, :, k \right]} \gets \tilde{\Gamma}^{\star} P H$
    \EndFor
    \State $\hat{B}^{\star}_{\text{Lower}} \gets 0^{p \times K}$
    \State $\hat{B}^{\star}_{\text{Upper}} \gets 0^{p \times K}$
    \For{$i \gets 1, p$}
      \For{$j \gets 1, K$}
        \State $\hat{B}^{\star}_{\text{Lower}} \left[ i, j \right] \gets \Call{Quantile}{\frac{\alpha}{2}, \hat{B}^{\star}[i, j, :]}$
        \State $\hat{B}^{\star}_{\text{Upper}} \left[ i, j \right] \gets \Call{Quantile}{1 - \frac{\alpha}{2}, \hat{B}^{\star}[i, j, :]}$
      \EndFor
    \EndFor
    \State $\hat{\Gamma}^{\star}_{\text{Lower}} \gets 0^{q \times K}$
    \State $\hat{\Gamma}^{\star}_{\text{Upper}} \gets 0^{q \times K}$
    \For{$i \gets 1, q$}
      \For{$j \gets 1, K$}
        \State $\hat{\Gamma}^{\star}_{\text{Lower}} \left[ i, j \right] \gets \Call{Quantile}{\frac{\alpha}{2}, \hat{\Gamma}^{\star}[i, j, :]}$
        \State $\hat{\Gamma}^{\star}_{\text{Upper}} \left[ i, j \right] \gets \Call{Quantile}{1 - \frac{\alpha}{2}, \hat{\Gamma}^{\star}[i, j, :]}$
      \EndFor
    \EndFor
    \State \textbf{return} $\hat{B}^{\star}_{\text{Lower}}, \hat{B}^{\star}_{\text{Upper}}, \hat{\Gamma}^{\star}_{\text{Lower}}, \hat{\Gamma}^{\star}_{\text{Upper}}$
  \end{algorithmic}
\end{algorithm}

\section{Empirical Results on Synthetic Data}
\label{sec:combootcca-empirical-results}

In this section, we apply combootcca and several alternative methods described in Section \ref{sec:combootcca-methods} below to three different simulation settings, with data drawn from different generative models.
Since in simulations we have the ground truth available, we will compare the different confidence intervals  on coverage, length, and rejection rate if used to test the hypothesis of a parameter being equal to zero (rejected if zero is not in the interval).    The confidence level is fixed at $95\%$ in all cases.  

All of our synthetic data will be drawn from a multivariate normal distribution with means  $\mu_x = 0_p$ and $\mu_y = 0_q$, and 
the (joint) covariance $\Sigma$ given by 
\begin{equation*}
  \Sigma =
  \begin{bmatrix}
    \Sigma_{x} & \Sigma_{xy} \\
    \Sigma_{yx} & \Sigma_{y}
  \end{bmatrix}.
\end{equation*}

When determining coverage, we have to confront the fundamental sign ambiguity of CCA.
Suppose the true canonical correlations $\rho$ are all distinct
and that we know the ``true'' canonical directions $\left(B, \Gamma\right)$.
Any pair $\left( B H, \Gamma H \right)$, where $H$ is a signature matrix, \emph{also} qualify as the ``true'' canonical directions.
This ambiguity can lead to low coverage if not accounted for, so when evaluating coverage, we maximize it over all such signature matrices $H$ (sign flips).  
The sign ambiguity is of no consequence when evaluating rejection rate, since that depends only on whether the interval contains zero.   
Note that we do not allow for reordering of directions when considering coverage, length, or rejection decisions.

\subsection{Alternative Confidence Intervals for Canonical Directions}
\label{sec:combootcca-methods}
We next present several methods for obtaining confidence intervals corresponding to the elements of the canonical directions.

\paragraph{Asymptotic confidence intervals. }
\label{sec:combootcca-asymptotic}
\citet{anderson1999AsymptoticTheoryCanonical} derived the asymptotic distribution of the canonical directions in the case where
$x$ and $y$ are jointly multivariate normal, $p = q$ and $\rho_1 > \rho_2 > \ldots > \rho_p > 0$.   We shall later see empirically this result does not generalize to the setting $p \neq q$.
The key result is a limiting normal distribution for the entries of $\hat{B}$ and $\hat{\Gamma}$ which can be used to construct asymptotic confidence intervals for each entry.
For $\hat{B}$, this takes the form of
\begin{equation*}
  \sqrt{n} \left( \hat{B}_{i j} - B_{i j} \right) \overset{d}{\to} \mathcal{N} \left( 0, \sigma^2_{B_{i j}} \right),
\end{equation*}
where
\begin{equation*}
  \sigma^2_{B_{i j}} = \frac{1}{2} B_{i j}^2 + \left( 1 - \rho_j^2 \right) \sum_{\substack{k = 1 \\ k \neq j}}^p \frac{\rho_k^2 + \rho_j^2 - 2 \rho_k^2 \rho_j^2}{\left( \rho_j^2 - \rho_k^2 \right)^2} B_{i k}^2,
\end{equation*}
and analogous results can be obtained for the elements of $\hat{\Gamma}$.
In practice, estimates have to be substituted for parameters in the expression for variance in order to obtain asymptotic confidence intervals.
Extending Anderson's results to the case $p \neq q$ is non-trivial and is outside the scope of this work.

\paragraph{Regression-based confidence intervals.}
\label{sec:combootcca-regression}

Given the intimate connection between regression and CCA discussed in Section \ref{sec:combootcca-intro},
it is natural to consider adapting tools from regression to provide inference for CCA.
Recall the estimation strategy discussed in Section \ref{sec:combootcca-cca-pop-model}.
Suppose that we have already obtained the estimated directions $\hat{\Gamma}$,
and want to obtain estimates of the directions $\hat{B}$ using regression.
First, suppose without loss of generality that all columns of data matrices below have been centered.
Let
\begin{equation*}
  \tilde{\beta}_k = \operatorname*{argmin}_{\beta} \left\lVert Y \hat{\gamma}_k - X \beta \right\rVert_2^2.
\end{equation*}
This is the ordinary least squares estimator,
which as we noted in Section \ref{sec:combootcca-intro} maximizes the correlation between $Y \hat{\gamma}_k$ and $X \beta$,
and thus is a solution to the (unconstrained) CCA problem.
In order to satisfy the constraint that $\tilde{\beta}_k^{\intercal} \hat{\Sigma}_x \tilde{\beta}_k = 1$,
we can simply rescale and obtain $\hat{\beta}_k =  \left( \tilde{\beta}_k^{\intercal} \hat{\Sigma}_x \tilde{\beta}_k  \right)^{-1/2} \tilde{\beta}_k$,
which will coincide with the solution we would have obtained with regular CCA;
similar results can be obtained for $\hat{\Gamma}$ by symmetry.
This least squares formulation of CCA was noted in \citet{gao2017SparseCCAAdaptive}.
Because we will depend upon the distribution of the regression-based estimators,
we need them to be independent of the estimated directions which we treat fixed.
To accomplish this, we propose to use this approach with sample splitting.
First, partition the observations into two disjoint sets
$X_1, Y_1$ and $X_2, Y_2$ and perform CCA on $\left( X_1, Y_1 \right)$ to obtain estimated directions $\hat{B}_1, \hat{\Gamma}_{1}$.
Next, in the held out-data fix $\hat{\Gamma} = \hat{\Gamma}_1$ and use the above procedure to obtain $\hat{B}_2$,
then fix $\hat{B} = \hat{B}_1$ and use the above procedure to obtain $\hat{\Gamma}_2$.
Because they were obtained using regression,
conditional on $\hat{\Gamma}_1$,
the entries of $\hat{B}_2$ will each follow a (scaled) $t$-distribution,
which we use to construct confidence intervals, e.g.,
\begin{equation*}
  \left( \hat{\beta}_k \right)_i \pm t^{(\alpha/2)}_{N/2 - q} \hat{\sigma}_k \left( X_2^{\intercal} X_2 \right)^{-1}_{i, i} , 
\end{equation*}
where $t^{(\alpha/2)}_{N/2 - q}$ is the $\alpha/2$th quantile of the $t$ distribution with $N/2 - q$ degrees of freedom,
and $\hat{\sigma}_k$ is the square root of the estimated error variance in the $k$th regression model after rescaling.
By symmetry, analogous confidence intervals can be obtained for $\Gamma$.

\paragraph{Debiased (Sparse) CCA}
\label{sec:combootcca-laha-method}
Recent work by \citet{laha2021StatisticalInferenceHigh} introduces a method for obtaining asymptotically exact inference for canonical directions in a high-dimensional regime.  It works with sparse CCA, although to make a direct comparison in our setting we simply do not regularize (i.e., take the regularization parameter $\lambda = 0$), which may be outside the scope of their theoretical results.  This approach relies on the characterization of the first canonical directions as the unique maximizers (modulo sign flipping) of a smooth function
and uses a one-step correction to de-bias the (regularized) estimators;
this de-biasing step is carefully accounted for in obtaining a limiting distribution.
One limitation of this approach, however, is that it only provides results for the leading canonical directions ($\beta_1$ and $\gamma_1$), and does not offer any inference for subsequent canonical directions.   We use the function \texttt{give\_SCCA} with default settings,  available from the authors' package on GitHub at \url{https://github.com/nilanjanalaha/de.bias.CCA}, to obtain estimates of the variances and use these to construct confidence intervals.
This function requires that we provide an estimate of the canonical directions:
in their example, they use the result of the sparse CCA method of \citet{mai2019IterativePenalizedLeast},
but since we do not require sparsity we simply use the (rescaled) estimates from \texttt{R}'s \texttt{cancor} function.

\subsection{Simulation I: Synthetic Data with One Canonical Correlation}
\label{sec:combootcca-sim-i}

In our first simulation study, we consider the setting where there is a single non-zero canonical correlation, i.e., $K = 1$.
We vary $(p, q) \in \left\{ (10, 10), (100, 10) \right\}$, and $\rho_1 \in \left\{ 0.9, 0.5, 0.2 \right\}$.
In line with the simulation studies of \citet{laha2021StatisticalInferenceHigh},
we construct a sparse precision matrix for both $x$ and $y$ and then invert it to obtain (dense) covariance matrices
$\Sigma_x, \Sigma_y$.
The sparse precision matrix takes the initial form
$\Omega_{i, j} = 1_{\left\{ i = j \right\}} +
0.5 \cdot 1_{\left\{ \left\lvert i - j \right\rvert = 1 \right\}} +
0.4 \cdot 1_{\left\{ \left\lvert i - j \right\rvert = 2 \right\}}$.
We then apply a modification to $\Omega$ to make specification of the canonical directions in Simulation II simpler.
Specifically, for the $\Omega$ associated with $x$
we place $0$'s everywhere but the diagonal in the
$\operatorname{floor} \left( p \right)$ and $\operatorname{floor} \left( p \right) + 1$ rows and columns,
and we make an analogous modification of the $\Omega$ associated with $y$.
This has the effect of breaking the marginal dependence between the first half of the coordinates and the latter half of the coordinates,
and without this it is difficult to specify subsequent canonical directions (as we do in Simulation II, see Section \ref{sec:combootcca-sim-ii}) without running afoul of the orthogonality constraints.
We also repeated this simulation study with identity covariance matrices and found similar results to those given below.

We consider both a ``dense'' and a ``sparse'' regime for the canonical directions.
In the dense regime, the canonical directions are proportional to
\begin{align*}
  \check{\beta}_1 &= \begin{bmatrix} \mathbf{1}_{p/2}^{\intercal} & \mathbf{0}_{p/2}^{\intercal} \end{bmatrix}^{\intercal} \\
  \check{\gamma}_1 &= \begin{bmatrix} \mathbf{1}_{q/2}^{\intercal} & \mathbf{0}_{q/2}^{\intercal} \end{bmatrix}^{\intercal},
\end{align*}
whereas in the sparse regime they are proportional to 
\begin{align*}
  \check{\beta}_1 &= \begin{bmatrix} \mathbf{1}_{2}^{\intercal} & \mathbf{0}_{p - 2}^{\intercal} \end{bmatrix}^{\intercal} \\
  \check{\gamma}_1 &= \begin{bmatrix} \mathbf{1}_{2}^{\intercal} & \mathbf{0}_{q - 2}^{\intercal} \end{bmatrix}^{\intercal},
\end{align*}
i.e.,
in the dense regime the first half of the coordinates are nonzero,
whereas in the sparse regime only the first two coordinates are nonzero.
In both cases, we then normalize to obtain
$\beta_1 = \left( \check{\beta}_1^{\intercal} \Sigma_x \check{\beta}_1 \right)^{-1/2} \check{\beta}_1$ and
$\gamma_1 = \left( \check{\gamma}_1^{\intercal} \Sigma_y \check{\gamma}_1 \right)^{-1/2} \check{\gamma}_1$.
We fix $N = 1000$.
In line with \citet{chen2013SparseCCAPrecision},
we construct the cross-covariance as $\Sigma_{xy} = \rho_1 \Sigma_x \beta_1 \gamma_1^{\intercal} \Sigma_y$.
We draw \num{1000} replicates for each setting for each of the methods.
Since there are only two possible values for coordinates in our setup, for simplicity we examine statistical properties associated with the confidence intervals at just two coordinates for each vector:  the last coordinate ($\left( \beta_1 \right)_p$ and $\left( \gamma_1 \right)_q$) which is always zero, and the first coordinate 
($\left( \beta_1 \right)_1$ and $\left( \gamma_1 \right)_1$), which is always non-zero.

\begin{figure}[htb!]
  \centering
  \includegraphics[width=\textwidth]{{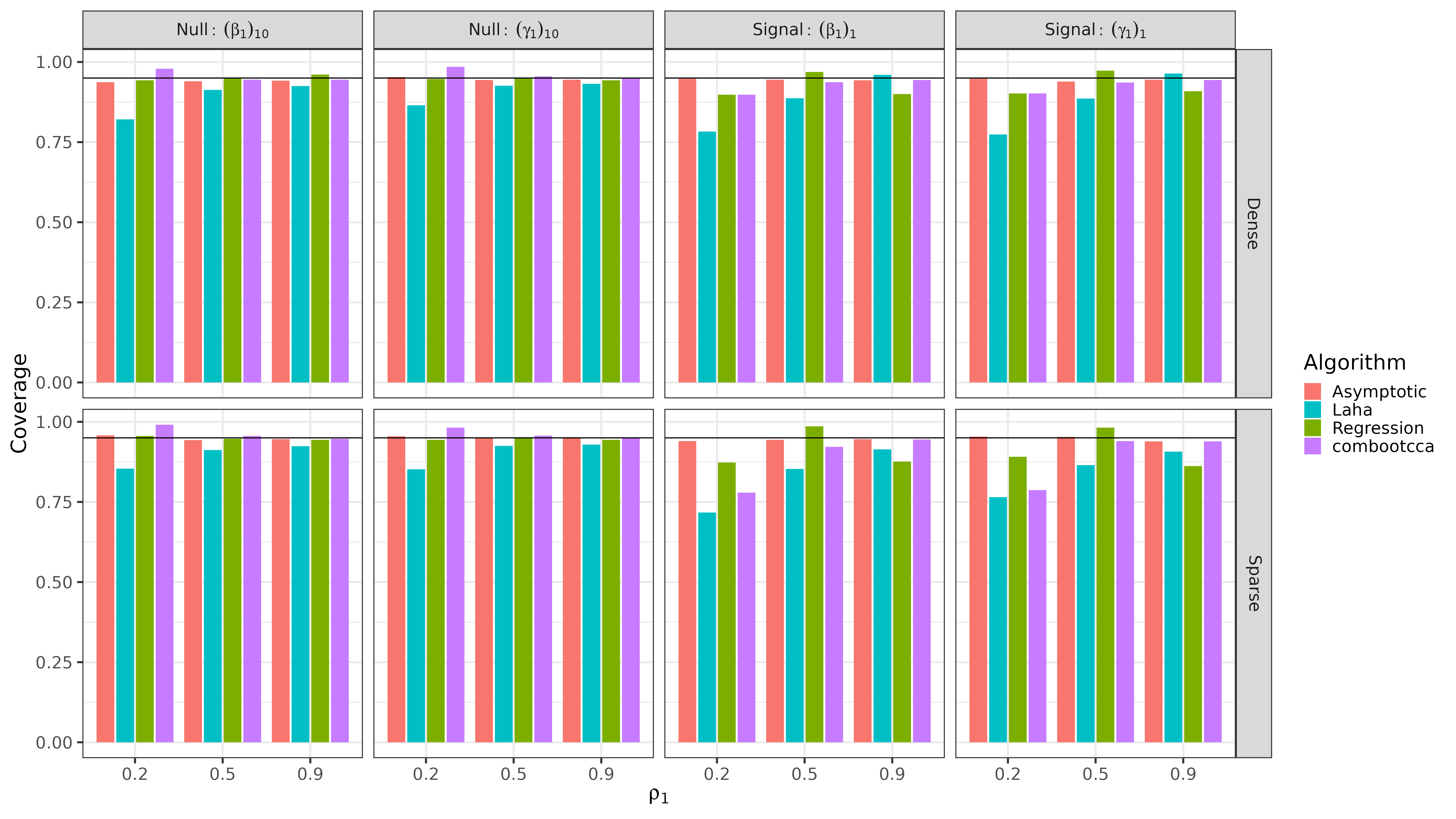}}
  \caption{Coverage rates in simulation I for $p = q = 10$. The horizontal line indicates nominal 95\% coverage.}
  \label{fig:sim-sprec-i-coverage-low}
\end{figure}

\begin{figure}[htb!]
  \centering
  \includegraphics[width=\textwidth]{{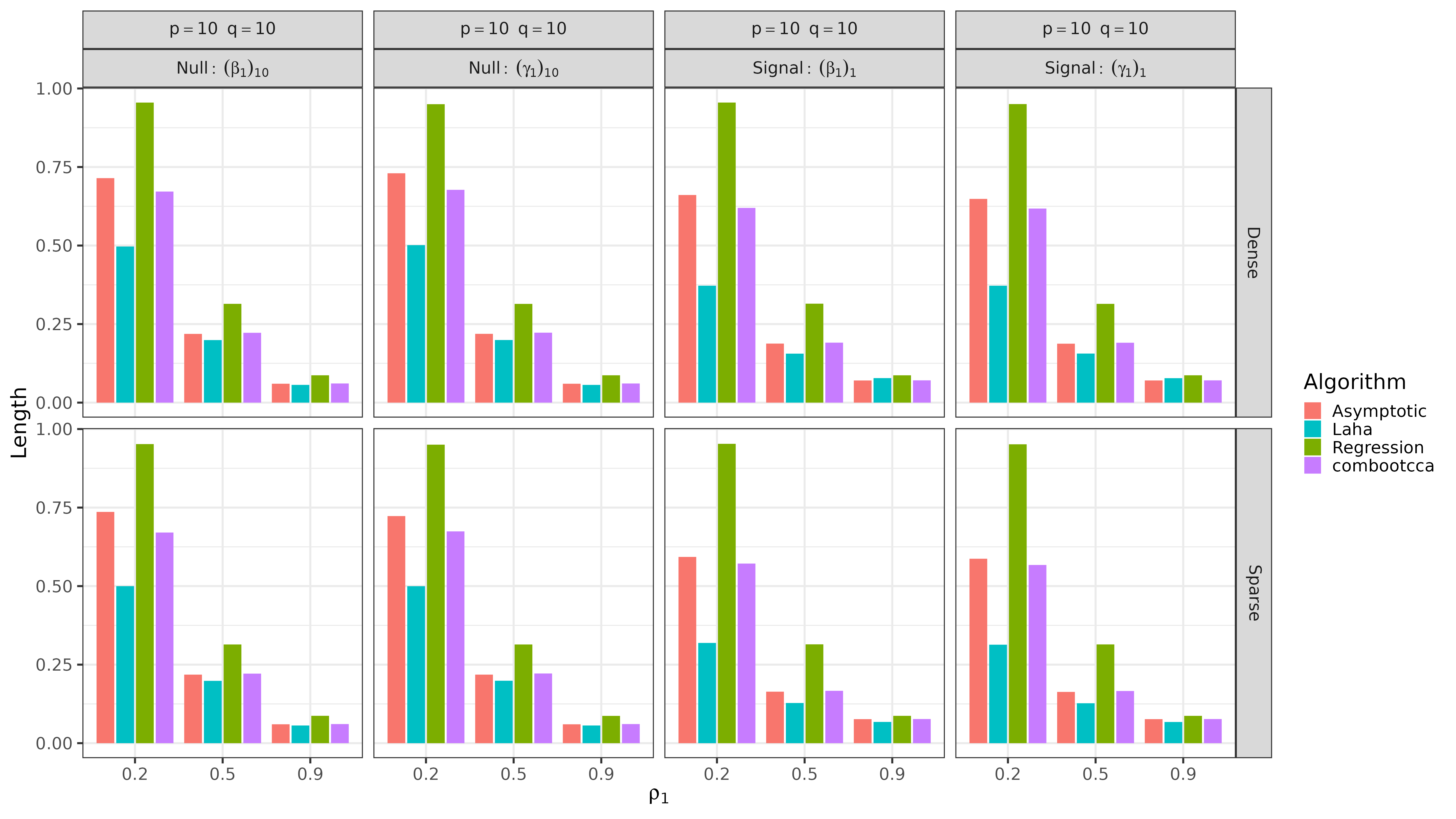}}
  \caption{Lengths of confidence intervals in simulation I for $p = q = 10$.}
  \label{fig:sim-sprec-i-length-low}
\end{figure}

\begin{figure}[htb!]
  \centering
  \includegraphics[width=\textwidth]{{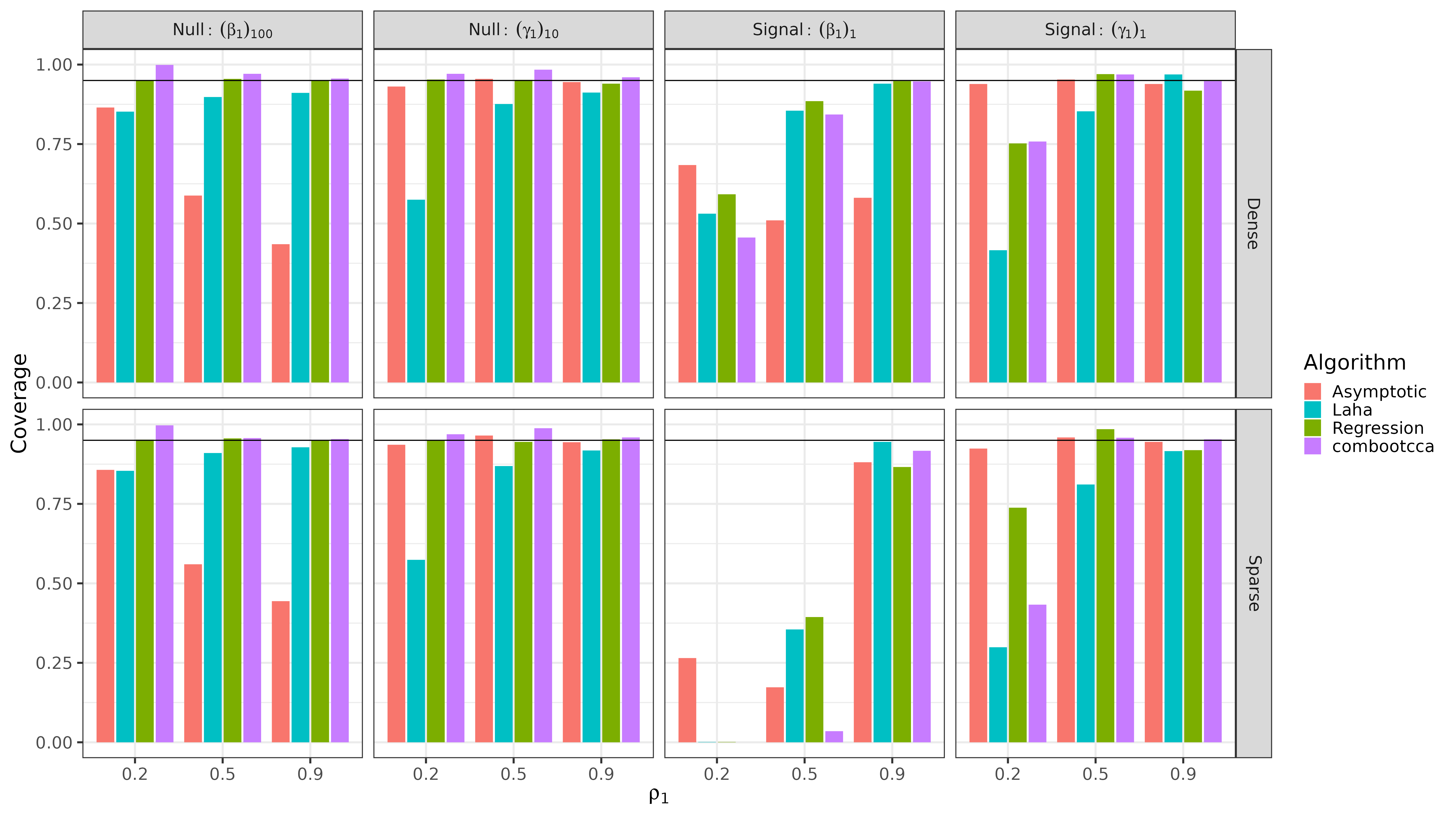}}
  \caption{Coverage rates in simulation I for $p = 100, q = 10$. The horizontal line indicates nominal 95\% coverage.}
  \label{fig:sim-sprec-i-coverage-high}
\end{figure}

\begin{figure}[htb!]
  \centering
  \includegraphics[width=\textwidth]{{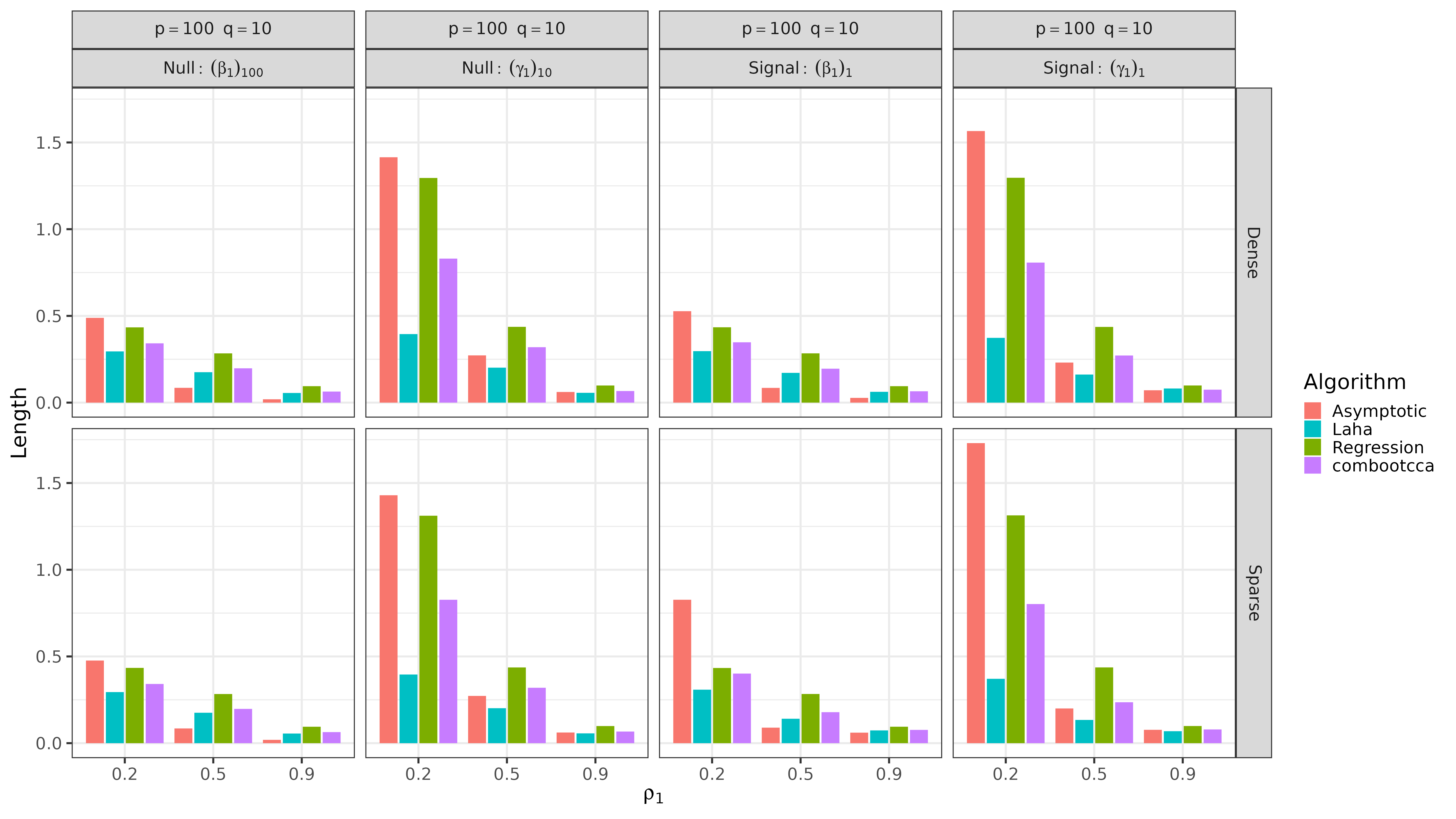}}
  \caption{Lengths of confidence intervals in simulation I for $p = 100, q = 10$.}
  \label{fig:sim-sprec-i-length-high}
\end{figure}

We plot coverage and lengths of intervals at these representative coordinates for $p = q = 10$ in Figures \ref{fig:sim-sprec-i-coverage-low} and \ref{fig:sim-sprec-i-length-low},
while coverage and lengths for $p = 100, q = 10$ are depicted in Figures \ref{fig:sim-sprec-i-coverage-high} and \ref{fig:sim-sprec-i-length-high}, respectively.
With balanced dimensions $p = q = 10$, the methods generally perform well,
although all but the asymptotic approach fall short in their coverage of signals when $\rho_1$ is small.
Notably, the asymptotic, regression, and combootcca methods attain nominal coverage of null coordinates,
which is tantamount to valid control of Type I error,
whereas the method of \citet{laha2021StatisticalInferenceHigh} does not achieve Type I error control when $\rho_1 = 0.2$.
The good performance of the asymptotic approach was expected since the $p = q$ regime satisfies its assumptions.
The pattern for the high-dimensional case where $p = 100$ and $q = 10$ is generally similar but with some differences.
Since $p \neq q$, the theoretical justification for the asymptotic approach fails,
and indeed the coverage for the high-dimensional $\beta_1$ is poor.   
This provides empirical evidence that the theoretical results developed in \citet{anderson1999AsymptoticTheoryCanonical} are indeed not applicable beyond the setting they were obtained for.     Interestingly, the asymptotic method does appear to provide generally nominal coverage for the entries of the low-dimensional $\gamma_{1}$,
The method of \citet{laha2021StatisticalInferenceHigh} continues to struggle with coverage of null coordinates except when $\rho = 0.9$; interestingly this is worse for coverage of the low-dimensional $\gamma_1$.
The sub-nominal coverage of combootcca for signals when $\rho_1$ is small is exacerbated,
especially in the sparse regime and for $\beta_1$.

In general, it appears that achieving nominal coverage for a non-null coordinate is generally more challenging than for a null coordinate,
and that this difficulty is greater when the canonical correlation $\rho_1$ is smaller.
A heuristic explanation for this is as follows.
While $\hat{\beta_{1}}$ is a consistent estimator for $\beta_{1}$, it is not necessarily unbiased.
Because $\hat{\beta_1}$ must satisfy $\hat{\beta}_1^{\intercal} \hat{\Sigma}_x \hat{\beta}_1 = 1$,
this is approximately a constraint on the norm of $\hat{\beta}_1$.
Thus, if the direction is perfectly estimated, the leading coordinates of $\hat{\beta}_1$ will approach their true value,
but if the direction is misestimated (as is more likely with small $\rho_1$ and larger dimensions),
then there will be mass in other coordinates which will shrink the true non-zero coordinates towards $0$,
and our confidence intervals will reflect this.   
Moreover, we expect that this bias will be exacerbated in the sparse regime, when most coordinates are in fact zero.

\begin{figure}[htb!]
  \centering
  \includegraphics[width=\textwidth]{{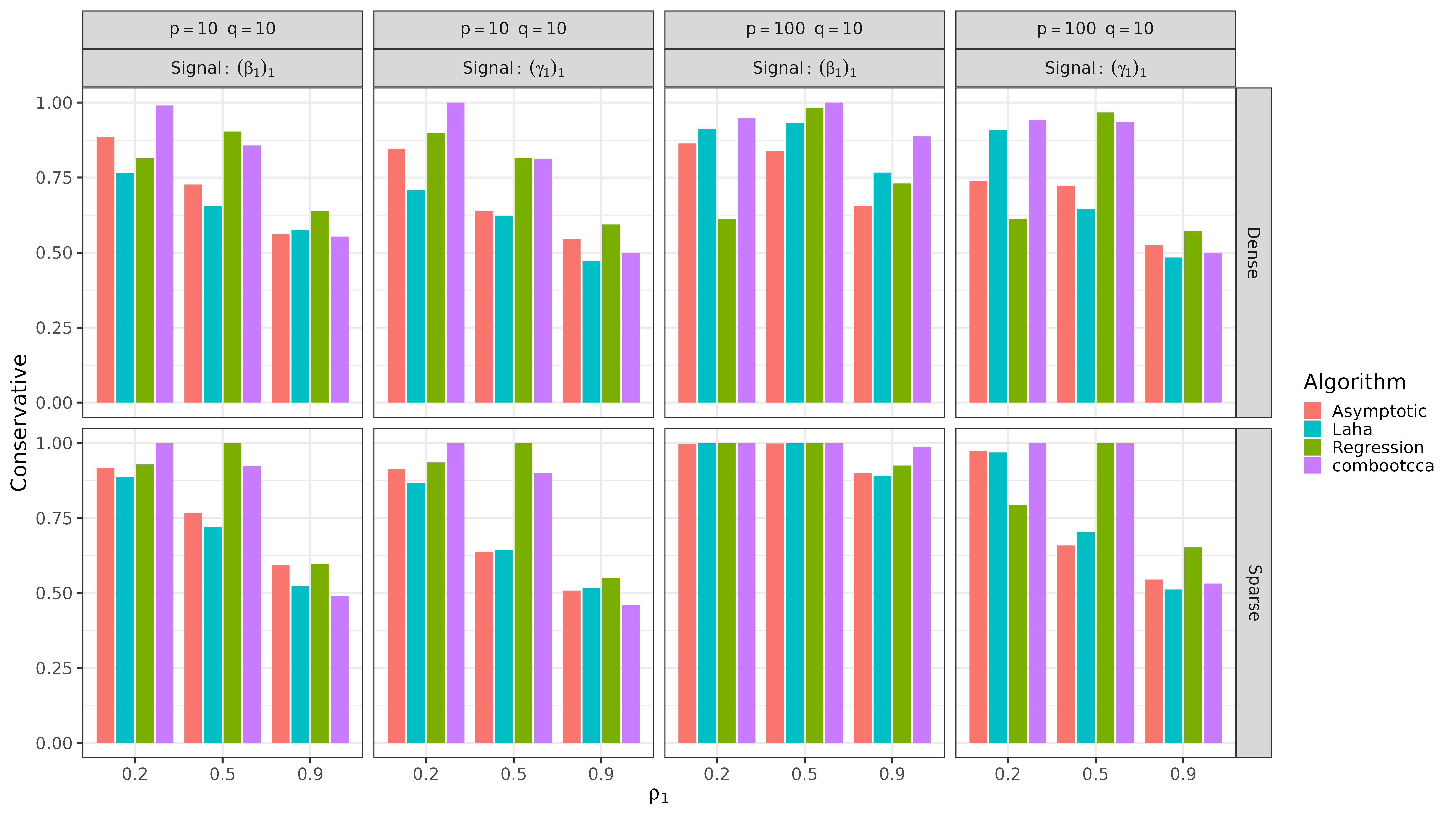}}
  \caption{Bias in simulation I: the proportion of confidence intervals that failed to cover non-null signals that are ``conservative'' (the true value is greater in magnitude than any value in the confidence interval).}
  \label{fig:sim-sprec-i-bias}
\end{figure}

In order to investigate this, we examined confidence intervals that failed to cover a non-zero signal, checking whether the (absolute
value of) the upper bound of the interval was less than (the absolute value of) the truth.
If so, we considered that confidence interval ``conservative.''
Figure \ref{fig:sim-sprec-i-bias} depicts the proportion of non-covering intervals that were conservative,
and indeed shows that when $\rho_1$ is small, and especially in the sparse regime,
the intervals from all methods are generally conservative, meaning that when they fail to cover the true non-zero value, it is likely because the estimate was shrunk towards zero.  

\begin{figure}[htb!]
  \centering
  \includegraphics[width=\textwidth]{{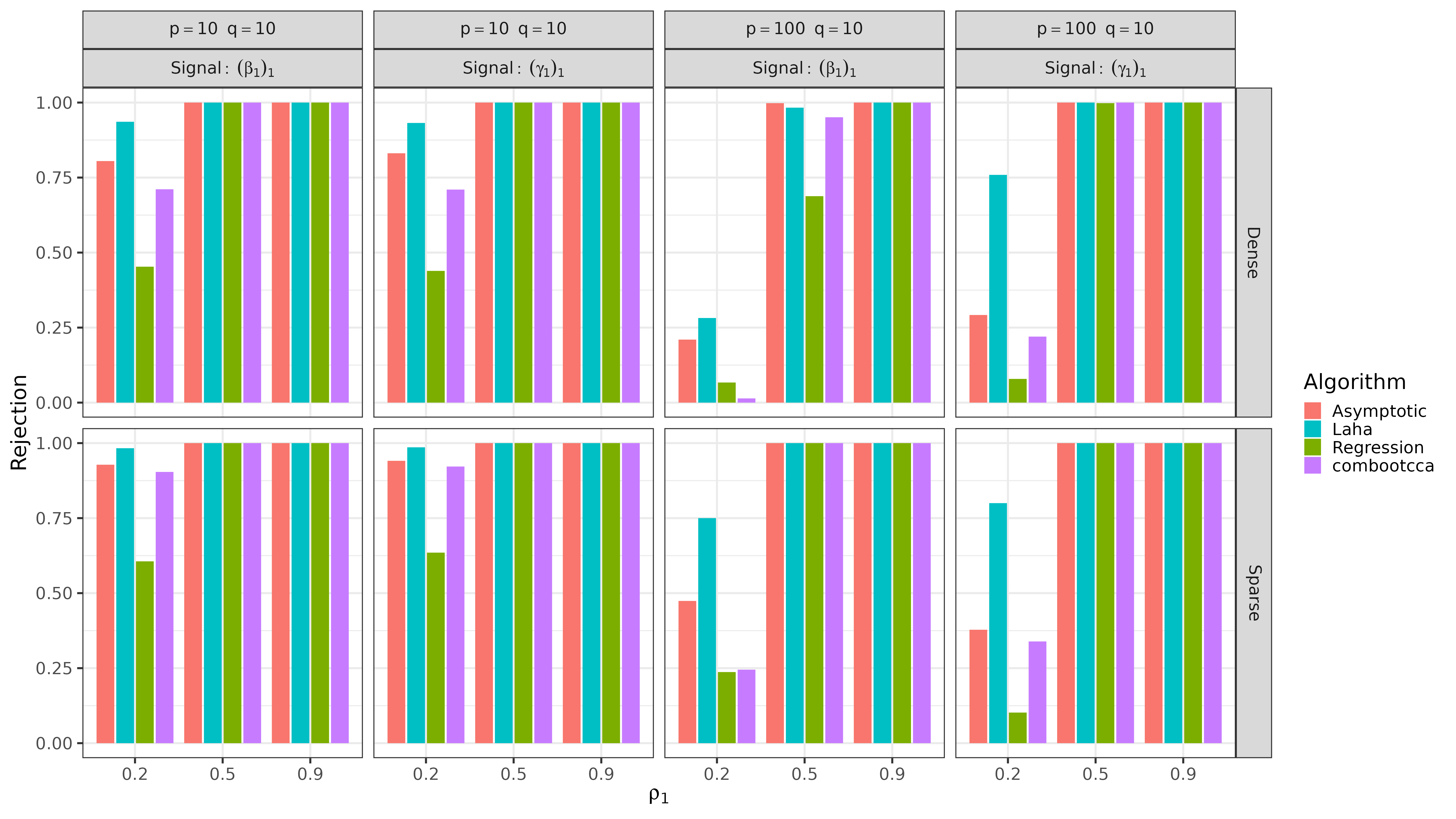}}
  \caption{Power (correct rejection rates) in simulation I.}
  \label{fig:sim-sprec-i-power}
\end{figure}

Even confidence intervals that fail to achieve nominal coverage can lead to correct inference when the question is whether a given coordinate is equal to zero.   Type I error for this hypothesis test is simply $1$ minus coverage at null coordinates (already depicted in Figures \ref{fig:sim-sprec-i-coverage-low} and \ref{fig:sim-sprec-i-coverage-high}), and in Figure \ref{fig:sim-sprec-i-power}, we show the power of the test, i.e.,
the proportion of times confidence intervals for non-zero signals do not contain \num{0}.
Here we see that combootcca is generally the most powerful method among the three that achieve nominal control of Type I error.

\subsection{Simulation II: Synthetic Data with Two Canonical Correlations}
\label{sec:combootcca-sim-ii}

Simulation II is similar to Simulation I in all respects except that we
fix $\rho_1 = 0.9$ and introduce a second nonzero canonical correlation.
We again consider both a dense and sparse regime,
where the first canonical directions are the same as in Section \ref{sec:combootcca-sim-i},
and the second canonical directions in the dense regime are proportional to 
\begin{align*}
  \check{\beta}_1 &= \begin{bmatrix} \mathbf{0}_{p/2}^{\intercal} & \mathbf{1}_{p/2}^{\intercal} \end{bmatrix}^{\intercal} \\
  \check{\gamma}_1 &= \begin{bmatrix} \mathbf{0}_{q/2}^{\intercal} & \mathbf{1}_{q/2}^{\intercal} \end{bmatrix}^{\intercal},
\end{align*}
and in the sparse regime to 
\begin{align*}
  \check{\beta}_1 &= \begin{bmatrix} \mathbf{0}_{2}^{\intercal} & \mathbf{1}_{p - 2}^{\intercal} \end{bmatrix}^{\intercal} \\
  \check{\gamma}_1 &= \begin{bmatrix} \mathbf{0}_{2}^{\intercal} & \mathbf{1}_{q - 2}^{\intercal} \end{bmatrix}^{\intercal}.
\end{align*}
Canonical directions are subsequently normalized with respect to their associated covariances.
Thanks to the structure of the covariance,
these new directions satisfy $\beta_1^{\intercal} \Sigma_x \beta_2 = \gamma_1^{\intercal} \Sigma_y \gamma_2 = 0$.
We fix $\rho_1 = 0.9$ and vary $\rho_2 \in \left\{ 0.8, 0.5, 0.2 \right\}$.
As in Simulation I,
we examine statistical properties associated with the confidence intervals for the first canonical directions
at the last coordinates
$\left( \beta_1 \right)_p$ and $\left( \gamma_1 \right)_q$ (always zero)
and at the first coordinates 
$\left( \beta_1 \right)_1$ and $\left( \gamma_1 \right)_1$ (always non-zero).  
We also consider statistical properties for the second canonical directions 
at the first coordinates
$\left( \beta_2 \right)_1$ and $\left( \gamma_2 \right)_1$ (always zero)
and the last coordinates
$\left( \beta_2 \right)_p$ and $\left( \gamma_2 \right)_q$ (always non-zero).  
The method of \citet{laha2021StatisticalInferenceHigh} is only applicable to the first canonical directions;
it gives no results for the second canonical direction.
As with Simulation I, we repeated this experiment with identity covariances for
$\Sigma_x$ and $\Sigma_y$,
and we found a generally similar pattern of results to those presented below.

\begin{figure}[htb!]
  \centering
  \includegraphics[width=\textwidth]{{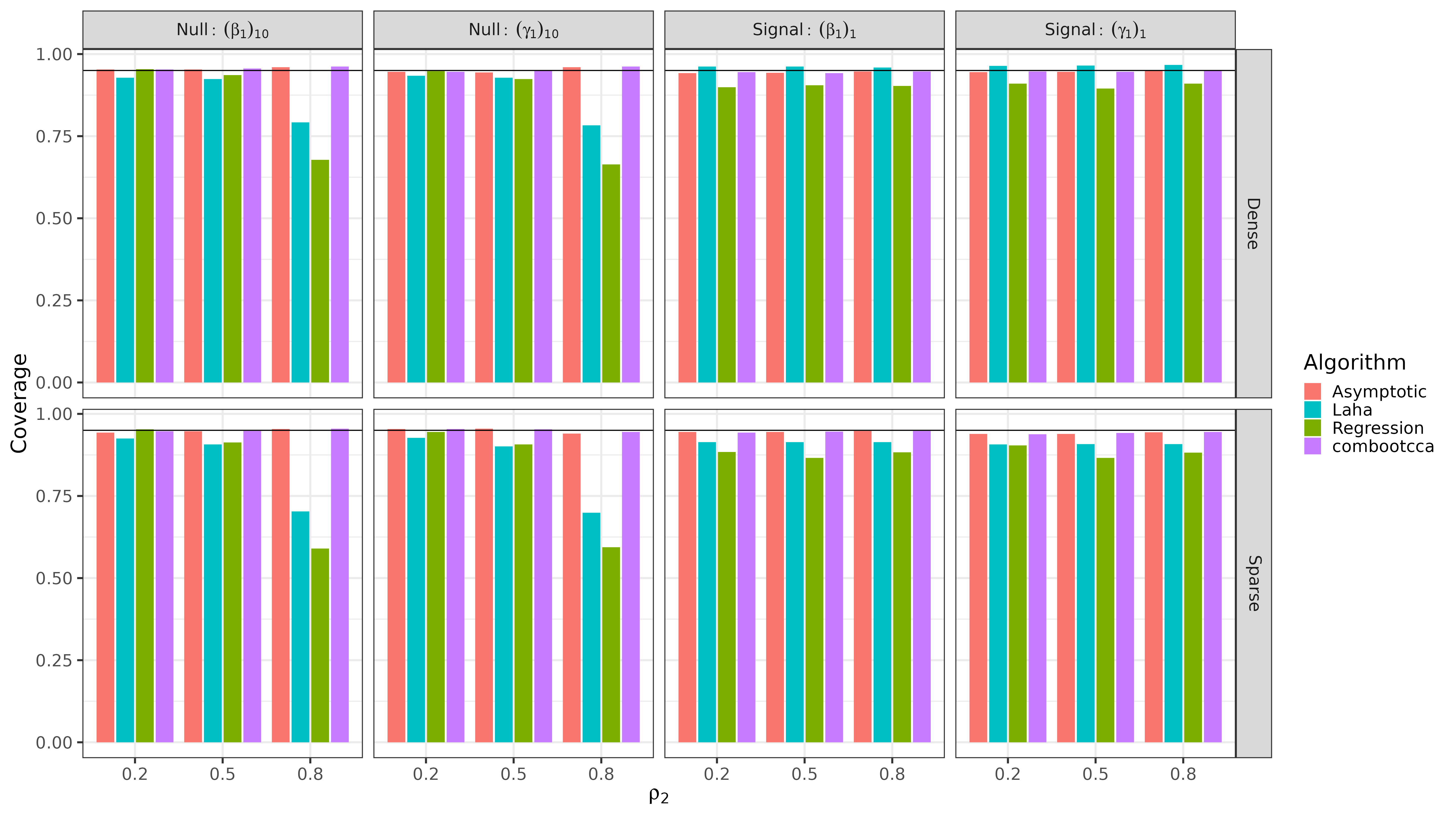}}
  \caption{Coverage rates for first canonical directions in simulation II for $p = q = 10$. The horizontal line indicates nominal 95\% coverage.}
  \label{fig:sim-sprec-ii-coverage1-low}
\end{figure}

\begin{figure}[htb!]
  \centering
  \includegraphics[width=\textwidth]{{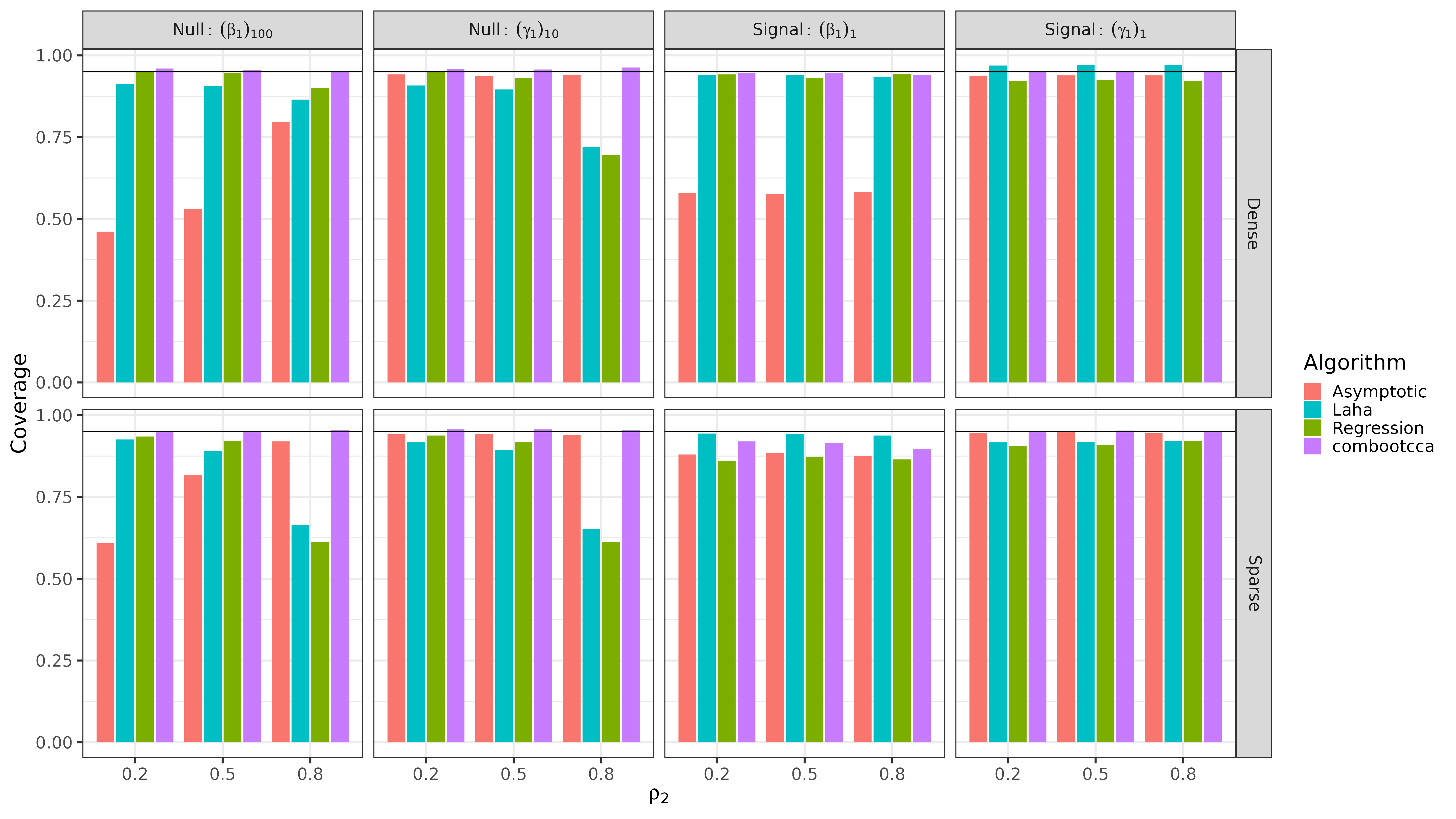}}
  \caption{Coverage rates for first canonical directions in simulation II for $p = 100, q = 10$. The horizontal line indicates nominal 95\% coverage.}
  \label{fig:sim-sprec-ii-coverage1-high}
\end{figure}

\begin{figure}[htb!]
  \centering
  \includegraphics[width=\textwidth]{{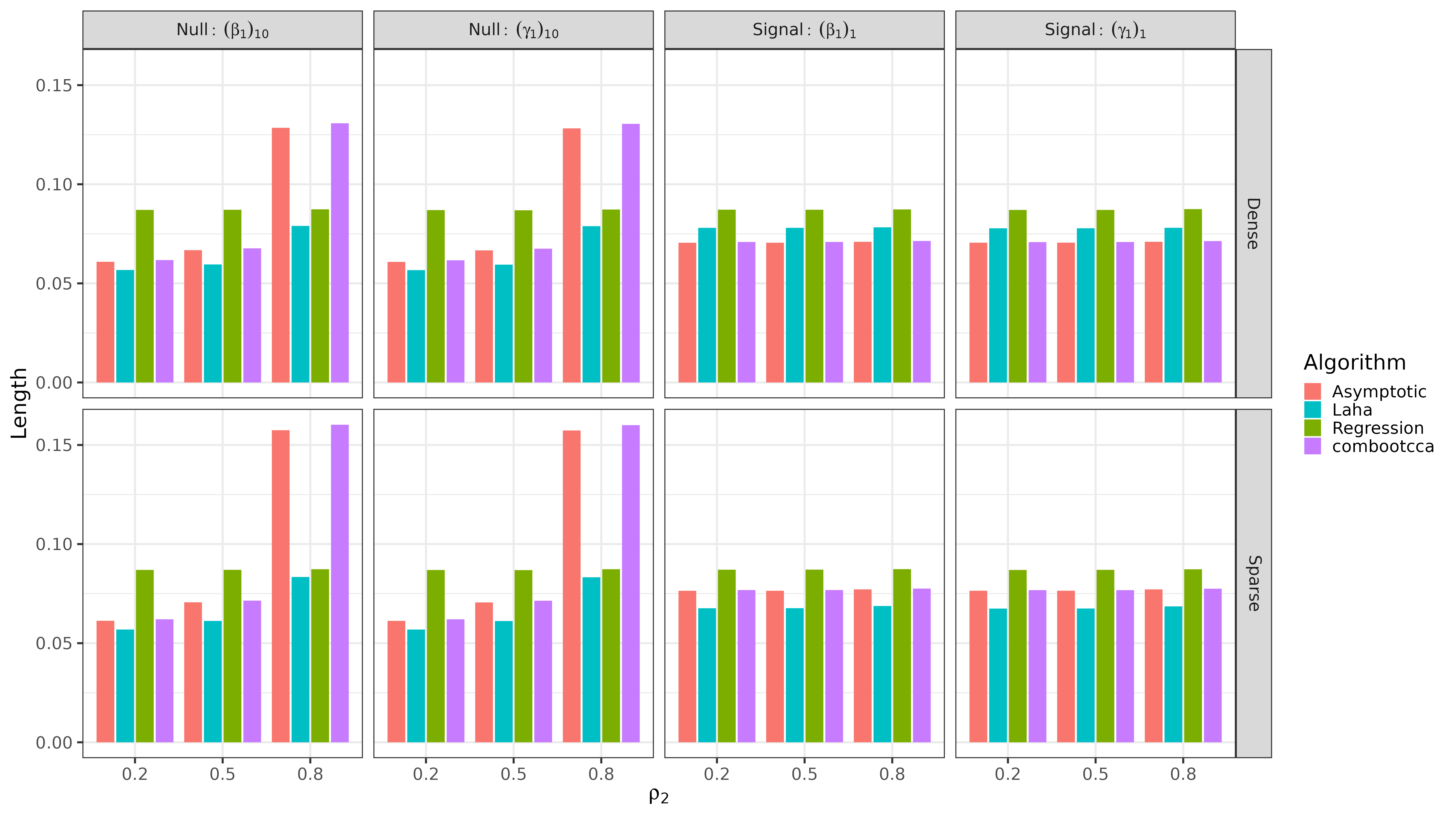}}
  \caption{Lengths of confidence intervals for first canonical directions in simulation II for $p = q = 10$.}
  \label{fig:sim-sprec-ii-length1-low}
\end{figure}

\begin{figure}[htb!]
  \centering
  \includegraphics[width=\textwidth]{{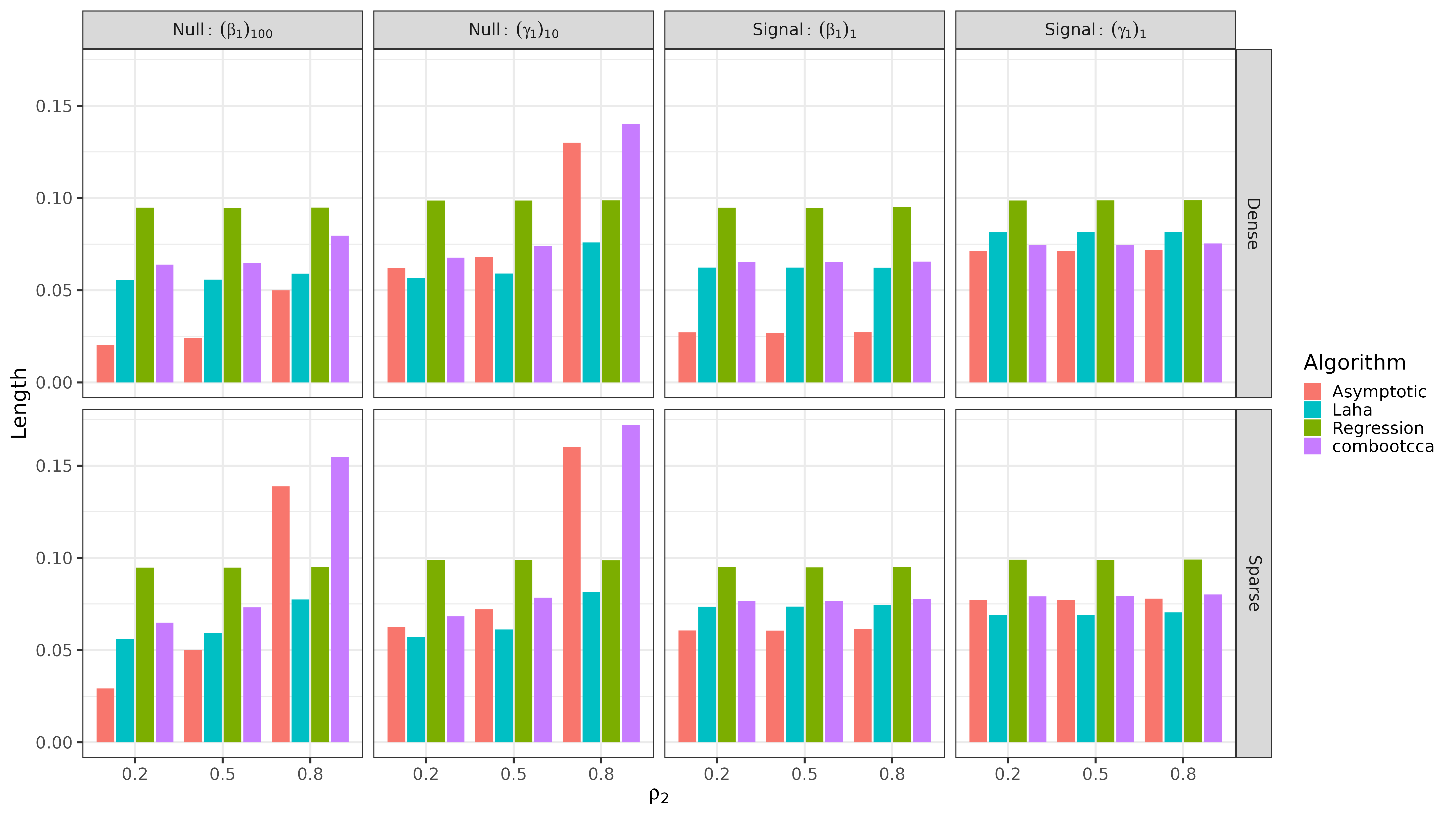}}
  \caption{Lengths of confidence intervals for first canonical directions in simulation II for $p = 100, q = 10$.}
  \label{fig:sim-sprec-ii-length1-high}
\end{figure}

\begin{figure}[htb!]
  \centering
  \includegraphics[width=\textwidth]{{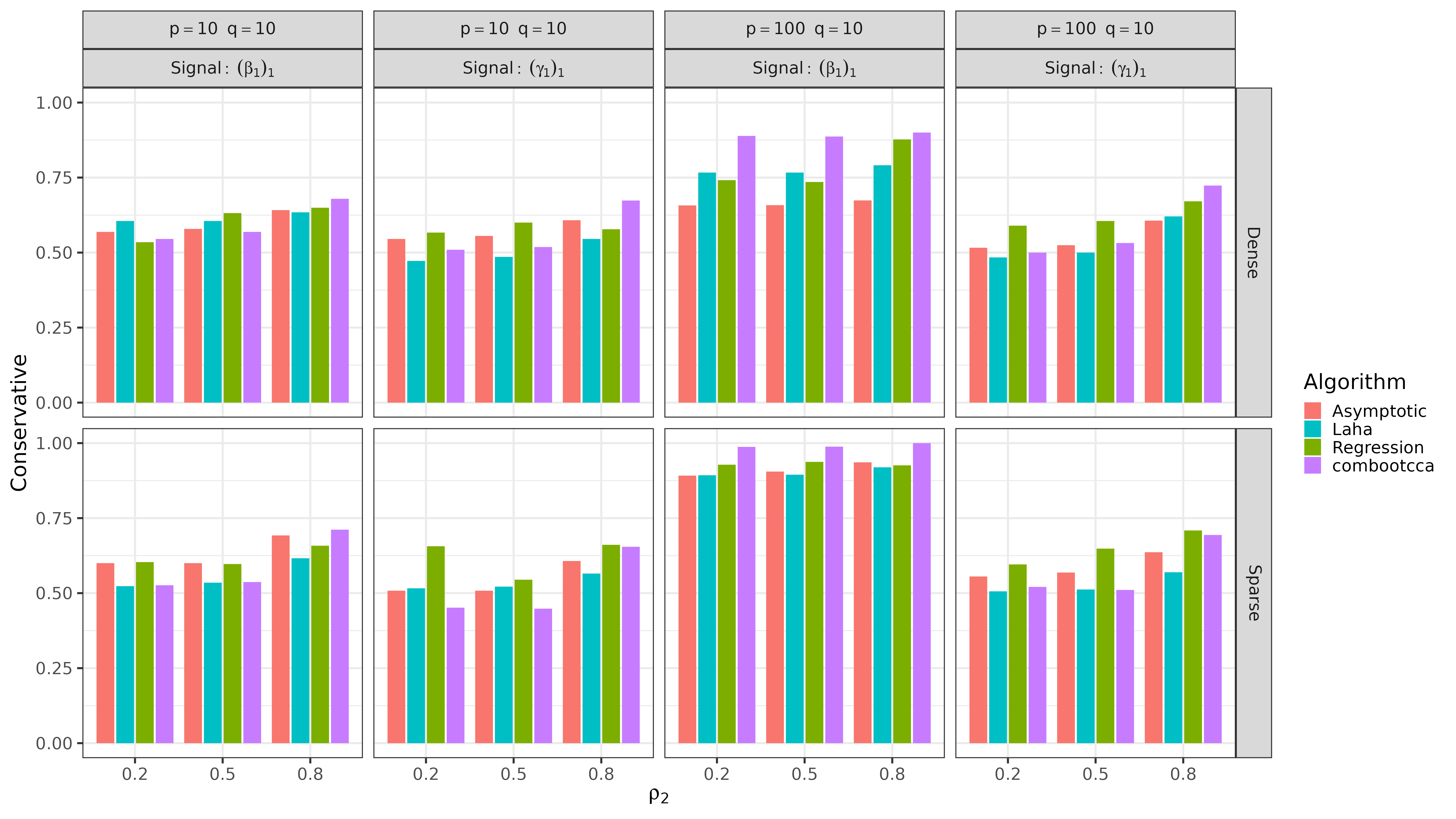}}
  \caption{Bias in simulation II for first canonical directions: the proportion of confidence intervals that failed to cover non-null signals that are ``conservative'' (the true value is greater in magnitude than any value in the confidence interval).}
  \label{fig:sim-sprec-ii-bias1}
\end{figure}

\begin{figure}[htb!]
  \centering
  \includegraphics[width=\textwidth]{{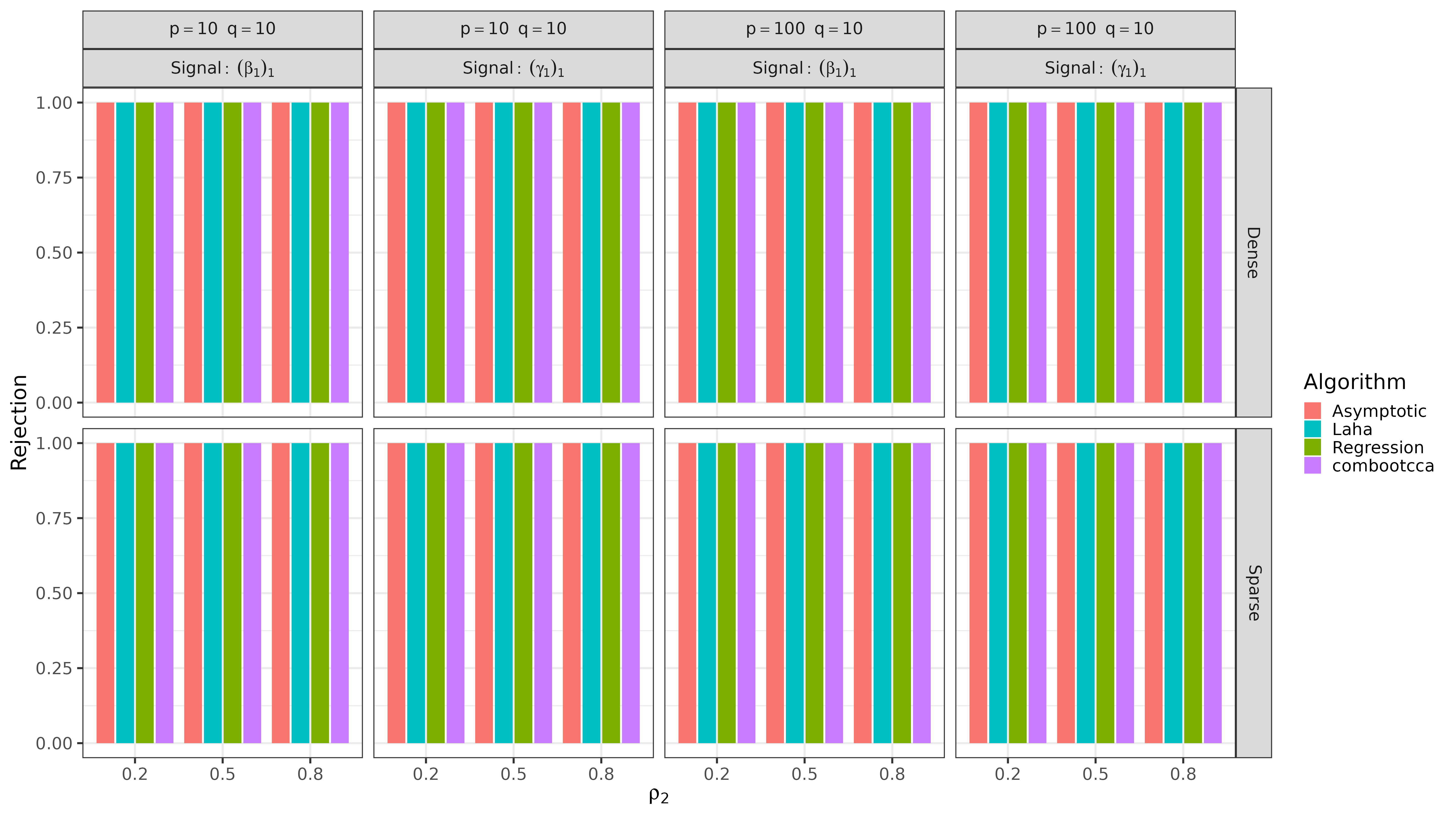}}
  \caption{Power (correct rejection rates) for first canonical directions in simulation II.}
  \label{fig:sim-sprec-ii-power1}
\end{figure}

We first evaluate how coverage for the first canonical direction (with $\rho_1$ fixed at \num{0.9}) varies when the strength of the second canonical correlation varies.
Coverage is shown in Figure \ref{fig:sim-sprec-ii-coverage1-low} for $p = q = 10$ and in Figure \ref{fig:sim-sprec-ii-coverage1-high} for $p = 100, q = 10$,
with corresponding lengths shown in Figures \ref{fig:sim-sprec-ii-length1-low} and \ref{fig:sim-sprec-ii-length1-high}.
Results are generally similar to Simulation I when $\rho_1$ was set to \num{0.9},
although both the method of \citet{laha2021StatisticalInferenceHigh} and regression-based method have poor coverage of null coordinates (i.e., inflated Type I error) when $\rho_2$ is large,
which suggests that a narrower gap between the canonical correlations is especially detrimental for this method.
For regression, we conjecture that in this setting, the initial estimate of the canonical directions is more likely to be inaccurate,
and because the regression method effectively performs inference conditional on this value,
it provides ``valid'' inference but for the wrong quantity.
Examining the lengths makes clear that the combootcca and asymptotic methods are appropriately sensitive to this small gap in the canonical correlations and make their confidence intervals wider than the other methods.

We repeat the investigation of bias in confidence intervals described in Section \ref{sec:combootcca-sim-i} and arrive at a similar conclusion: sparse signals yield more conservative (biased) intervals,
which can harm coverage while retaining good power.
These results are depicted in Figures \ref{fig:sim-sprec-ii-bias1} and \ref{fig:sim-sprec-ii-power1}.
All methods have very good power for the first canonical direction,
which is unsurprising as it is associated with a large canonical correlation ($\rho_1 = 0.9$).

\begin{figure}[htb!]
  \centering
  \includegraphics[width=\textwidth]{{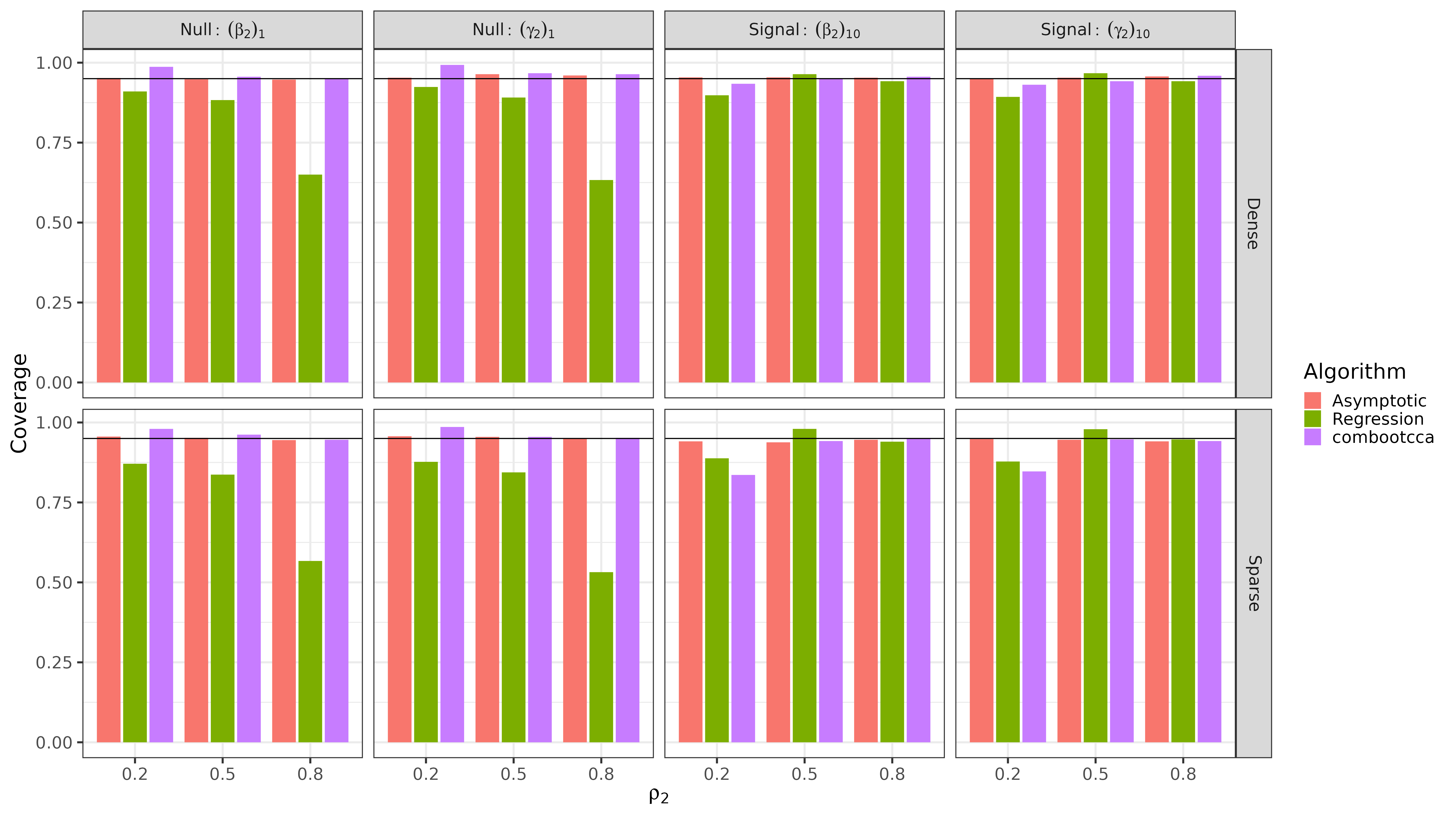}}
  \caption{Coverage rates for second canonical directions in simulation II for $p = q = 10$. The horizontal line indicates nominal 95\% coverage.}
  \label{fig:sim-sprec-ii-coverage2-low}
\end{figure}

\begin{figure}[htb!]
  \centering
  \includegraphics[width=\textwidth]{{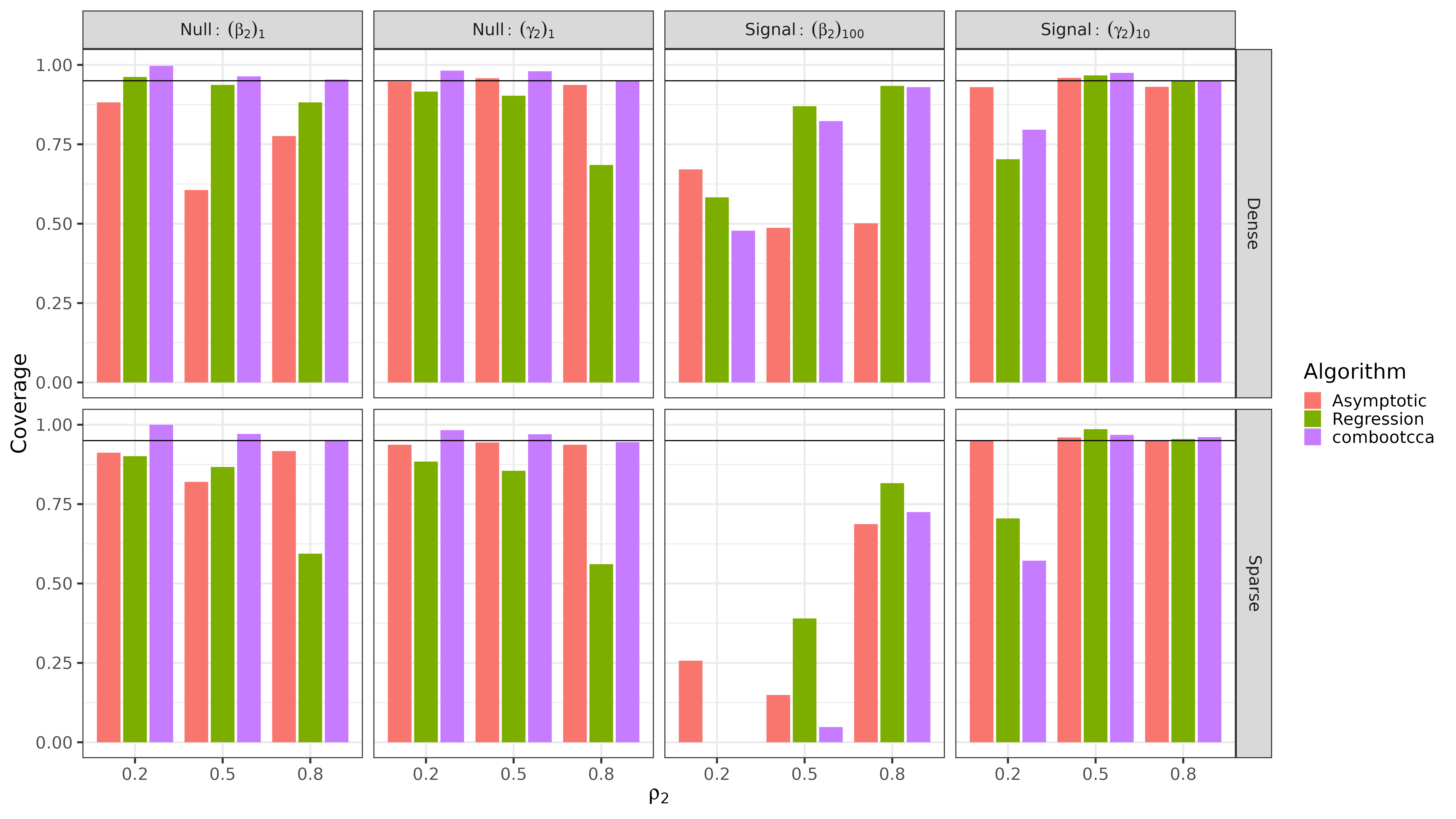}}
  \caption{Coverage rates for second canonical directions in simulation II for $p = 100, q = 10$. The horizontal line indicates nominal 95\% coverage.}
  \label{fig:sim-sprec-ii-coverage2-high}
\end{figure}

\begin{figure}[htb!]
  \centering
  \includegraphics[width=\textwidth]{{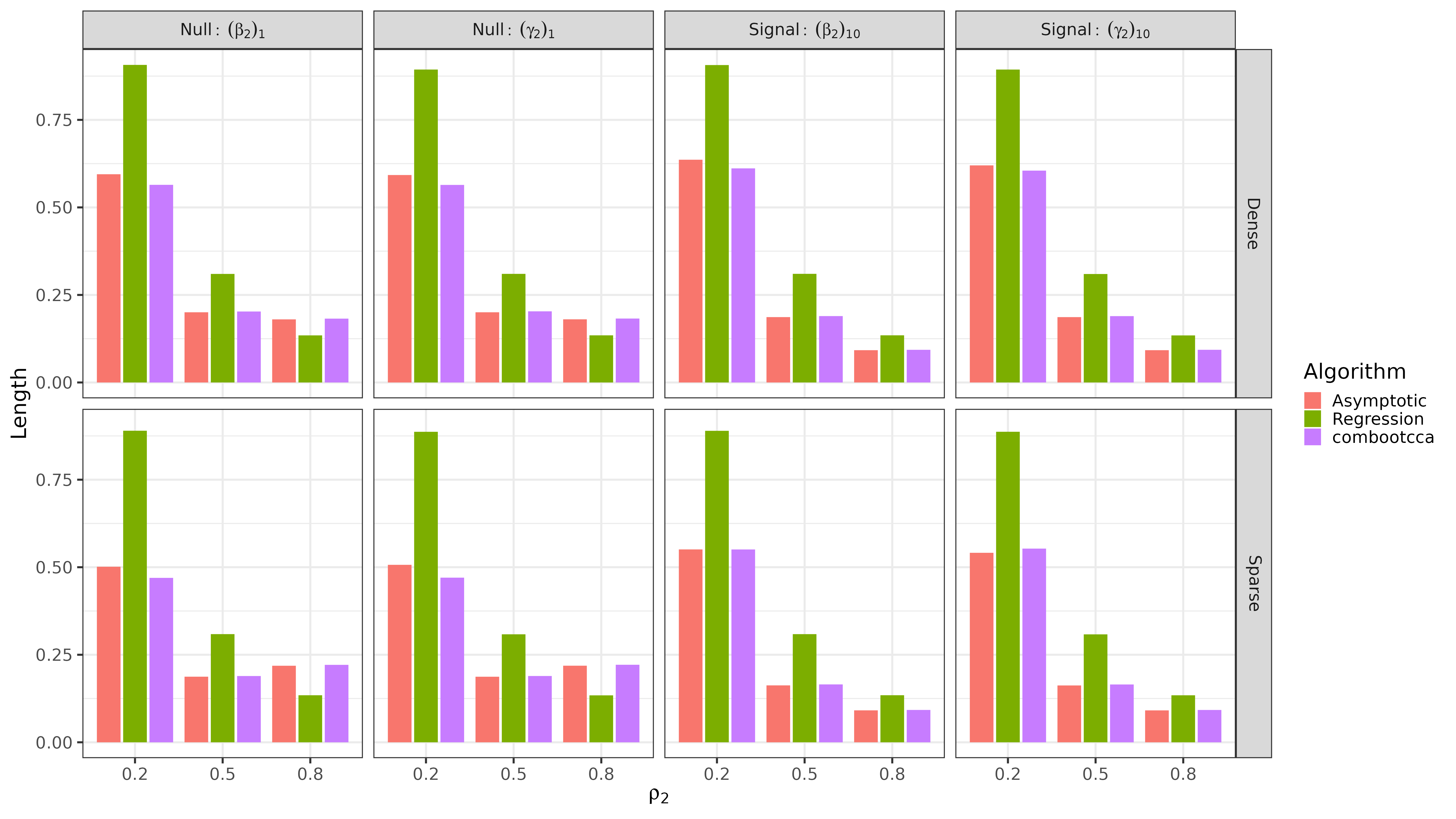}}
  \caption{Lengths of confidence intervals for second canonical directions in simulation II for $p = q = 10$.}
  \label{fig:sim-sprec-ii-length2-low}
\end{figure}

\begin{figure}[htb!]
  \centering
  \includegraphics[width=\textwidth]{{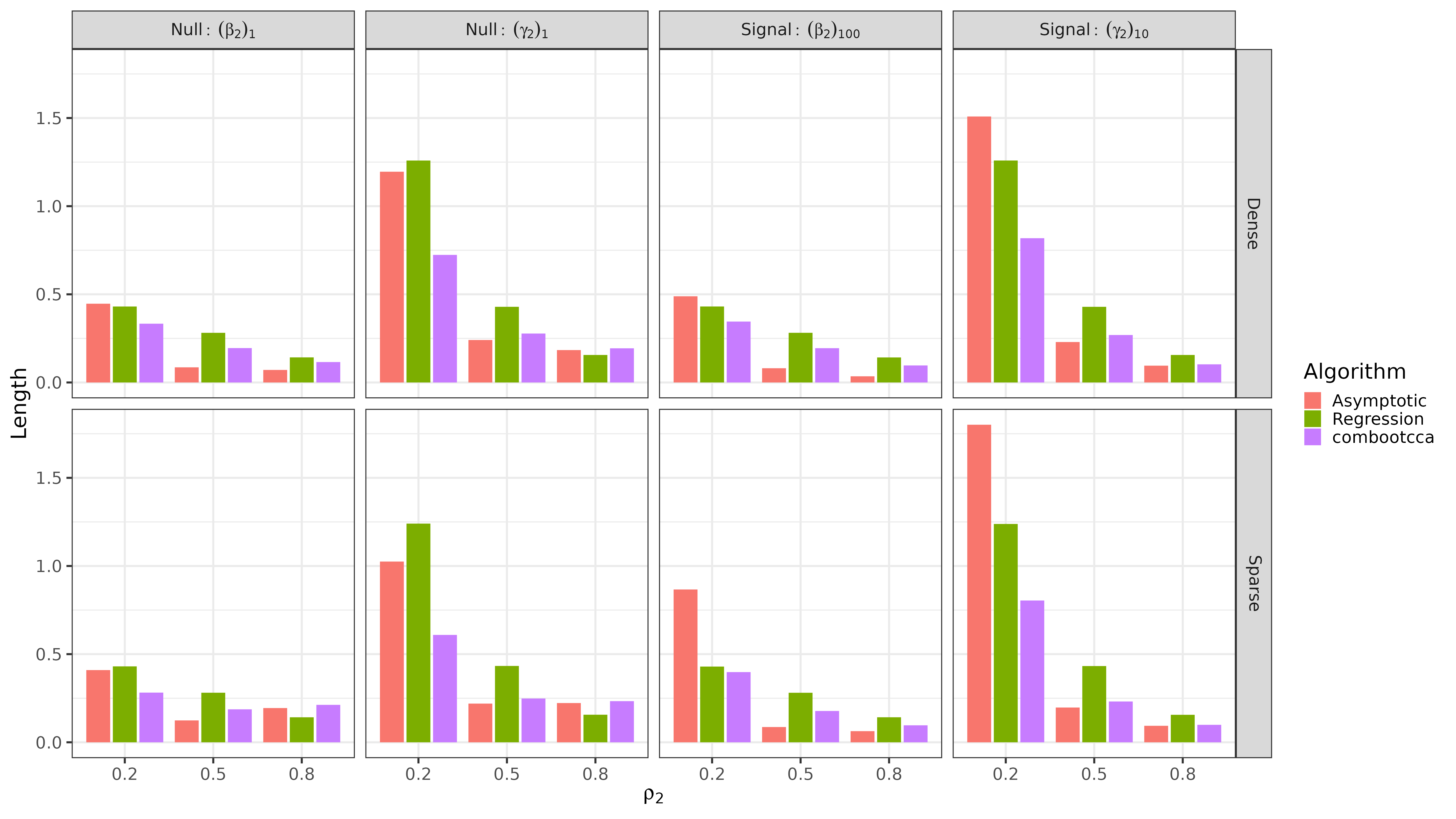}}
  \caption{Lengths of confidence intervals for second canonical directions in simulation II for $p = 100, q = 10$.}
  \label{fig:sim-sprec-ii-length2-high}
\end{figure}

\begin{figure}[htb!]
  \centering
  \includegraphics[width=\textwidth]{{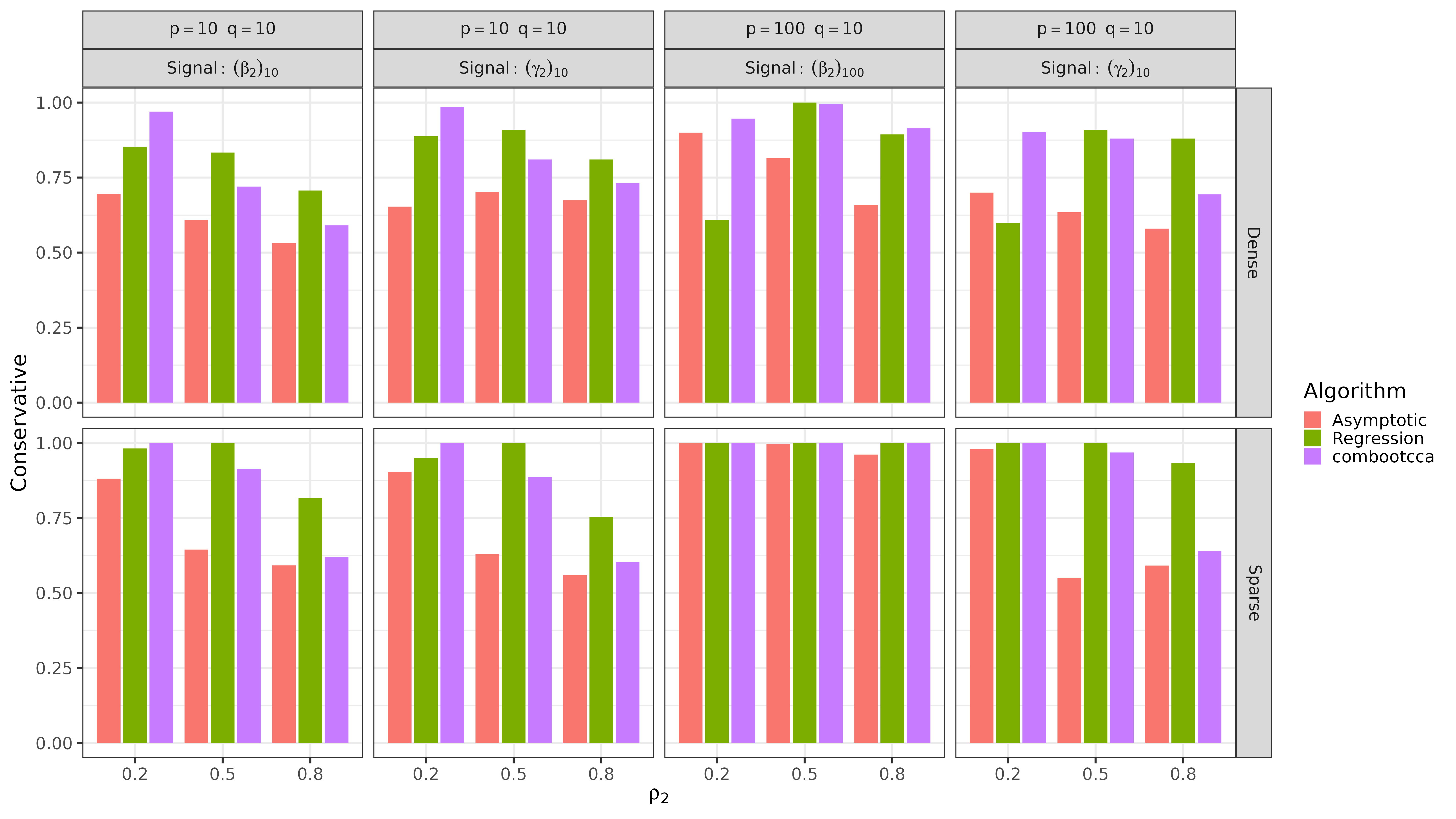}}
  \caption{Bias in simulation II for second canonical directions: the proportion of confidence intervals that failed to cover non-null signals that are ``conservative'' (the true value is greater in magnitude than any value in the confidence interval).}
  \label{fig:sim-sprec-ii-bias2}
\end{figure}

\begin{figure}[htb!]
  \centering
  \includegraphics[width=\textwidth]{{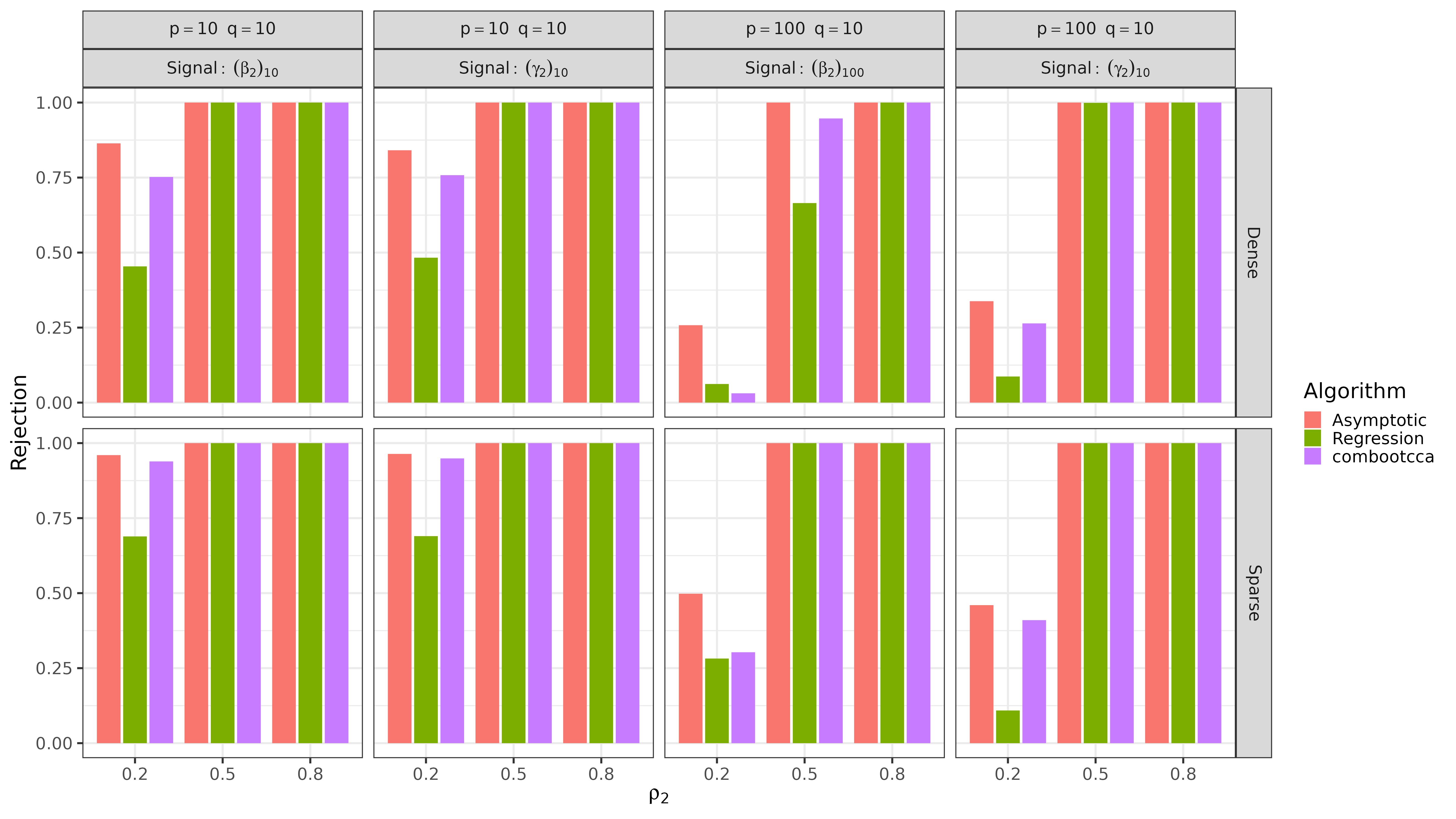}}
  \caption{Power (correct rejection rates) for second canonical directions in simulation II.}
  \label{fig:sim-sprec-ii-power2}
\end{figure}

Next, we examine coverage of the second canonical direction as $\rho_2$ varies.
Coverage is depicted in Figure \ref{fig:sim-sprec-ii-coverage2-low} for $p = q = 10$ and in Figure \ref{fig:sim-sprec-ii-coverage2-high} for $p = 100, q = 10$,
with associated lengths depicted in Figures \ref{fig:sim-sprec-ii-length2-low} and \ref{fig:sim-sprec-ii-length2-high}.
Recall that the method of \citet{laha2021StatisticalInferenceHigh} is only applicable for the \emph{first} canonical directions,
so it offers no results here.
The only method with nominal coverage of null coordinates (and thus valid control of Type I error) is combootcca;
the asymptotic approach works for $p = q$ but fails when $p \neq q$.
When the signal is dense, combootcca has nominal coverage of signal coordinates when $p = q$,
and when $p \neq q$ it approaches nominal coverage for signal coordinates except when $\rho_2 = 0.2$.
When the signal is sparse and $p \neq q$, combootcca again suffers from poor coverage of signal coordinates,
and we see in Figure \ref{fig:sim-sprec-ii-bias2} that this again reflects overly conservative (in magnitude) confidence intervals,
but that combootcca still has non-trivial power as depicted in Figure \ref{fig:sim-sprec-ii-power2}.

\subsection{Simulation III: Data-Based Simulation}
\label{sec:combootcca-sim-data-based}

While Simulation I and II were useful for studying the behavior of various methods in simple settings,
Simulation III offers a more realistic setting.
Rather than specifying $\Sigma_x, \Sigma_y$ along with $\rho, B$, and $\Gamma$ a priori,
we instead specify them based on our neuroimaging data set as described below in Section \ref{sec:combootcca-application-abcd}.
Specifically, we take the estimated covariance $\hat{\Sigma}$ for our processed and cleaned data,
i.e., the empirical covariances of $X_2$ and $Y_2$,
as well as their cross-covariance.
Using the empirical covariance as the ground truth,
we solve the population version of CCA as described in Section \ref{sec:combootcca-cca-pop-model},
and we arrive at corresponding values for $\rho, B$, and $\Gamma$.
However, the directions are fully dense with variable magnitudes,
and so modifications are necessary in order to carefully study the empirical statistical properties of confidence intervals.
Specifically, we modify the last coordinate of one of $\beta_1, \beta_2, \gamma_1$, or $\gamma_2$.
We set it to take one of the following three values:
(i) 0 (in which case it corresponds to a true null),
(ii) the mean of the absolute values of the other entries of the direction,
(iii) the max of the absolute values of the other entries of the direction.
In both cases (ii) and (iii), we preserve the sign of the coordinate.
After this modification, we need to reconstruct the generative covariance $\Sigma$,
since it will no longer correspond to this modified solution.
We invert the CCA model as described in Section \ref{sec:combootcca-cca-inverse-model}
to recover a $\Sigma$ that fits the desired CCA solution,
and then we generate data from it.
Our choice of $\Sigma$ requires that we leave $(p, q)$ fixed at $(250, 11)$ (to correspond to the application),
and we similarly set $N = 2969$ to correspond to the application.

\begin{figure}[htb!]
  \centering
  \includegraphics[width=\textwidth]{{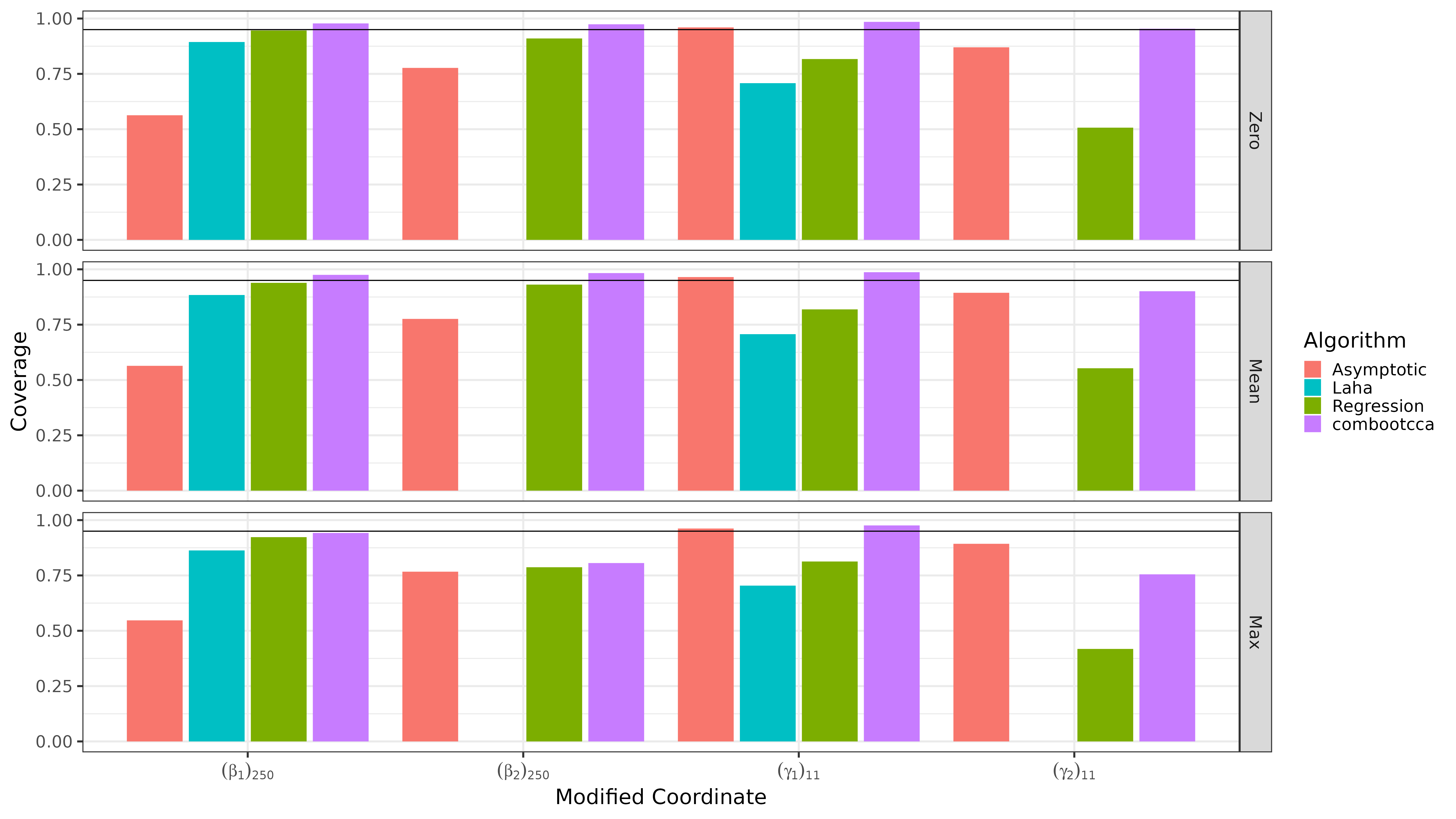}}
  \caption{Coverage rates in simulation III.}
  \label{fig:sim-iii-coverage}
\end{figure}

\begin{figure}[htb!]
  \centering
  \includegraphics[width=\textwidth]{{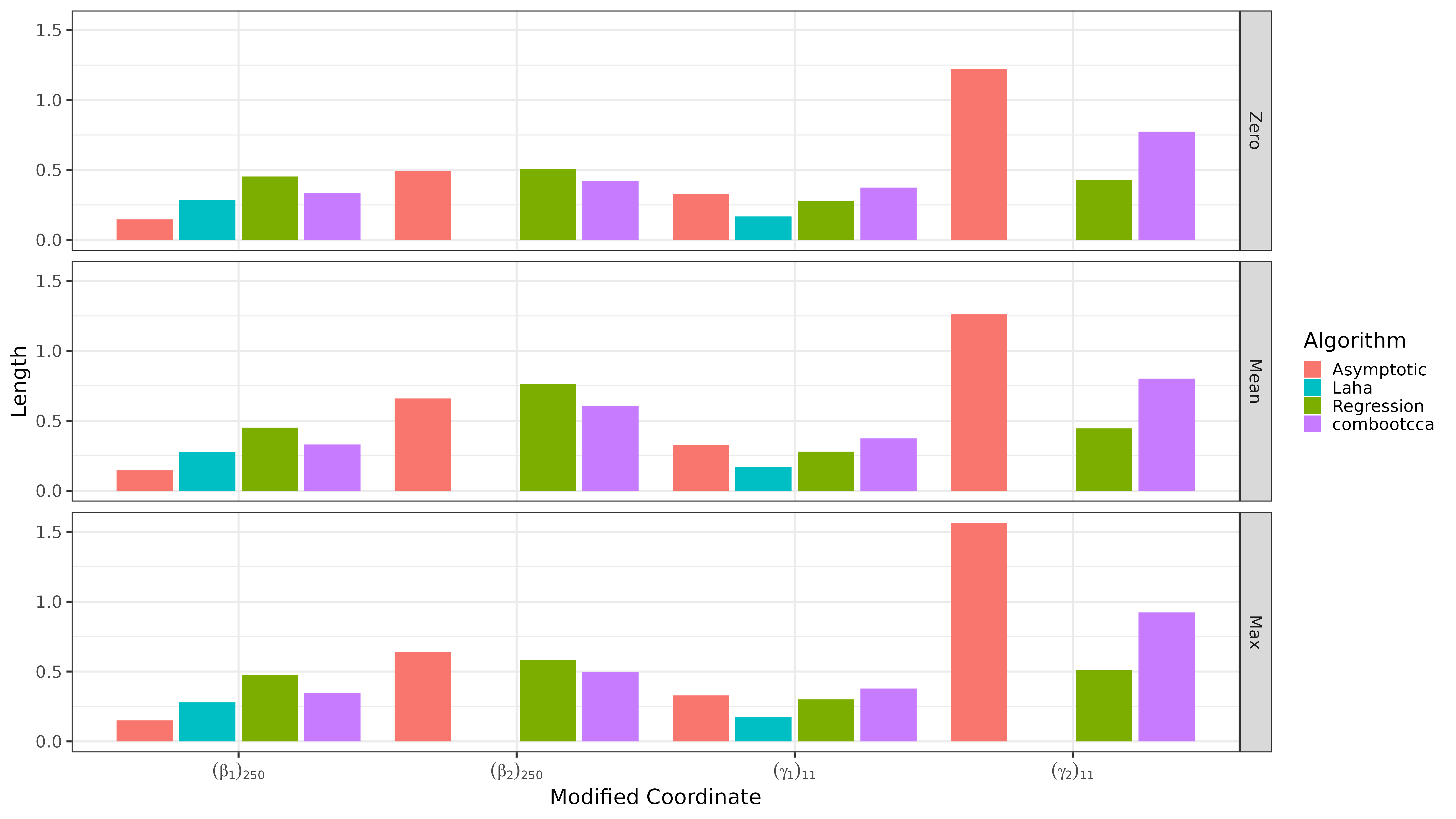}}
  \caption{Lengths of confidence intervals in simulation III.}
  \label{fig:sim-iii-length}
\end{figure}

\begin{figure}[htb!]
  \centering
  \includegraphics[width=\textwidth]{{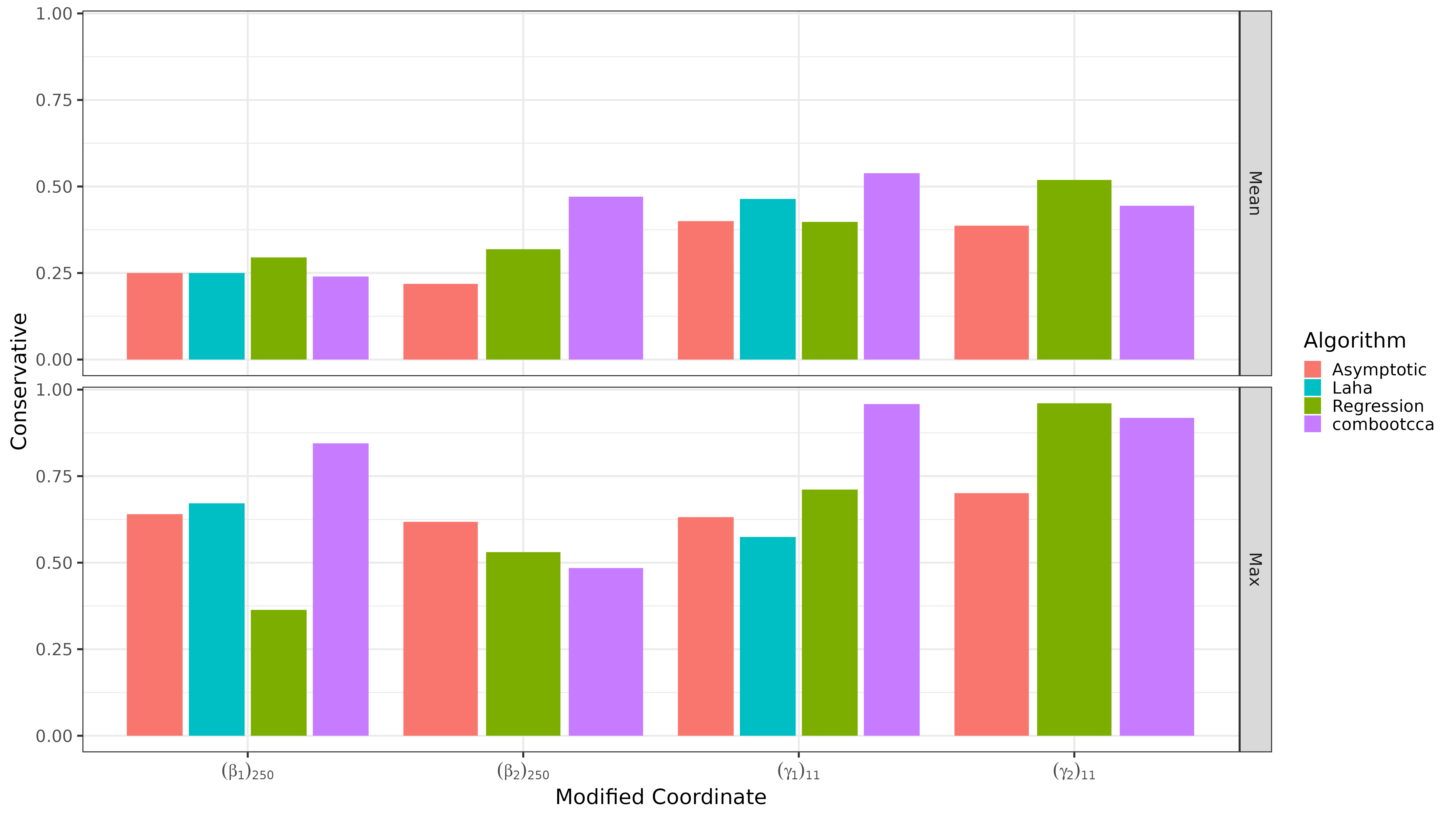}}
  \caption{Bias in simulation III.  The proportion of confidence intervals that failed to cover non-null signals that are ``conservative'' (the true value is greater in magnitude than any value in the confidence interval).}
  \label{fig:sim-iii-bias}
\end{figure}

\begin{figure}[htb!]
  \centering
  \includegraphics[width=\textwidth]{{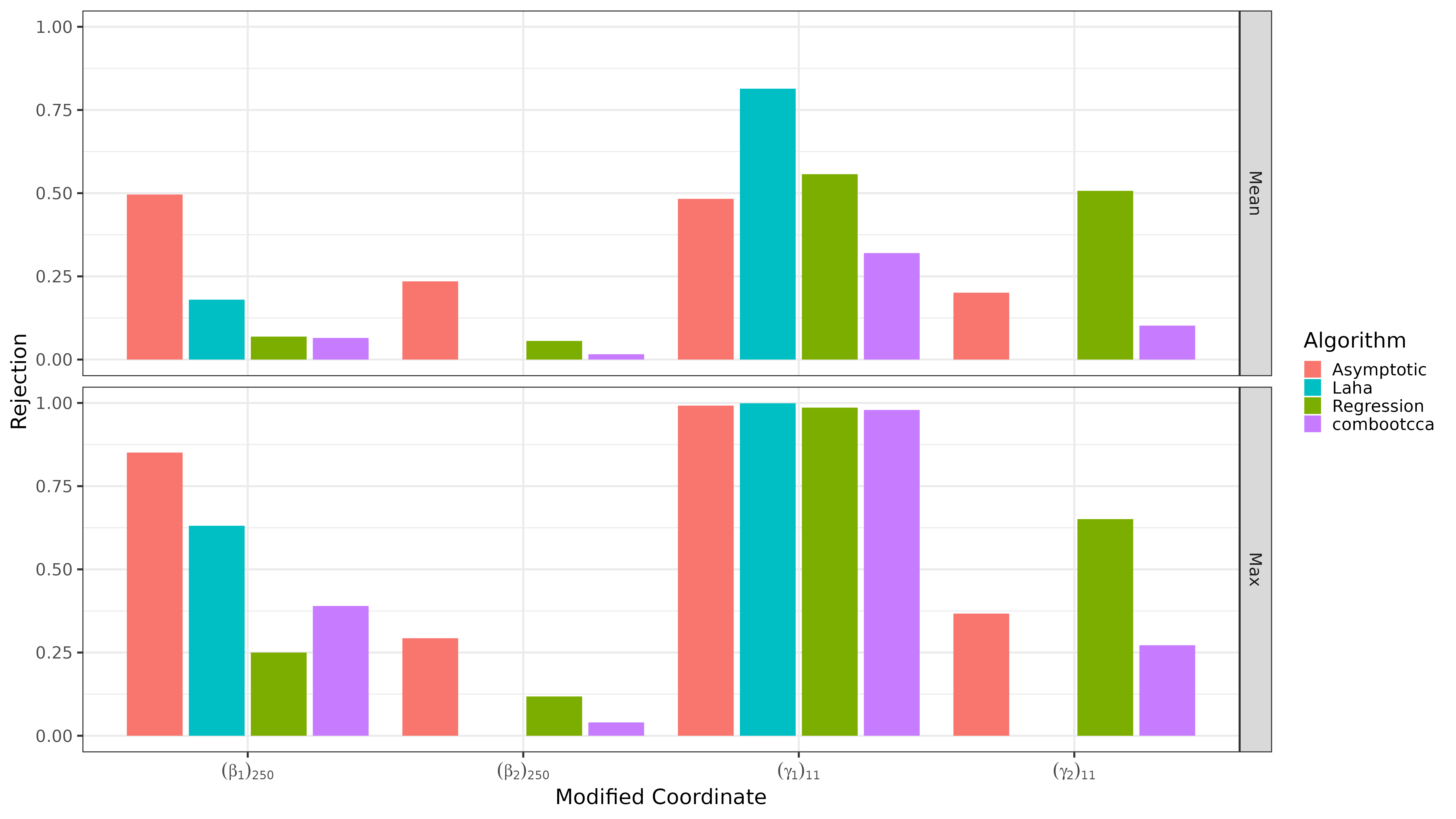}}
  \caption{Power (correct rejection rates) in simulation III.}
  \label{fig:sim-iii-power}
\end{figure}

We present coverage for all methods in Figure \ref{fig:sim-iii-coverage}
with associated lengths in Figure \ref{fig:sim-iii-length}.
When considering only the first direction, only combootcca attains nominal coverage in all settings,
while the asymptotic method does so for the low-dimensional $\gamma_1$.
The method of \citet{laha2021StatisticalInferenceHigh} has coverage close to (but short of) nominal for $\beta_1$,
but poor coverage for $\gamma_{1}$,
and offers no results for the second directions $\beta_2$ and $\gamma_2$.
When examining results for the second canonical directions $\beta_2$ and $\gamma_2$, combootcca
has nominal (or near nominal) coverage for null and moderate signal,
but falls short of nominal coverage when the signal is set to its maximum.
This is consistent with our earlier simulation studies where confidence intervals for coordinates with large values are conservatively biased, which can be confirmed in Figure \ref{fig:sim-iii-bias}.
In Figure \ref{fig:sim-iii-power}, we show the power for each method.
Although not the most powerful method,
combootcca is unique in that it had nominal Type I error control,
but still enjoys non-trivial power, especially in the presence of coordinates with large values.

\subsection{Comparison of Bootstrap Strategies}
\label{sec:combootcca-bootstrap-consequences}

In Section \ref{sec:combootcca-bootstrap},
we outlined different alignment strategies for the bootstrap as well as two approaches to constructing confidence intervals.
Here, we provide empirical evidence to justify our choices in the combootcca method,
namely, performing alignment with a weighted Hungarian algorithm and using percentile  bootstrap.  

\begin{figure}[htb!]
  \centering
  \includegraphics[width=\textwidth]{{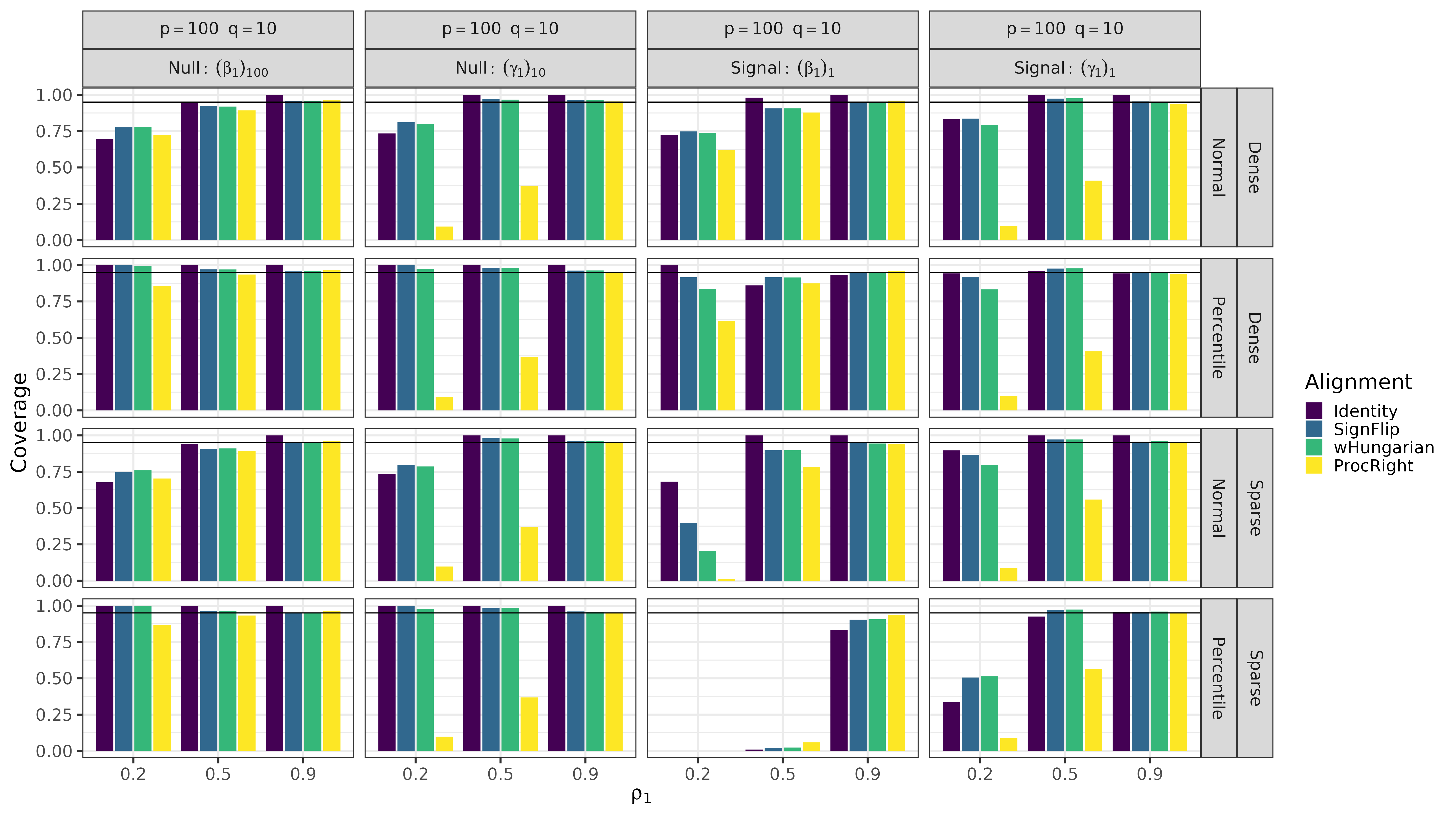}}
  \caption{Comparison of coverage rates for different types of bootstraps and alignment strategies in the setting of simulation I.}
  \label{fig:sim-bootcomp-i-coverage}
\end{figure}

\begin{figure}[ht]
  \centering
  \includegraphics[width=\textwidth]{{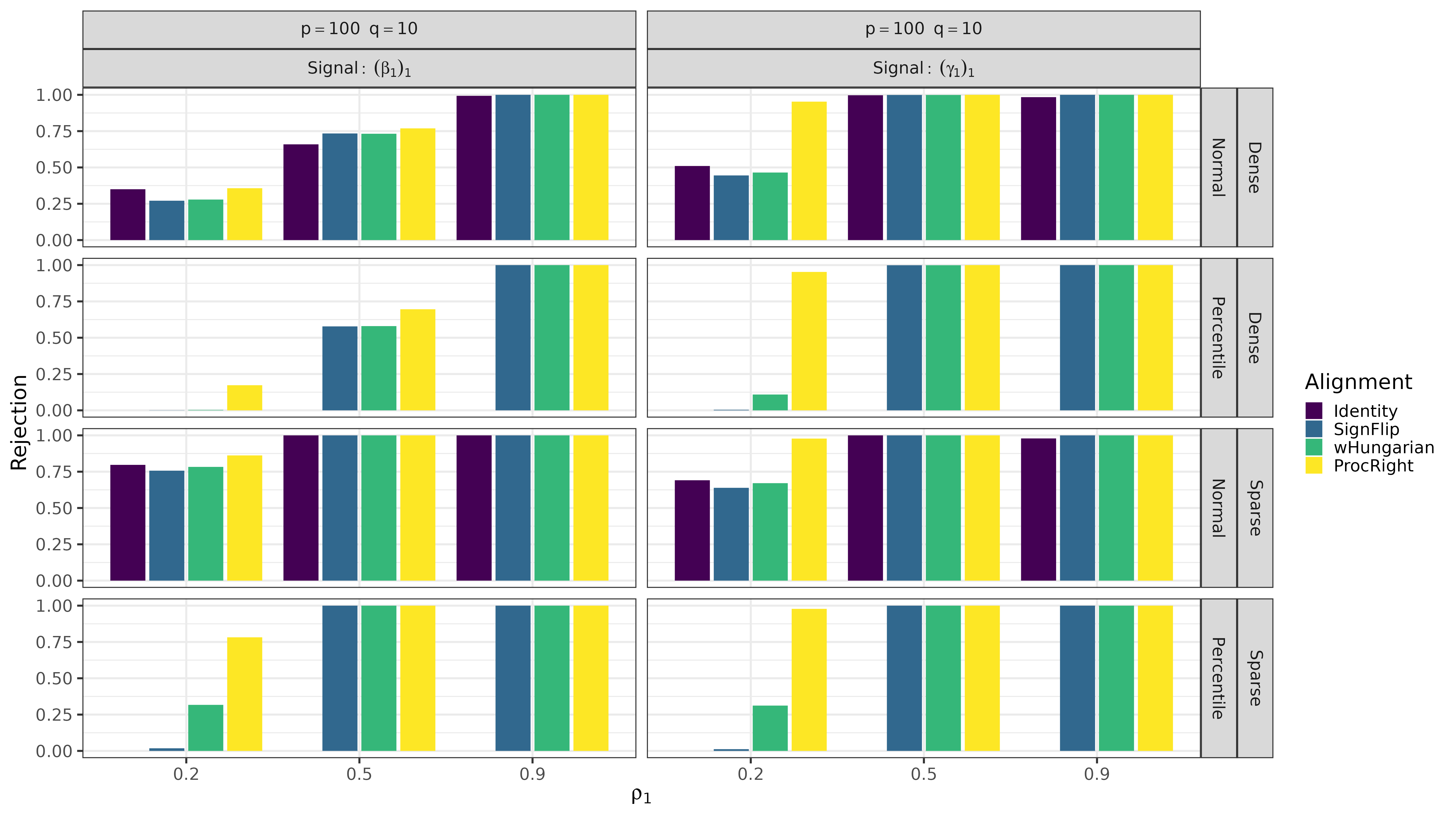}}
  \caption{Comparison of power (correct rejection rates) for different types of bootstraps and alignment strategies in the setting of simulation I.}
  \label{fig:sim-bootcomp-i-power}
\end{figure}

Figure \ref{fig:sim-bootcomp-i-coverage} shows coverage rates in the setting of Simulation I (see Section \ref{sec:combootcca-sim-i}) for all four alignment strategies considered as well as the two different types of confidence intervals.
We can easily see that the bootstrap that uses the normal approximation (in rows 1 and 3) generally fails to achieve nominal coverage when $\rho$ is small, regardless of alignment strategy.
This is a particularly noteworthy observation, given the popularity of this type of bootstrap in the applied literature.
Coverage is generally better for the percentile-based bootstrap,
although as we have seen in the preceding sections, its coverage is poor when the signal is sparse and $\rho_1$ is small,
however this generally reflects a conservative bias in the intervals that still leaves non-trivial power.
Considering alignment strategies, the Procrustes-based alignment fails to achieve nominal coverage in many settings for both the normal-approximation and percentile bootstrap;
this is consistent with our characterization of it as an overly aggressive alignment strategy.
In order to choose between the remaining alignment strategies, we consider power as presented in Figure \ref{fig:sim-bootcomp-i-power}.
We have already eliminated the normal-approximation bootstrap from consideration due to its poor coverage,
and when using the percentile-based bootstrap we see that the weighted Hungarian alignment has the best power.

\begin{figure}[ht]
  \centering
  \includegraphics[width=\textwidth]{{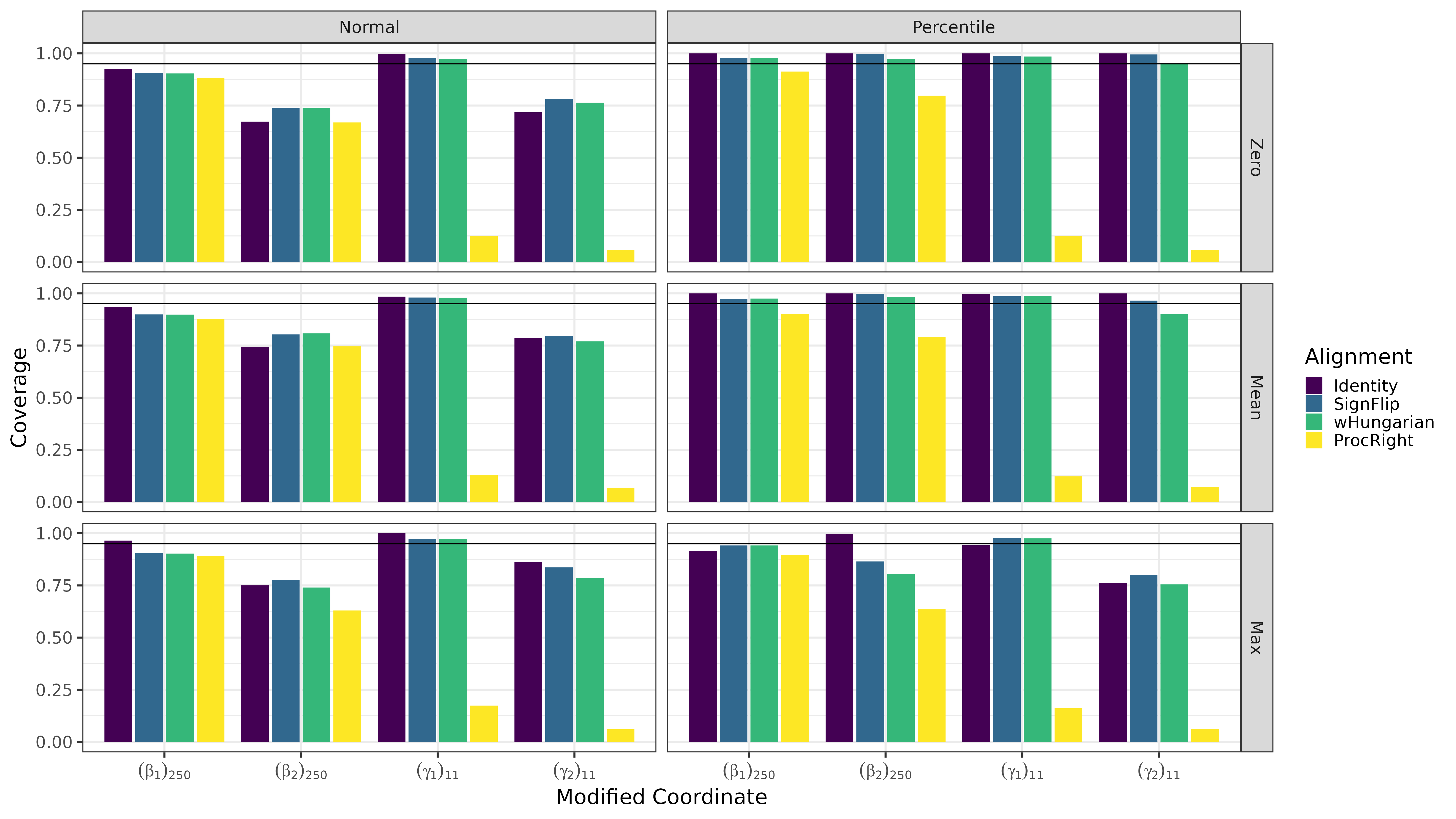}}
  \caption{Comparison of coverage rates for different types of bootstraps and alignment strategies in the setting of simulation III.}
  \label{fig:sim-bootcomp-iii-coverage}
\end{figure}

\begin{figure}[ht]
  \centering
  \includegraphics[width=\textwidth]{{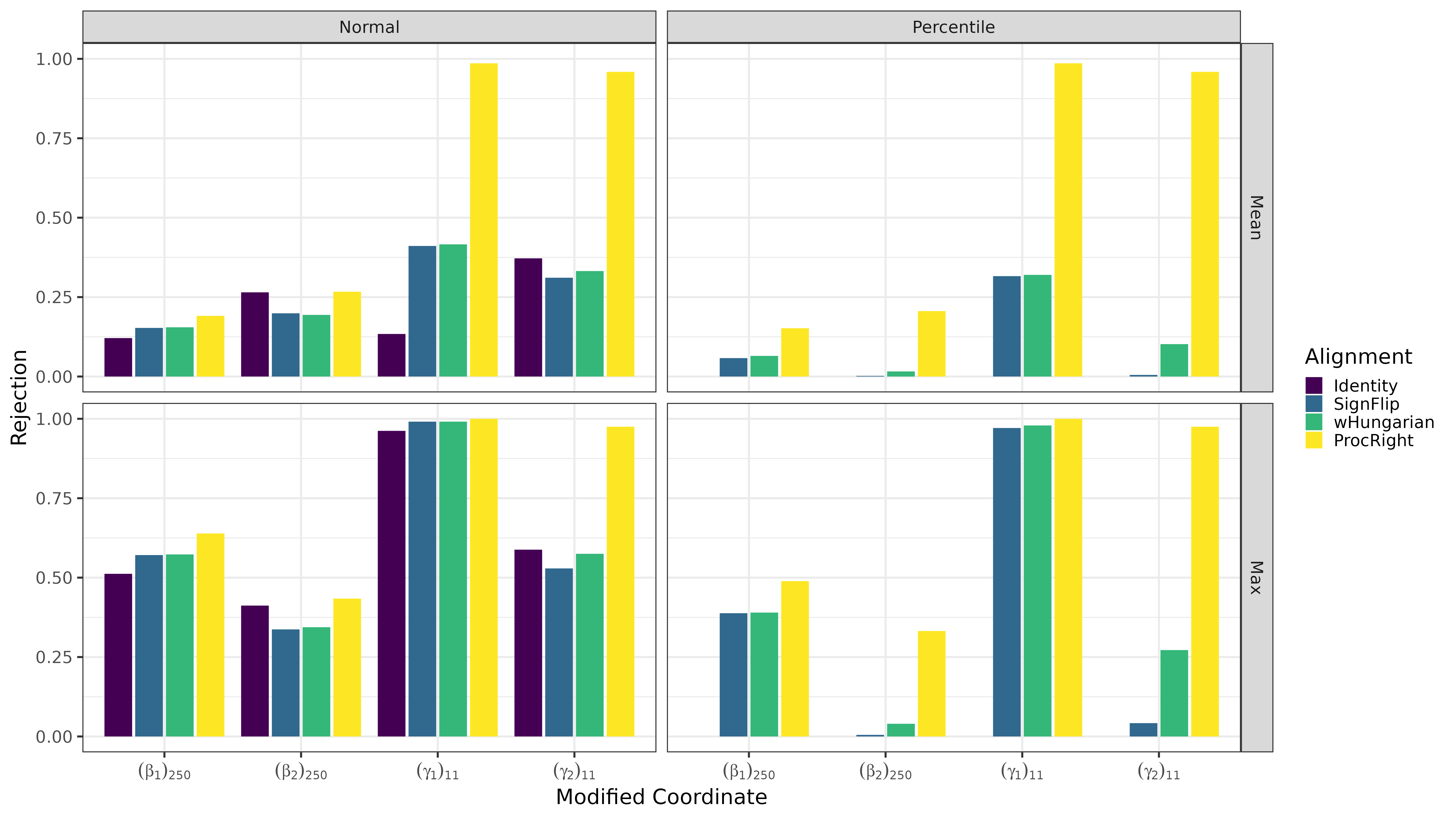}}
  \caption{Comparison of power (correct rejection rates) for different types of bootstraps and alignment strategies in the setting of simulation III.}
  \label{fig:sim-bootcomp-iii-power}
\end{figure}

In order to verify that this recommendation is not specific to our toy simulation study,
we also study the choice of alignment strategy and type of confidence interval in the more realistic setting of Simulation III (see Section \ref{sec:combootcca-sim-data-based}).
In Figure \ref{fig:sim-bootcomp-iii-coverage}, we can again see that the normal approximation is problematic for coverage,
and that Procrustes-based alignment can yield very poor coverage, especially for the smaller dimensional $\gamma_1$ and $\gamma_2$.
Turning to power as depicted in Figure \ref{fig:sim-bootcomp-iii-power},
the weighted Hungarian alignment coupled with percentile-based bootstraps remains the winning combination.

\section{Application to ABCD Dataset}
\label{sec:combootcca-application-abcd}
We apply our methods to data taken from the ABCD study \citep{casey2018AdolescentBrainCognitive} processed by the lab of our collaborator,
Dr.~Chandra Sripada.
We work with an initial corpus of $5937$ individuals who have complete data on 
(i) usable resting state functional magnetic resonance imaging (fMRI) data,
(ii) behavioral performance scores on $11$ tasks, and
(iii) nuisance covariates.
This same subset of the ABCD data was analyzed in \citet{sripada2021BrainwideFunctionalConnectivity},
which describes the data and processing in more detail.
Interestingly, our analysis which aims to tie behavioral tests to biological measurements,
is very much in the spirit of Hotelling's seminal work which introduced CCA \citep{hotelling1936RelationsTwoSets},
which suggested on its first page that with CCA,
``the scores on a number of mental tests may be compared with physical measurements on the same persons.''

The initial data matrix $\tilde{X} \in \mathbb{R}^{5937 \times \binom{418}{2}}$ holds functional connectivity data.
Each row corresponds to a vectorized correlation matrix taken pairwise over time between $418$ parcels in the brain according to the parcellation of \citet{gordon2016GenerationEvaluationCortical}.
The initial data matrix $\tilde{Y} \in \mathbb{R}^{5937 \times 11}$ holds behavior scores for the same participants on a corpus of 11 tasks taken from the neurocognition assessment from the ABCD study and described in more detail in \citet{luciana2018AdolescentNeurocognitiveDevelopment}.
In brief,
seven of the tasks are taken from the NIH Toolbox \citep{hodes2013NIHToolboxSetting}:
(i) Picture Vocabulary (Vocabulary),
(ii) Oral Reading Recognition (Reading),
(iii) Pattern Comparison Processing Speed (Processing Speed),
(iv) List Sorting Working Memory (Working Memory),
(v) Picture Sequence Memory (Episodic Memory).
(vi) Flanker Inhibitory Control \& Attention (Flanker), and
(vii) Dimensional Change Card Sort (Card Sort).
From the Rey Auditory Verbal Learning Test,
we use performance in both the
(viii) Short Delay (Memory: Short Delay) and
(ix) Long Delay (Memory: Long Delay)
conditions.
Finally, we also used performance in the following tasks:
(x) Matrix Reasoning, and
(xi) Little Man Task (Spatial Rotation).

We corrected for six nuisance covariates for each participant, namely age, age$^2$, sex, meanFD, meanFD$^2$, and race/ethnicity.  
MeanFD is a summary measure of how much the participant moved their head during the resting state scanning session.
After adding an intercept column of ones and dummy-coding categorical nuisance covariates,
we obtained the nuisance matrix $W \in \mathbb{R}^{5937 \times 10}$.
Before performing CCA, we first remove variation associated with the nuisance covariates and then reduce the dimension of the neuroimaging data.
We randomly partitioned our data into two roughly equally-sized  sets, 
\begin{equation*}
  \tilde{X} =
  \begin{bmatrix}
    \tilde{X}_1 \\
    \tilde{X}_2
  \end{bmatrix}, \quad
  \tilde{Y} =
  \begin{bmatrix}
    \tilde{Y}_1 \\
    \tilde{Y}_2
  \end{bmatrix}, \quad
  W =
  \begin{bmatrix}
    W_1 \\
    W_2
  \end{bmatrix}.
\end{equation*}
Using the training data $(\tilde{\M{X}}_1, \tilde{\M{Y}}_1, \tilde{\M{W}}_1)$ , we regress the variables of interest on the nuisance covariates, obtaining the coefficients 
\begin{equation*}
  \hat{\M{A}}_{X} = \left( \M{W}_1^{\intercal} \M{W}_1 \right)^{-1} \M{W}_1^{\intercal} \M{X}_1,
  \quad \hat{\M{A}}_{Y} = \left( \M{W}_1^{\intercal} \M{W}_1 \right)^{-1} \M{W}_1^{\intercal} \M{Y}_1.
\end{equation*}
We then remove the contributions of nuisance covariates using the coefficients learned in the training data by setting
\begin{align*}
  \check{\M{X}}_1 &= \M{X}_1 - \M{W}_1 \hat{\M{A}}_X,& \check{\M{X}}_2 &= \M{X}_2 - \M{W}_2 \hat{\M{A}}_X, \\
  \check{\M{Y}}_1 &= \M{Y}_1 - \M{W}_1 \hat{\M{A}}_Y,& \check{\M{Y}}_2 &= \M{Y}_2 - \M{W}_2 \hat{\M{A}}_Y .
\end{align*}
Next, we reduced the dimension of the neuroimaging data matrix $\check{\M{X}}$ using PCA.
This is a common processing step upstream of CCA in the neuroimaging literature (see e.g., \citet{helmer2020StabilityCanonicalCorrelation,fernandez-cabello2022AssociationsBrainImaging}).
We learned the PCA transformation on the training data:
because $\check{X}_1$ was already column-centered (since an intercept was included in the nuisance matrix $W_1$),
we performed PCA via SVD and decomposed
$\check{\M{X}}_1 = \M{U} \M{S} \M{V}^{\intercal}$.
Based on input from our collaborators, we retained the leading $250$ principal components and truncated $V$ accordingly.
Then, we projected the held-out, but nuisance-corrected neuroimaging data onto this basis with
$X_2 = \check{X}_2 V_{\left[ :, 1:250 \right]}$.     The phenotypic data is already low-dimensional, so we simply set
$Y_2 = \check{Y}_2$.
As a final preprocessing step, we standardized the columns of $X_2$ and $Y_2$ to have mean 0 and variance 1.

\begin{figure}
  \centering
  \includegraphics[width=\textwidth]{{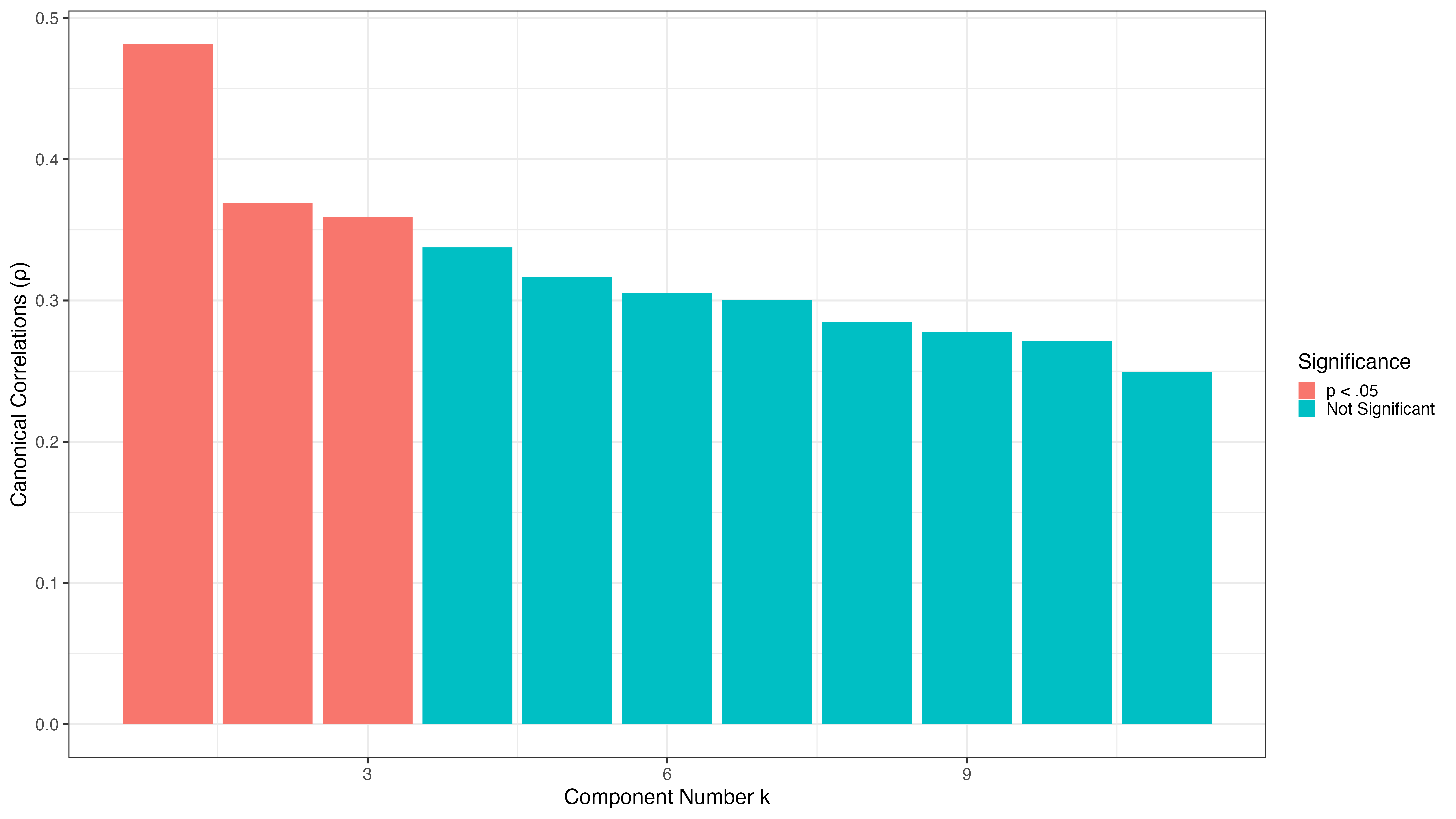}}
  \caption{Results for ABCD: canonical correlations.}
  \label{fig:abcd-rho}
\end{figure}

We then perform CCA on $X_2$ and $Y_2$.  
We use \texttt{R}'s \texttt{cancor} function (which uses QR decomposition internally),
but we rescale the canonical directions as discussed in Section \ref{sec:combootcca-cca-pop-model}.
In Figure \ref{fig:abcd-rho} we plot the canonical correlations.
Although inference for $\rho$ is not the main focus of the present manuscript,
for completeness we report the results of inference on $\rho$ obtained from the \texttt{yacca} package's \texttt{F.test.cca} function \citep{butts2022YaccaAnotherCanonical}.
As depicted in Figure \ref{fig:abcd-rho},
we find that three canonical correlations are significantly nonzero at $\alpha = 0.05$, and we will therefore restrict our attention to inference on first three canonical directions.  
Notably, the first canonical correlation is well-separated from the rest,
but subsequent canonical correlations are not.

\begin{figure}
  \centering
  \includegraphics[width=\textwidth]{{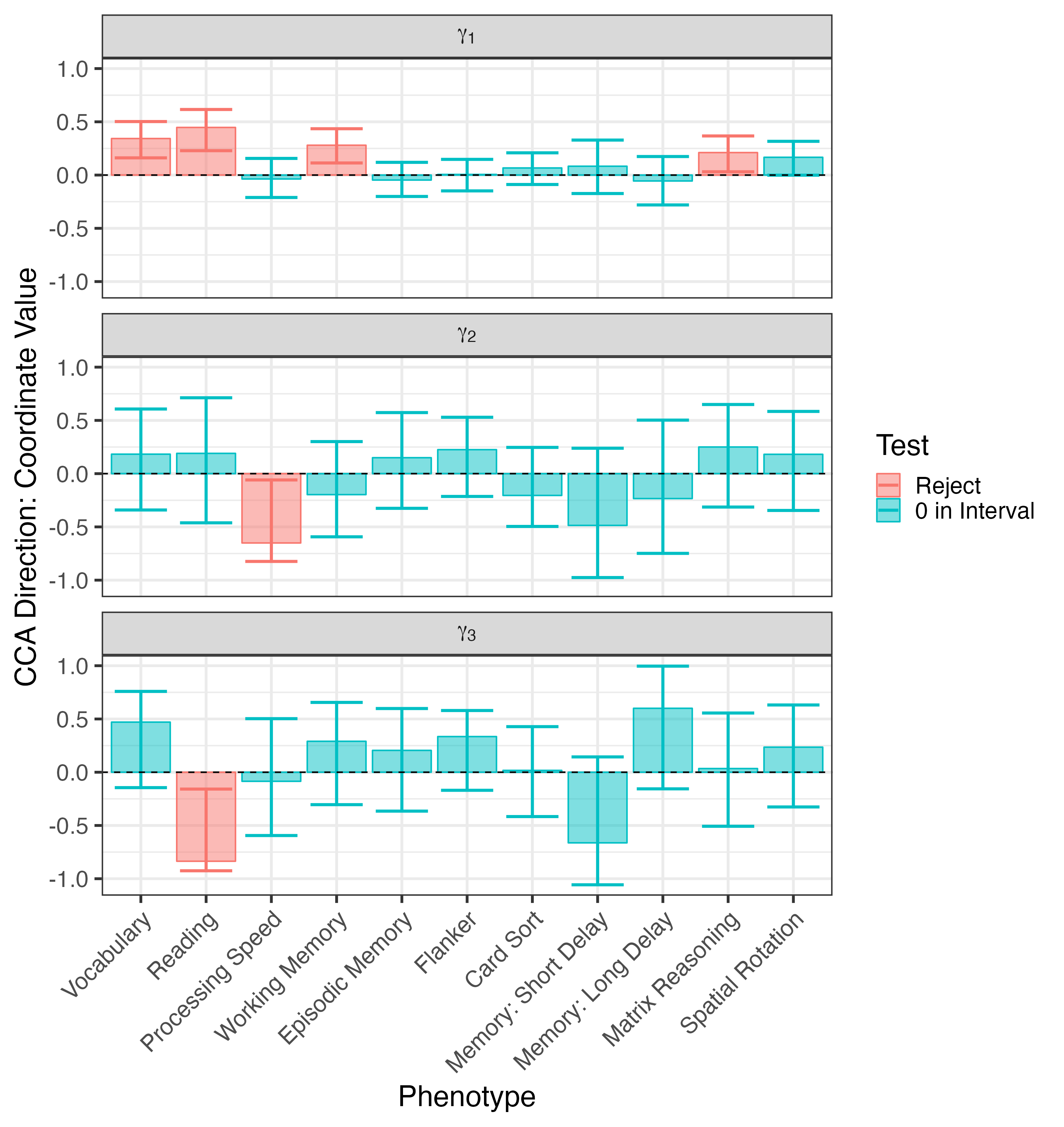}}
  \caption{Results for ABCD: point estimates and confidence intervals for first three canonical directions of $\Gamma$.}
  \label{fig:com-abcd-gamma-inf}
\end{figure}

\begin{figure}
  \centering
  \includegraphics[width=\textwidth]{{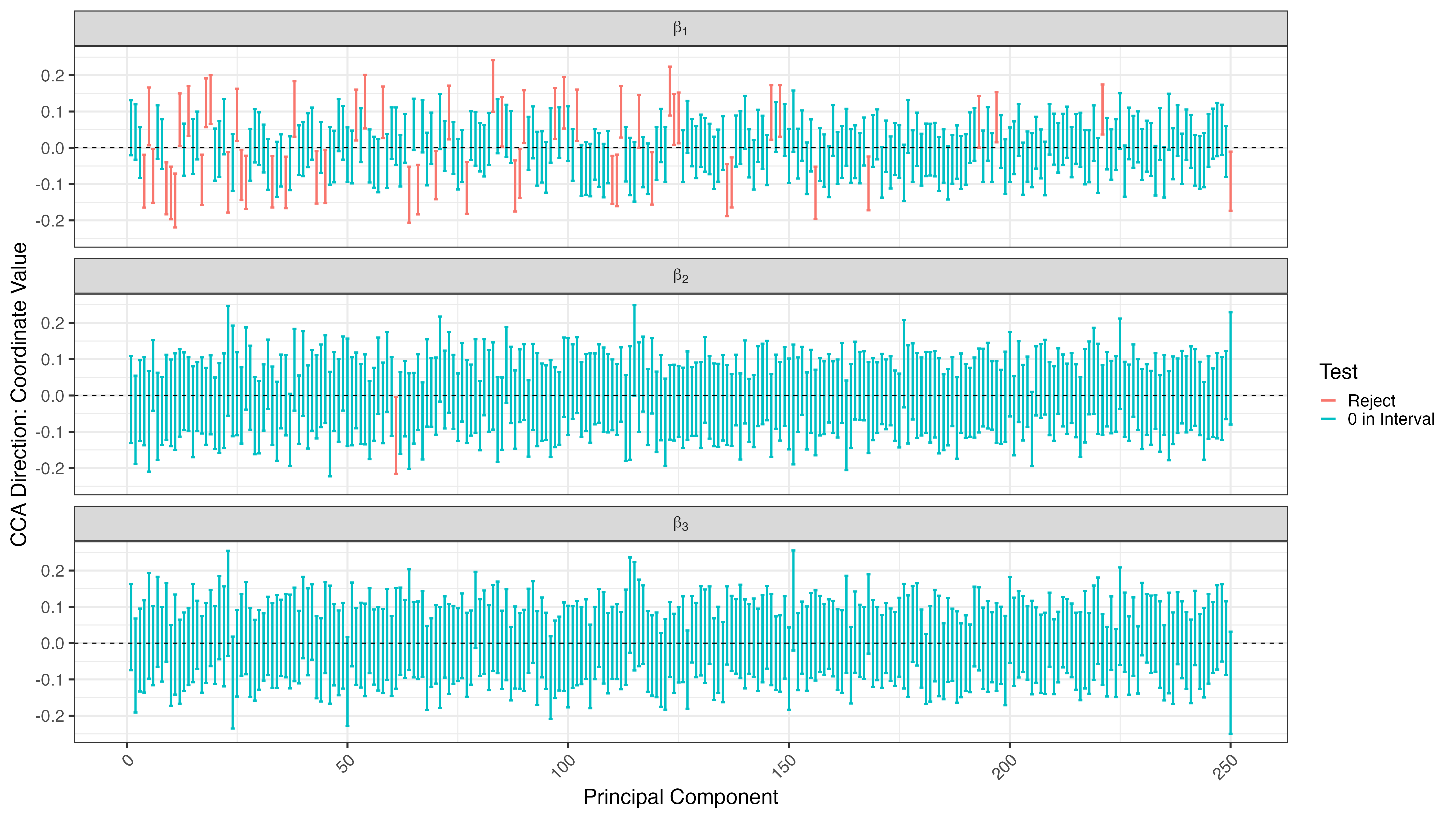}}
  \caption{Results for ABCD: confidence intervals for first three canonical directions of $B$.}
  \label{fig:com:abcd-beta-inf}
\end{figure}

We perform inference on the canonical directions using the combootcca method.
Figure \ref{fig:com-abcd-gamma-inf} shows point estimates and associated confidence intervals for the first three directions of $\Gamma$.
Here, interpretability is aided by the data standardization, as all of the variables are on the same scale.  
The intervals for the first direction are markedly shorter than for the subsequent directions;
this is consistent with greater uncertainty due to the poor separation of canonical correlations beyond the first.
While there is a fundamental sign ambiguity in CCA, the relative signs of the directions can be meaningfully interpreted.
For example, in the first direction,
the confidence intervals for Vocabulary, Reading, Working Memory, and Matrix Reasoning
do not include 0 and all have the same sign.
The fact that these four tasks appear in a single canonical direction is noteworthy:
the Vocabulary and Reading tests are classic hallmarks of ``crystallized intelligence,''
whereas Working Memory and Matrix Reasoning are considered strong indicators of ``fluid intelligence,''
and there is disagreement as to whether these are really two distinct capacities or if they reflect a single general ability.
Indeed, \citet{panigrahi2023SelectiveInferenceSparse} deployed a multi-task learning approach with the same data but restricted their behavioral measures to these four tasks in an investigation of the shared versus distinct neural bases for these two types of cognitive abilities.
Our results can also be compared with those obtained from an independent CCA analysis of the ABCD data in \citet{goyal2022PositiveNegativeMode},
where their post-hoc analysis of their second canonical variate implicates many of the same tasks that we identified.

\begin{figure}[htb!]
  \centering
  \includegraphics[width=\textwidth]{{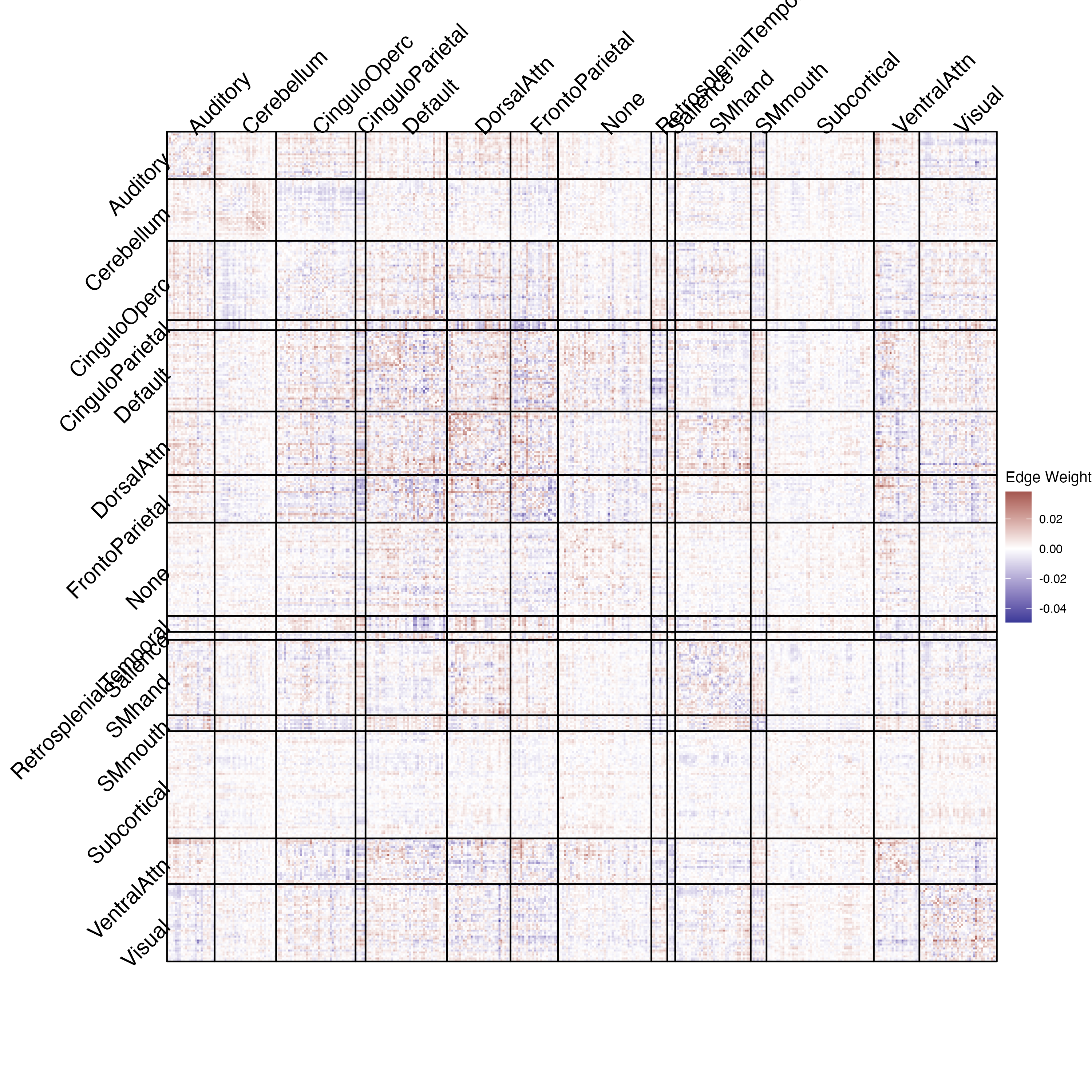}}
  \caption{Brain connectivity features recovered from first canonical direction $\beta_1$.}
  \label{fig:com:betacheck-1-all}
\end{figure}

\begin{figure}[htb!]
  \centering
  \includegraphics[width=\textwidth]{{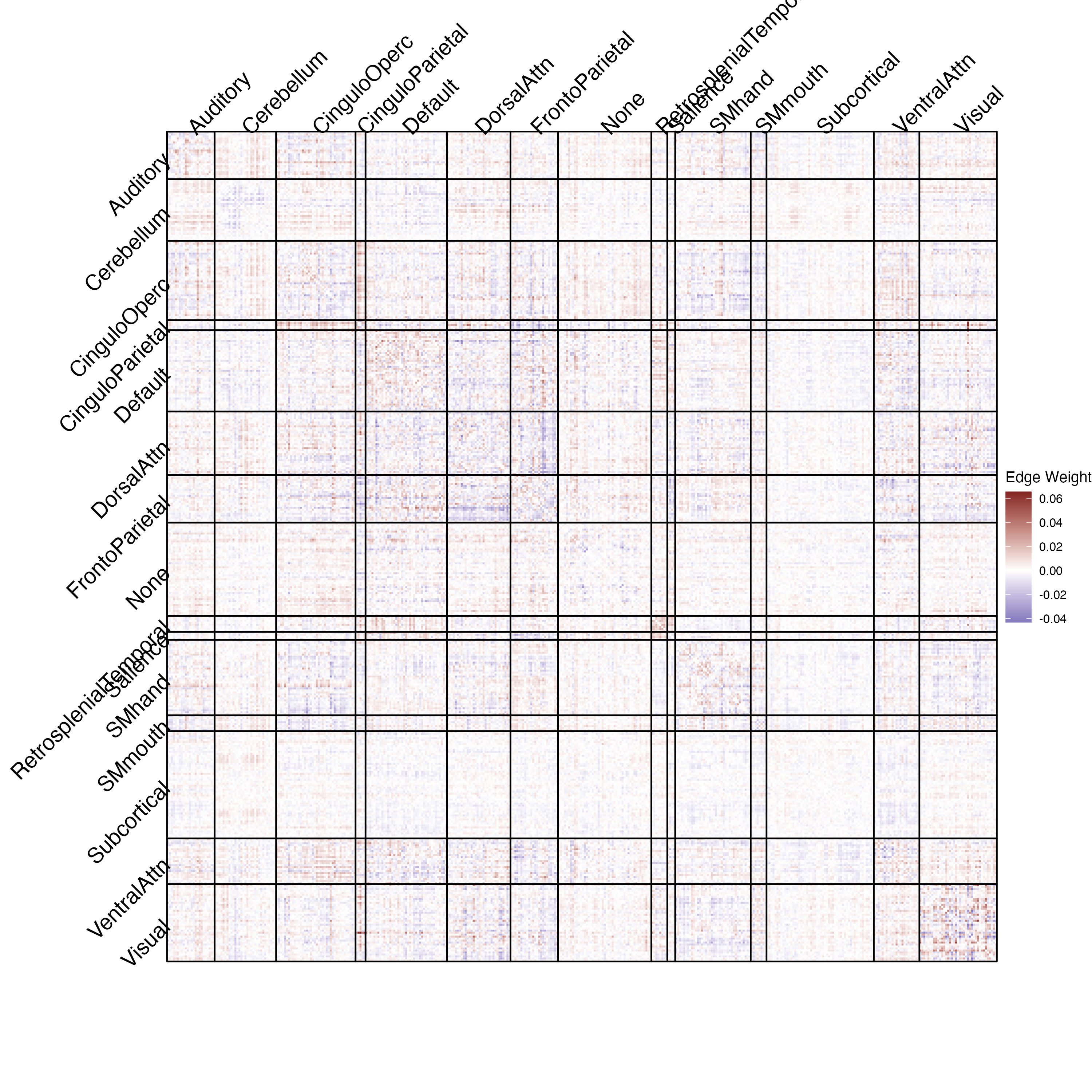}}
  \caption{Brain connectivity features recovered from second canonical direction $\beta_2$.}
  \label{fig:com:betacheck-2-all}
\end{figure}

\begin{figure}
  \centering
  \includegraphics[width=\textwidth]{{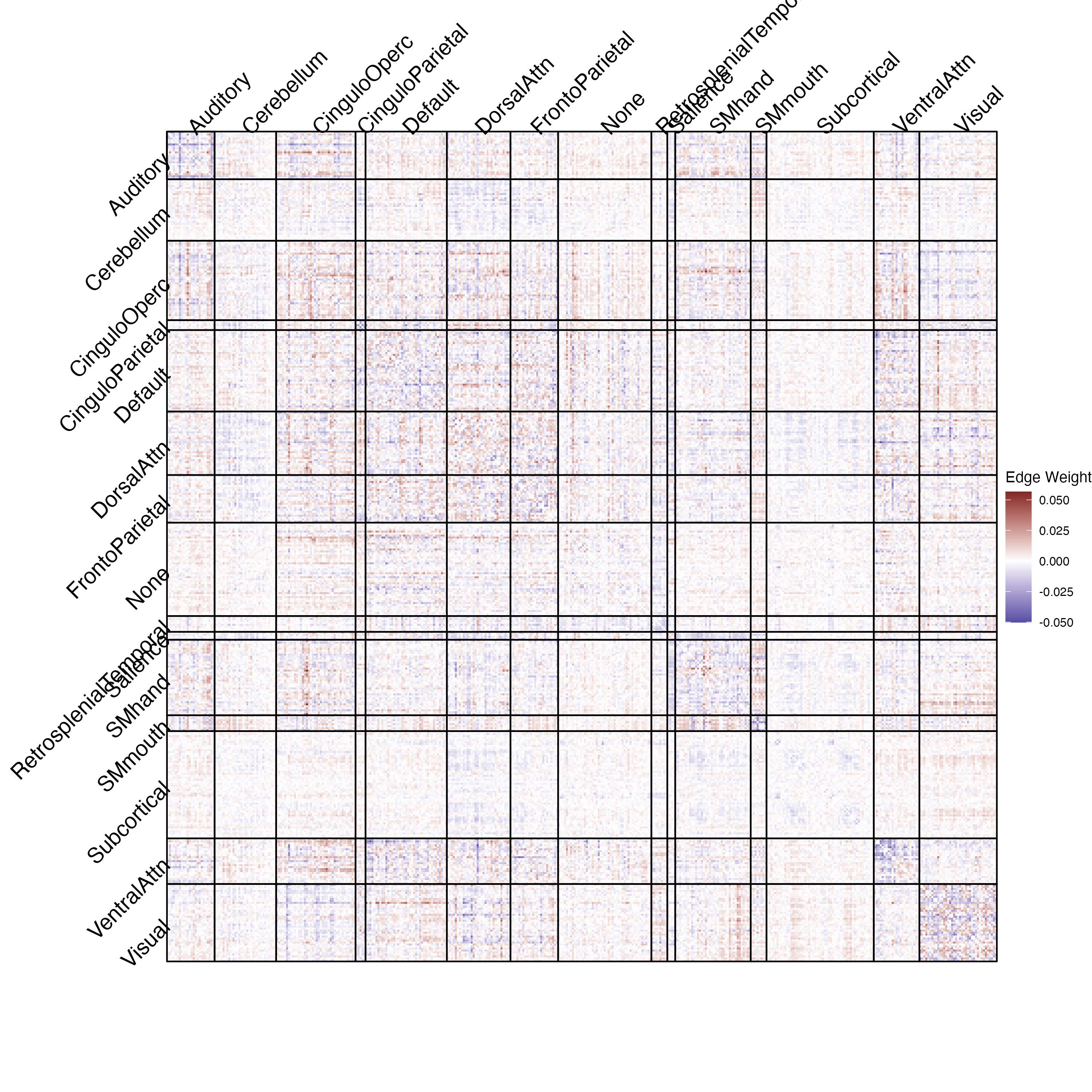}}
  \caption{Brain connectivity features recovered from third canonical direction $\beta_3$.}
  \label{fig:com:betacheck-3-all}
\end{figure}

\begin{figure}
  \centering
  \includegraphics[width=\textwidth]{{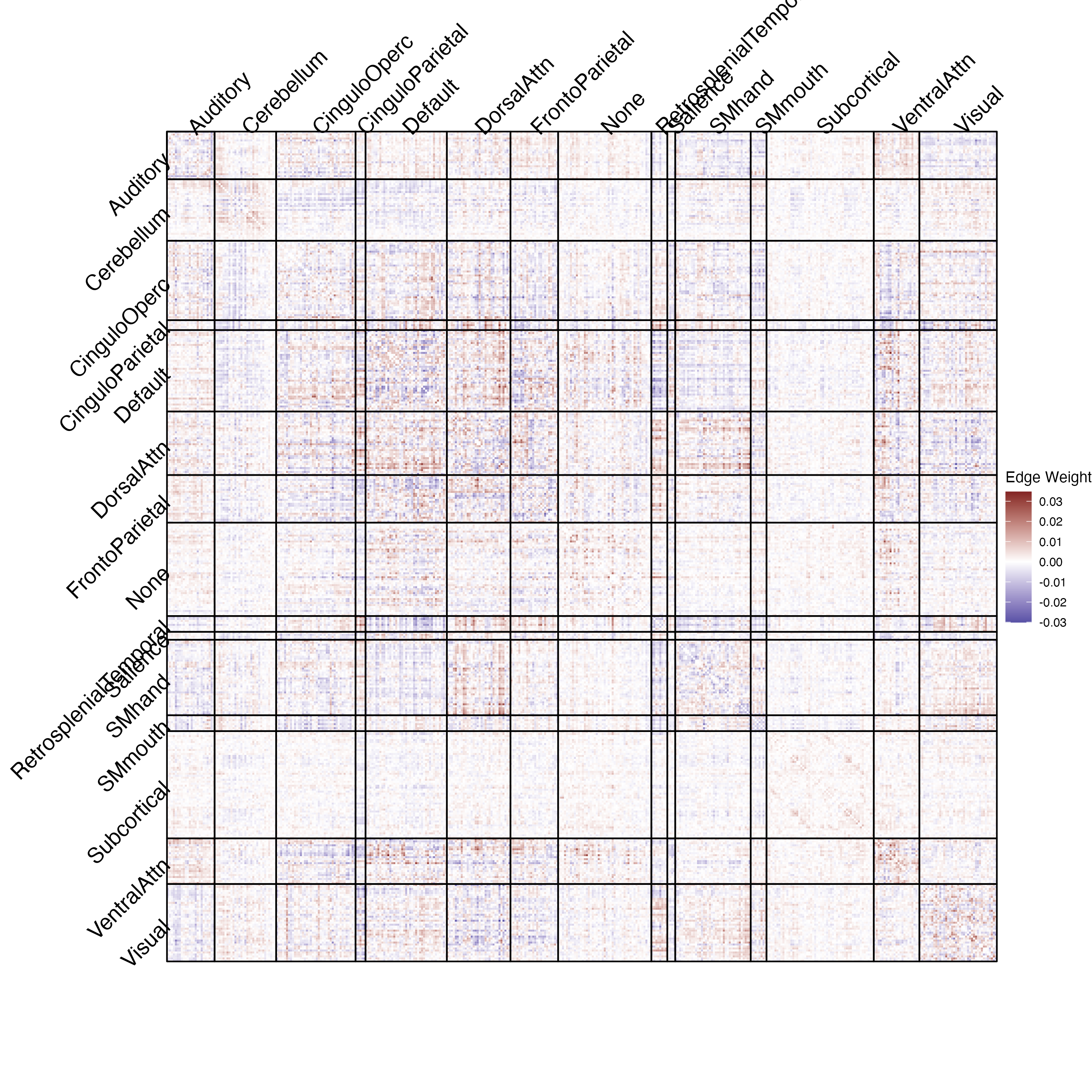}}
  \caption{Brain connectivity features recovered from first canonical direction $\beta_1$ using only significantly nonzero coordinates.}
  \label{fig:com:betacheck-1-sig}
\end{figure}

\begin{figure}
  \centering
  \includegraphics[width=\textwidth]{{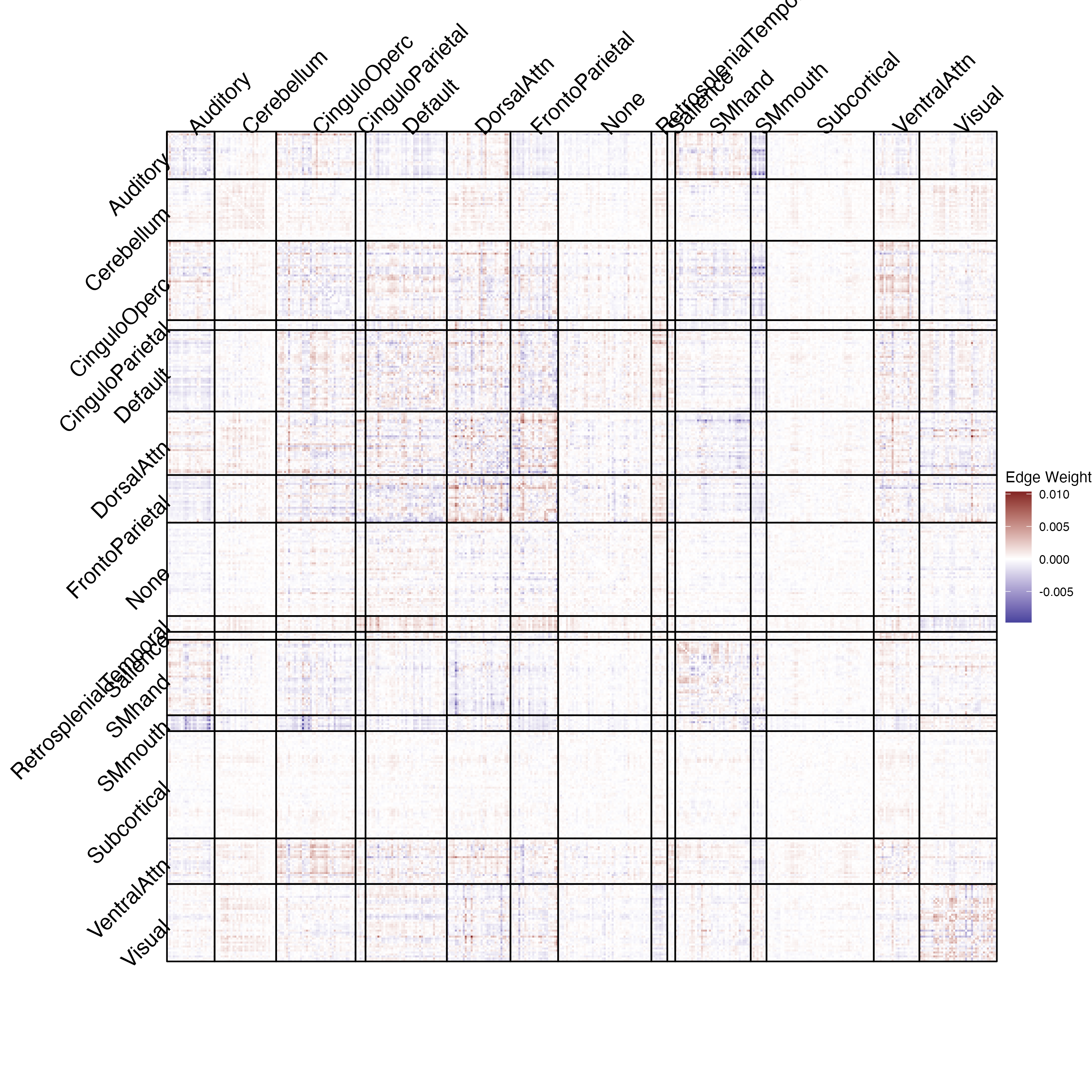}}
  \caption{Brain connectivity features recovered from second canonical direction $\beta_2$ using only significantly nonzero coordinates.}
  \label{fig:com:betacheck-2-sig}
\end{figure}

Figure \ref{fig:com:abcd-beta-inf} shows confidence intervals for the first three directions of $B$.
Notably, the first direction has a number of coordinates that are significantly nonzero,
whereas there is just one for the second direction and none for the third direction.
This parallels our findings in Simulation III (Section \ref{sec:combootcca-sim-data-based}),
wherein we saw that combootcca had non-trivial power for the leading direction of $B$
but little power for the second direction,
perhaps due to the poor separation of the second canonical correlation from subsequent canonical correlations.
Recall that the coordinates of $B$ correspond to PCA scores from the higher-dimensional brain imaging data:
while they are not directly interpretable, it is possible to invert the data reduction step.
If we wish to recover the canonical directions in the original feature space, we can compute
$\check{B} = V_{\left[ 1:250 \right]} D^{-1/2} \hat{B}$,
where $D$ is a diagonal matrix with the empirical variances of $X_2$
(the adjustment by $D$ is necessary to invert the standardization we applied as a preprocessing step).
Alternately, we can set to $0$ any entries of $\hat{B}$ whose corresponding confidence intervals included 0,
and construct an analogous quantity with this thresholded version of $\hat{B}$.
These reconstructed directions can be reshaped into matrix form and organized according to the assignment of parcels to putative brain systems in the parcellation of \citet{gordon2016GenerationEvaluationCortical}.
Figures \ref{fig:com:betacheck-1-all}, \ref{fig:com:betacheck-2-all}, and \ref{fig:com:betacheck-3-all}
depict $\check{\beta}_1$, $\check{\beta}_2$, and $\check{\beta}_3$, respectively,
whereas Figures \ref{fig:com:betacheck-1-sig} and \ref{fig:com:betacheck-2-sig}
show the analogous quantities obtained after thresholding $\hat{\beta}_1$ and $\hat{\beta}_2$
(there are no significantly non-zero coordinates in $\hat{\beta}_3$, so we do not depict it).
Of particular interest given the tasks associated with the first direction,
we note that when examining the brain features associated with $\hat{\beta}_1$ in Figures \ref{fig:com:betacheck-1-all} and \ref{fig:com:betacheck-1-sig},
there qualitatively appears to be far more mass in edges linking the Default, FrontoParietal, Dorsal Attention, Salience, Cingulo-Opercular, Cingulo-Parietal, and Ventral Attention systems.
These systems were situated by \citet{margulies2016SituatingDefaultmodeNetwork} near the beginning of a gradient that transitions from transmodal to sensory cortices.
This pattern, coupled with the behavioral variables implicated in $\gamma_1$,
suggest that our leading pair of canonical directions may indeed reflect shared structure in both brain and behavior that undergird important and general cognitive ability.

\section{Discussion}
\label{sec:combootcca-discussion}

In this work, we have considered the problem of inference for the directions obtained from CCA.
While much statistical work has focused on inference for the canonical correlations,
the canonical directions in the classical setting have received less treatment,
except for \citet{anderson1999AsymptoticTheoryCanonical}'s treatment of the $p = q$ fixed, $N \rightarrow \infty$ case.
In the absence of clear guidance from the statistical literature,
practitioners often either ignore this part of the analysis, or use a variety of heuristic resampling methods.  Recent methodological work from applied groups
\citep[e.g.,][]{helmer2020StabilityCanonicalCorrelation,mcintosh2021ComparisonCanonicalCorrelation}
has focused on characterizing the ``stability'' or ``reproducibility'' of the canonical directions.
While this is informative, it is not the same as performing statistical inference on the canonical directions,
and often involves arriving at global conclusions (e.g., ``this vector is unstable'')
based on the angles between canonical directions under resampling
as opposed to local inference (e.g., ``this coordinate is significantly nonzero'').
While this line of work has generally approached the issue numerically,
very recent work in the statistical literature \citep{bykhovskaya2023HighdimensionalCanonicalCorrelation} has provided theoretical results in a similar vein in the setting where $p, q$, and $N$ all grow together.
That framework, too, still has a global rather than local focus:
in the example application,
the authors
obtain an estimate of the angle between estimated canonical variates and true canonical variates,
remark that it is low,
and then present point estimates for the canonical directions.
While these are useful studies, ultimately our goal is to go one step further and use the characterization of stability in order to deliver inference.
While this is indeed what is attempted in practice,
our carefully designed simulation studies,
which range from tightly controlled to highly realistic settings,
allow us to evaluate how well various methods perform with regard to statistical inference with known and non-trivial ground truth.
In addition, in contrast to most applications,
we carefully consider a number of design choices for bootstrap-based approaches and their relative consequences, merits, and pitfalls.
This is especially useful since various applied papers
often arrive at different procedures which
may have consequences for coverage as well as both Type I and Type II error control.
Moreover, our use of a data-based simulation study,
wherein we assess the statistical properties of our procedures on synthetic data designed to closely mimic our eventual application with realistic levels of signal,
is a useful example of how a statistical method can be evaluated prior to its application on a given data set.    

Based on our simulation studies, we specifically recommend the use of percentile-based bootstraps with the weighted Hungarian algorithm alignment as performed in combootcca, which overall delivers the best combination of coverage, error control, and power.
The recent approach of \citet{laha2021StatisticalInferenceHigh},
originally developed for sparse CCA,
seems to be a promising alternative,
although its control of Type I error is not sufficiently consistent for us to recommend its use in low dimensional settings with modest sample sizes,
and its lack of results for canonical directions beyond the first can be an important restriction.

While combootcca has good Type I and II error control,
 the simulation studies show it can fail to achieve nominal coverage of coordinates with large magnitude,
especially when the true direction is sparse.
We conjectured that this was due to bias in the estimator which was exacerbated by the bootstrap,
and we saw empirical evidence for this in Figures
\ref{fig:sim-sprec-i-bias},
\ref{fig:sim-sprec-ii-bias1},
\ref{fig:sim-sprec-ii-bias2}, and
\ref{fig:sim-iii-bias}.
It may be possible to mitigate this bias by estimating and correcting for it.
This could be done with the use of a double bootstrap \citep[p.\ 103]{davison1997BootstrapMethodsTheir} or related procedures such as bias-corrected and accelerated intervals
\citep[$\text{BC}_{\text{a}}$;][p.\ 184]{efron1993IntroductionBootstrap},
although this of course comes at additional computational cost and requires further study.

One noteworthy limitation of the approaches we have presented is that we do not correct for multiple comparisons.
While this is common in the applied literature,
where confidence intervals for the canonical directions are generally considered descriptive and not corrected \citep[e.g.,][]{misic2016NetworklevelStructurefunctionRelationships},
this is an important caveat that should be kept in mind particularly with respect to our findings from the ABCD data.
While control of the family-wise error rate could be achieved with Bonferroni correction,
this may be rather conservative and decrease power, which in some settings is already limited.
An alternative would be to instead control the false discovery rate \citep[FDR;][]{benjamini1995ControllingFalseDiscoverya}.
While the classical procedure is valid for independent tests and those that are positively dependent \citep{benjamini2001ControlFalseDiscovery},
it is not immediately clear what sort of dependence might exist among tests for the elements of $B$ and $\Gamma$.
However, because combootcca already uses a resampling-based procedure,
it may be possible to estimate the dependence and to adjust accordingly, as in \citet{yekutieli1999ResamplingbasedFalseDiscovery}.
We are not selective in terms of which confidence intervals we report for a given direction;
thus we should not suffer from decreased coverage rates due to selective reporting \citep{benjamini2005FalseDiscoveryRate},
but future work may be needed to account for the potentially sequential nature of inference,
wherein a subset of canonical directions are selected for further scrutiny based on hypothesis tests applied to the canonical correlations.
This could be viewed as testing hypotheses on a tree,
and so the approach of \citet{bogomolov2021HypothesesTreeNew} may be applicable.

Future work includes additional theoretical analysis.
Given the use of SVD in CCA as discussed in Section \ref{sec:combootcca-cca-pop-model},
recent results in \citet{agterberg2022EntrywiseEstimationSingular} concerning the limiting distribution of entries of singular vectors may be of use for this analysis.
In addition, our simulation studies suggest that when $p > q$ but $q \ll N$,
the results of \citet{anderson1999AsymptoticTheoryCanonical} may still give the correct limiting distribution for $\hat{\Gamma}$.
This appears to be in line with theoretical results in \citet{fine2003AsymptoticStudyCanonical},
which revisited the work of \citet{anderson1999AsymptoticTheoryCanonical}
from an operator- and tensor-focused perspective,
but appears to offer results for only the lower dimensional direction.
This approach to inference would be of particular utility: the coordinates of the lower-dimensional directions, which often correspond to phenotype, are often of primary interest,
and asymptotic confidence intervals are vastly faster to compute than the bootstrap.
Nonetheless, given the widespread use of resampling-based strategies in the applied literature,
our contribution of combootcca along with demonstration of the pitfalls of related strategies is still valuable.

Another direction for future work involves developing forms of matrix CCA, which would be more closely tailored to our neuroimaging application.
Recall that in our application, $X$ is the matrix of brain connectivity, but the structure of this matrix is ignored when it is vectorized and reduced using PCA as part of preprocessing.   While this transformation is reversible for the purposes of visualization (e.g., Figure \ref{fig:com:betacheck-1-sig}),  our methods do not take advantage of this rich structure.
There are thus opportunities to develop variants of CCA that involve structure-enforcing penalties (e.g., similar to that of \citet{relion2019NetworkClassificationApplications,kessler2022PredictionNetworkCovariates}).

\section*{Acknowledgments}
This research was supported by NSF grants DMS-1521551, DMS-1646108, DMS-1916222, and DMS-2052632 and a grant from the Dana Foundation.  D.~K.~was also supported by a Rackham Predoctoral Fellowship awarded by the University of Michigan. 
We thank Dr.~Chandra Sripada and his research group for providing us with a processed version of the data as well as helpful comments in interpreting our results.
We thank Dr.~Peter Bickel for helpful comments and input at various stages of this project.
This research was supported in part through computational resources and services provided by Advanced Research Computing (ARC), a division of Information and Technology Services (ITS) at the University of Michigan, Ann Arbor.
Data used in the preparation of this article were obtained from the Adolescent Brain Cognitive Development (ABCD) Study (\url{https://abcdstudy.org}), held in the NIMH Data Archive (NDA). This is a multisite, longitudinal study designed to recruit more than 10,000 children age 9-10 and follow them over 10 years into early adulthood. The ABCD Study® is supported by the National Institutes of Health and additional federal partners under award numbers U01DA041048, U01DA050989, U01DA051016, U01DA041022, U01DA051018, U01DA051037, U01DA050987, U01DA041174, U01DA041106, U01DA041117, U01DA041028, U01DA041134, U01DA050988, U01DA051039, U01DA041156, U01DA041025, U01DA041120, U01DA051038, U01DA041148, U01DA041093, U01DA041089, U24DA041123, U24DA041147. A full list of supporters is available at \url{https://abcdstudy.org/federal-partners.html}. A listing of participating sites and a complete listing of the study investigators can be found at \url{https://abcdstudy.org/consortium_members/}. ABCD consortium investigators designed and implemented the study and/or provided data but did not necessarily participate in the analysis or writing of this report. This manuscript reflects the views of the authors and may not reflect the opinions or views of the NIH or ABCD consortium investigators.
The ABCD data repository grows and changes over time. The ABCD data used in this report came from NDA Study 721, DOI: \href{http://dx.doi.org/10.15154/1504041}{10.15154/1504041}, which can be found at \url{https://nda.nih.gov/study.html?id=721}.
We use the same data as NDA Study 1364, DOI: \href{http://dx.doi.org/10.15154/1523385}{10.15154/1523385}.
The specific NDA study associated with our report is NDA Study 2278, DOI: \href{http://dx.doi.org/10.15154/pv4c-kb39}{10.15154/pv4c-kb39}.

\bibliographystyle{apalike}

\bibliography{references}

\end{document}